\documentclass[aps,prd,onecolumn,notitlepage,eqsecnum,nofootinbib,floatfix]{revtex4-2}
\usepackage{amssymb}
\usepackage{amsmath}
\usepackage{amsfonts} 
\usepackage{graphicx}
\usepackage{hyperref}
\DeclareSymbolFont{EulerScript}{U}{eus}{m}{n}
\SetSymbolFont{EulerScript}{bold}{U}{eus}{b}{n}
\DeclareSymbolFontAlphabet\scrpt{EulerScript}
\newcommand{\FF}{{\scrpt F}}
\newcommand{\PP}{{\scrpt P}} 
\newcommand{\SSS}{{\scrpt S}}
\allowdisplaybreaks
\begin{document}
\title{Self-gravitating thin shells are dynamically unstable on all angular scales}
\author{Tristan Pitre}
\email{tpitre@uoguelph.ca}
\author{Berend Schneider}
\email{berend@uoguelph.ca}
\author{Eric Poisson}
\email{epoisson@uoguelph.ca}
\affiliation{Department of Physics, University of Guelph, Guelph, Ontario, N1G 2W1, Canada}
\date{\today}

\begin{abstract}
We establish the dynamical instability of a static, spherically symmetric, and infinitesimally thin shell in general relativity. The shell is made up of a perfect fluid with a barotropic equation of state, and it produces a Schwarzschild spacetime in its exterior and a Minkowski spacetime in its interior. We introduce a linear perturbation of the matter and spacetime, decompose it in spherical harmonics, and compute the shell's spectrum of quasinormal modes. We reveal the existence of two modes with a purely imaginary frequency, one negative (which describes stable oscillations), the other positive (which describes an exponential growth); these modes occur for all sampled values of the shell's compactness and adiabatic index, and all sampled values of the multipolar order $\ell \geq 2$, in the even-parity sector of the perturbation. All other quasinormal modes describe damped oscillations and do not contribute to the instability. This study complements a recent analysis by Yang, Bonga, and Pen [Phys.\ Rev.\ Lett.\ {\bf 130}, 011402 (2023)], which also concluded in a dynamical instability, but was limited by an eikonal approximation to small angular scales ($\ell \gg 1$); our treatment applies to all angular scales. The eigenvalue problem for the mode frequencies is formulated by introducing a perturbation of Minkowski spacetime inside the shell, a perturbation of Schwarzschild spacetime outside the shell, and a perturbation of the shell matter. The metric perturbations are governed by the Einstein field equations, and they are matched across the shell with the help of Israel's junction conditions. The matter perturbation is governed by the equations of fluid mechanics, and it produces a source term in the junction conditions. All calculations are carried out in full general relativity, but we also examine a nonrelativistic formulation of the problem; we show that a Newtonian shell also is necessarily unstable to a time-dependent perturbation. Our conclusion suggests that a compact object that features a thin shell at its surface will be dynamically unstable; this makes it nonviable as a model of black-hole mimicker.      
\end{abstract}

\maketitle

\section{Introduction and summary}
\label{sec: intro}

\subsection{Self-gravitating thin shells in general relativity}

Observations of compact objects through gravitational waves, horizon-scale imaging, and observations of galactic nuclei have firmly established black holes as physical constituents of the Universe \cite{LIGO:first_BH, GWTC1, GWTC2, GWTC3, GWTC4, EHT-proposition, EHT-first, Gal_nuclei}. With current observations probing increasingly subtle features of black-hole dynamics and theoretical models reaching similar levels of precision, black holes provide a unique laboratory to test the strong-field regime of general relativity and the nature of spacetime. 

Alongside this progress, a variety of alternative compact objects have been proposed that reproduce many observational signatures of black holes while avoiding features such as an event horizon or a spacetime singularity. Such objects are often referred to as black-hole mimickers and are consistent with current observational constraints \cite{mimickers_1, mimickers_2, mimickers_3}. Any measurement (e.g. the observation of a quasinormal mode \cite{QNMs_Yang, QNMs_Berti}) that would favor the existence of mimickers over traditional black holes would cause a major upheaval in the field of gravitational physics. At present, black-hole mimickers cannot be ruled out; the next generation of gravitational-wave detectors might be able to do so \cite{Cosmic_E, Einstein_T, LISA}.  

One class of black-hole mimickers are gravastars, which were introduced in 2001 by Mazur and Mottola \cite{Gravastars_porposition}. They consist of an infinitesimally thin shell of matter lying just outside of where the event horizon would be, with an interior typically taken to be a de Sitter spacetime; the exterior spacetime is still described by the Schwarzschild metric. Since their introduction, substantial effort has been devoted to understanding these objects and determine how they could be distinguished from black holes in gravitational-wave measurements \cite{QNMs_Cardoso, mimickers_3}. So far, LIGO's observations have either found gravastars to be inconsistent with the data, or that available data does not favor them over black holes \cite{QNMs_Cardoso, Gravastar_meas1}.  

Recently, Yang, Bonga and Pen demonstrated that such self-gravitating thin shells are dynamically unstable under nonradial perturbations \cite{Yang_paper}. Their conclusion relies on an analysis of a shell's deformation in the eikonal limit, which corresponds to very short angular scales --- in terms of a decomposition of the perturbation in spherical harmonics, this is the $\ell \gg 1$ regime. They find that if the shell has a positive pressure, then the perturbations grows exponentially, so that the shell is unstable. An application to the specific cases of a gravastar and a thin-shell wormhole reveals that both are dynamically unstable: modes with small wavelengths grow exponentially. Models of black-hole mimickers that feature an unstable thin shell are nonviable as astrophysical objects. 

In this paper we generalize the approach of Yang, Bonga, and Pen by examining the stability of the shell on all angular scales. To achieve this, we rely on the fact that the object's dynamical instability is entirely associated with the shell itself; the nature of the interior does not matter. Exploiting this observation, we make the simplest choice of an interior: a flat spacetime. Our object is therefore a spherical thin shell that encloses a portion of Minkowski spacetime, with an exterior described by the Schwarzschild spacetime. We find that this object possesses unstable modes at all angular scales; in a decomposition of spherical harmonics, this is the entire $\ell \geq 2$ interval. This result appears to be in disagreement with those of Pani et al.~\cite{Pani_paper}, who carried out a similar study, but chose the interior to consist of a portion of de Sitter space; they did not find unstable modes. In view of our belief that the nature of the interior is immaterial to the dynamical instability, we are tempted to conclude that unstable modes should have been identified in their work.   

We summarize our results in more detail in the remaining subsections of this introduction. We conclude with a description of the paper's structure. 

\subsection{Unperturbed configuration}

We begin our discussion with an unperturbed, static and spherically symmetric, self-gravitating thin shell of mass $M$ and radius $R$ in general relativity (Secs.~\ref{sec:matter} and \ref{sec:unperturbed}). We take the shell to be infinitesimally thin, so that it traces a timelike hypersurface in spacetime. We take the matter constituting the shell to be a perfect fluid of areal mass density $\sigma$, areal energy density $\mu$, surface pressure $p$, and velocity field $u^a$ tangent to the hypersurface. We take the matter to satisfy a polytropic equation of state, 
\begin{equation}
    p = K\sigma^\Gamma, \qquad \mu = \sigma + \frac{p}{\Gamma-1}, 
\end{equation}
where $\Gamma$ is a constant adiabatic index and $K$ is a constant. 

The set of equations that govern the physics of the surface fluid consists of the continuity equation (conservation of mass), the relativistic Euler equation (conservation of momentum), the first law of thermodynamics (conservation of energy), and the Israel junctions condition \cite{israel:66}, which describe the discontinuity of the gravitational field across the shell. As specified previously, the spacetime is described by the Minkowski metric inside the shell, and by the Schwarzschild metric outside the shell. The governing equations supply the relations
\begin{equation}
    M = \frac{4p^2}{\pi\mu^3}\frac{1 + 2p/\mu}{(1+4p/\mu)^3}, \qquad R = \frac{p}{\pi\mu^2}\frac{1}{1+4p/\mu}
\end{equation}
for the mass and radius in relation to the pressure and energy density. The equations describe a sequence of equilibrium configurations parametrized by $\sigma$, with the equations of state determining $p$ and $\mu$. 

\subsection{Perturbation of the system: Theory}

To uncover the shell's spectrum of quasinormal modes, we introduce a small perturbation to the system, which is taken to oscillate in time with a frequency $\omega$ (Secs.~\ref{sec:even} and \ref{sec:odd}). The metric $g_{\alpha\beta}$ inside and outside the shell is written as
\begin{equation}
    g_{\alpha\beta} = \bar{g}_{\alpha\beta} + \gamma_{\alpha\beta},
\end{equation}
where $\bar{g}_{\alpha\beta}$ represents the unperturbed metric and $\gamma_{\alpha\beta}$ denotes the perturbation, which is taken to be regular at $r=0$ and to describe outgoing waves at $r=\infty$. The Einstein field equations are linearized in $\gamma_{\alpha\beta}$, which is decomposed into even-parity (polar) and odd-parity (axial) pieces \cite{BH_pert_theory}, 
\begin{equation}
    \gamma_{\alpha\beta} = \gamma_{\alpha\beta}^{\mathrm{even}} + \gamma_{\alpha\beta}^{\mathrm{odd}}. 
\end{equation}
The spectrum of quasinormal modes is therefore split into two categories: even-parity and odd-parity modes. It is well known that for a Schwarzschild black hole, the two classes of modes share the same spectrum of frequencies \cite{BH_iso}. As we shall see, this is not so for the shell spacetime.  

The perturbation also includes a deformation of the shell, so that the matter variables become
\begin{equation}
p \xrightarrow[]{} p + \delta p, \qquad
\mu \xrightarrow[]{} \mu + \delta \mu, \qquad
u^a \xrightarrow[]{} u^a + \delta u^a,
\end{equation}
with $\delta p$, $\delta\mu$, and $\delta u^a$ denoting the perturbation of the pressure, energy density, and velocity field, respectively. These also are taken to oscillate with a frequency $\omega$ and decomposed in even-parity and odd-parity pieces, and all calculations are linearized with respect to the perturbations. The equations of state provide a relation between $\delta p$ and $\delta \mu$.

\begin{figure}
\centering
\begin{minipage}{.5\textwidth}
  \centering
  \includegraphics[scale=0.5]{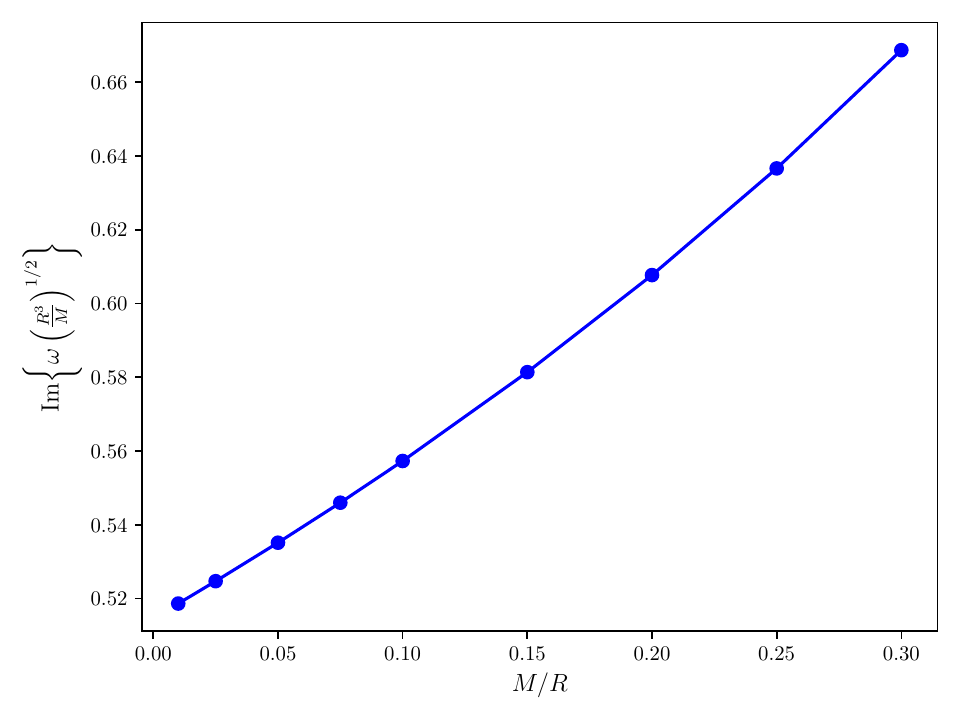}
\end{minipage}%
\begin{minipage}{.5\textwidth}
  \centering
  \includegraphics[scale=0.5]{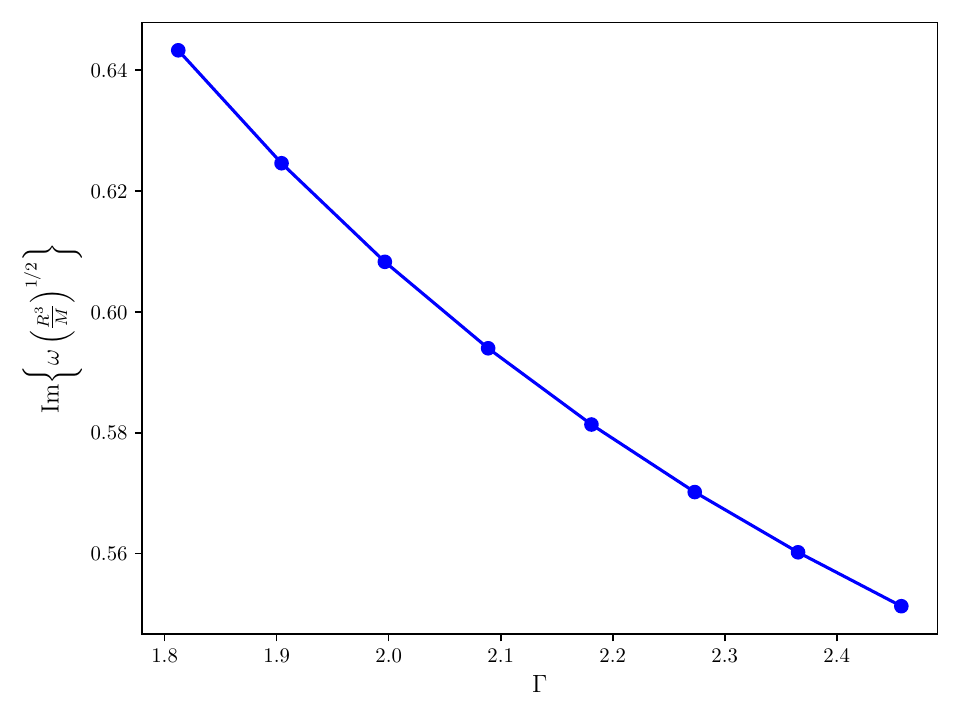}
\end{minipage}
\caption{Unstable even-parity matter mode: Mode frequency $\omega(R^3/M)^{1/2}$ for $\ell=2$. Left: Imaginary part of the frequency in relation to $M/R$ for $\Gamma = 2$. Right: Imaginary part of the frequency in relation to $\Gamma$ for $M/R = 0.2$. The real part of the frequency vanishes for this mode. The dots represent our numerical data, and they are linked by guiding lines.}
\label{fig:matter_imag, L=2, matter-modes even}
\end{figure}

\begin{figure}
\begin{minipage}[h]{0.47\linewidth}
\begin{center}
\includegraphics[width=1\linewidth]{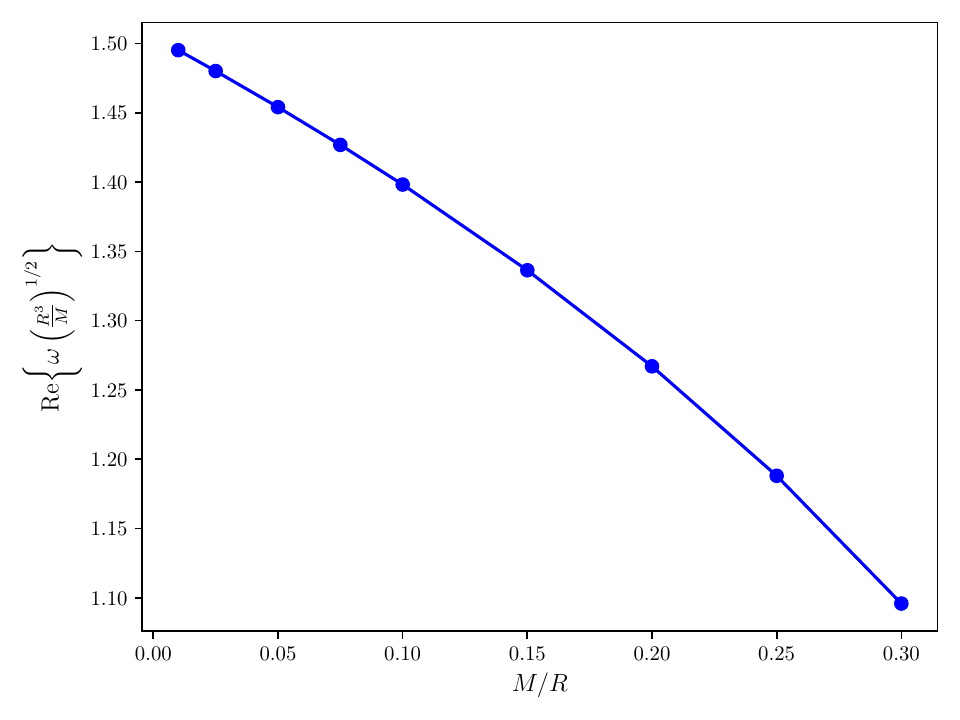} 
\end{center} 
\end{minipage}
\hfill
\begin{minipage}[h]{0.47\linewidth}
\begin{center}
\includegraphics[width=1\linewidth]{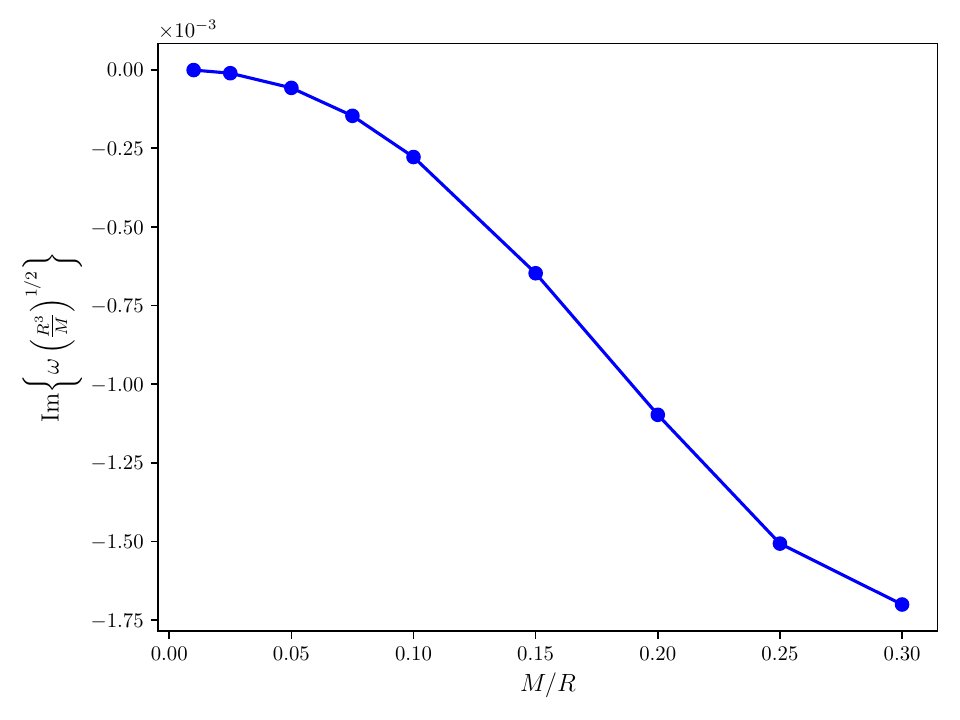} 
\end{center}
\end{minipage}
\vfill
\begin{minipage}[h]{0.47\linewidth}
\begin{center}
\includegraphics[width=1\linewidth]{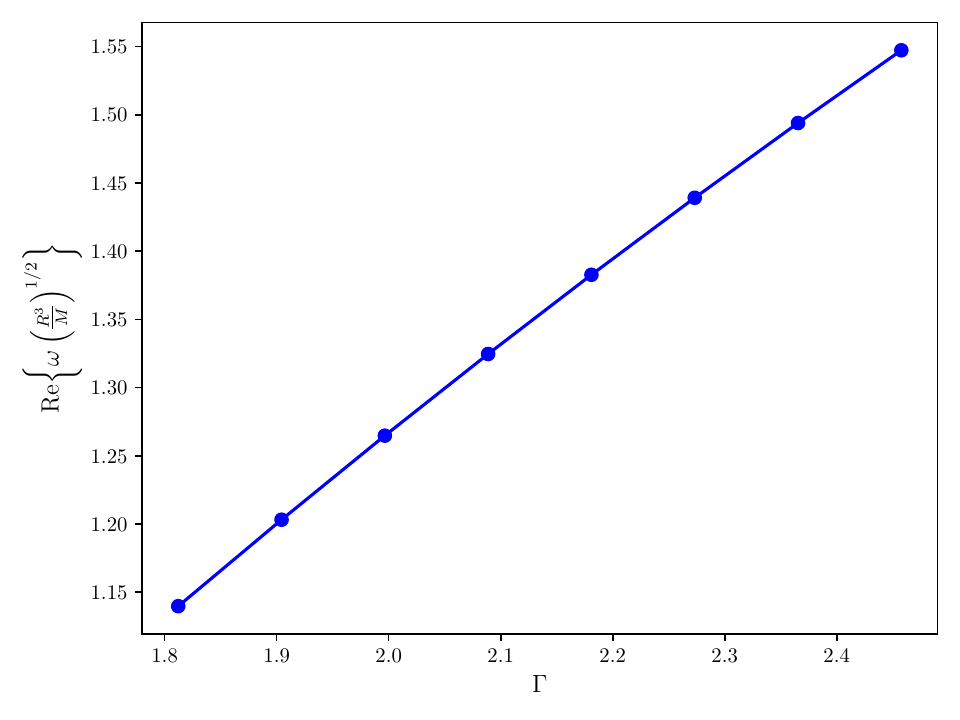} 
\end{center}
\end{minipage}
\hfill
\begin{minipage}[h]{0.47\linewidth}
\begin{center}
\includegraphics[width=1\linewidth]{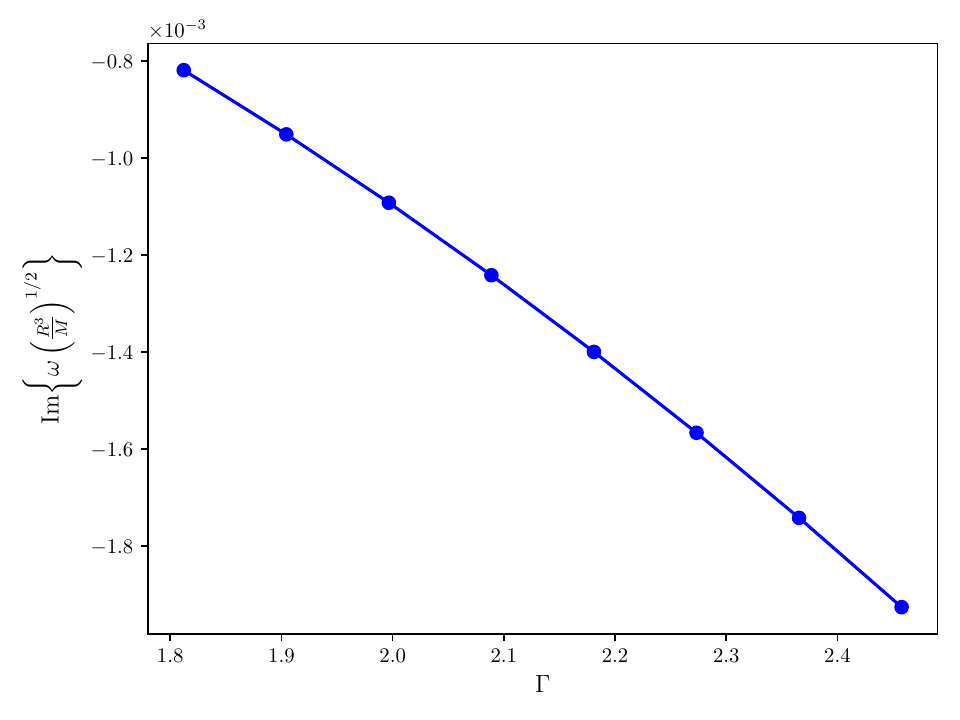} 
\end{center}
\end{minipage}
\caption{Stable pair of even-parity matter modes: Mode frequencies $\omega(R^3/M)^{1/2}$ for $\ell=2$. Upper left: Real part of the frequency in relation to $M/R$. Upper right: Imaginary part of the frequency in relation to $M/R$. Lower left: Real part of the frequency in relation to $\Gamma$. Lower right: Imaginary part of the frequency in relation to $\Gamma$. Upper plots: $\Gamma$ = 2. Lower plots: $M/R = 0.2$. The dots are linked by guiding lines.}
\label{fig:matter_mixed vs MR, L=2, matter-modes even}
\end{figure}

Throughout this paper we choose to work with the outgoing Eddington-Finkelstein coordinate system $[u,r,\theta,\phi]$, where the retarded-time coordinate $u$ is related to the standard Schwarzschild time $t$ by $u = t - r^*$, with $r^* := r + 2M\ln(r/2M-1)$ denoting the familiar tortoise coordinate. This choice allows us to impose outgoing-wave boundary conditions in the most direct and convenient way, without encountering divergences that typically occur when working with the usual time coordinate \cite{Hans-Peter}. A similar strategy is often used for black holes, by exploiting hyperboloidal slices \cite{hyper_slices_1, hyper_slices_2, hyper_slices_3}, which are asymptotically null. Our null, $u = \text{constant}$ slices would not work for a black hole, because of the requirement to impose ingoing-wave boundary conditions at the horizon, but they work beautifully for a material body. To eliminate the gauge freedom associated with the description of metric perturbations, we impose the Regge-Wheeler gauge conditions on $\gamma_{\alpha\beta}$ \cite{regge-wheeler:57, martel-poisson:05}. With our choice of coordinates, the perturbation is made proportional to $e^{-i\omega u}$. The perturbation equations that govern the shell and the spacetime form a closed system of equations that gives rise to an eigenvalue problem for $\omega$. The solutions are the quasinormal-mode frequencies of the shell.

The even-parity and odd-parity sectors of the perturbation decouple, and they can be treated separately. In the even-parity sector, the metric perturbations inside and outside the shell are described in terms of a master function $\psi$ that satisfies the Regge-Wheeler equation \cite{mukkamala-perenigues:25, poisson:25} instead of the Zerilli equation. The functions $\psi_{\rm in}$ and $\psi_{\rm out}$ (and their first derivative) are connected to each other by making use of the Israel junction conditions, which also implicates the matter variables; this gives rise to the eigenvalue problem for the mode frequencies. The solutions depend on the multipolar order $\ell$, the adiabatic index $\Gamma$, and the shell's compactness $M/R$. 

The treatment of the odd-parity sector is very similar. Here also the metric perturbation is encoded within a master variables $\psi$ that satisfies the Regge-Wheeler equation, and the junction conditions at the shell produce an eigenvalue problem for $\omega$. Because there is no independent matter variable in the odd-parity sector of the perturbation, the eigenvalues no longer depend on the adiabatic index $\Gamma$.   

The techniques that we employ to integrate the Regge-Wheeler equation are described in Sec.~\ref{sec:RWintegration}. We use a combination of numerical methods, as well as an analytical approach that can be applied inside the shell (where the background spacetime is flat), or when the magnitude of $2M\omega$ is small.  

\subsection{Quasinormal modes of a thin shell}
\label{sec: intro qnms}

\begin{figure}
\begin{minipage}[h]{0.47\linewidth}
\begin{center}
\includegraphics[width=1\linewidth]{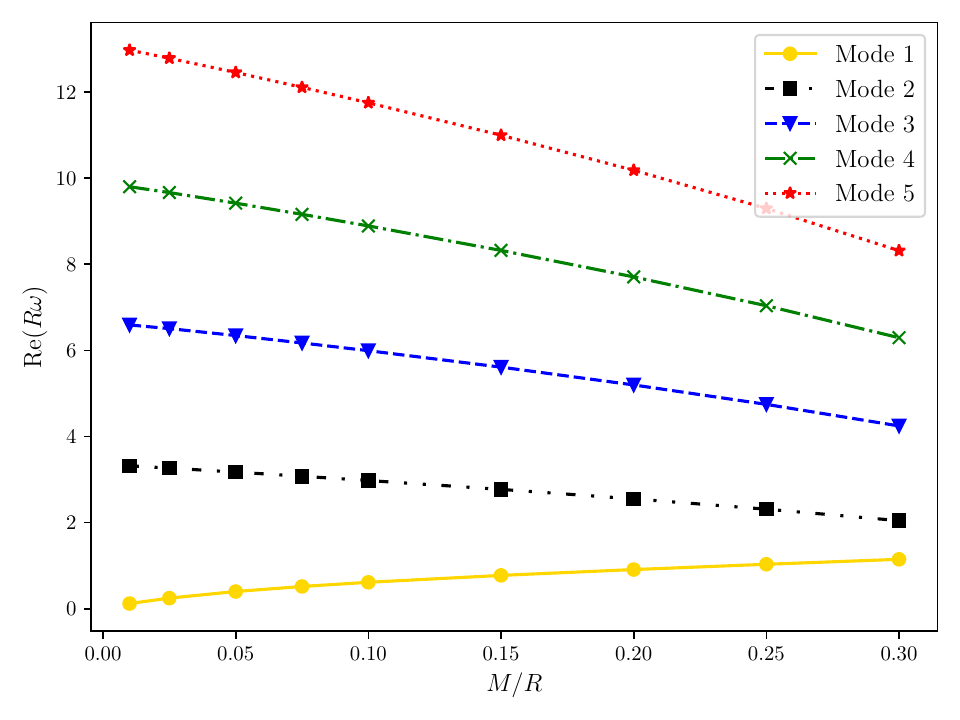} 
\end{center} 
\end{minipage}
\hfill
\begin{minipage}[h]{0.47\linewidth}
\begin{center}
\includegraphics[width=1\linewidth]{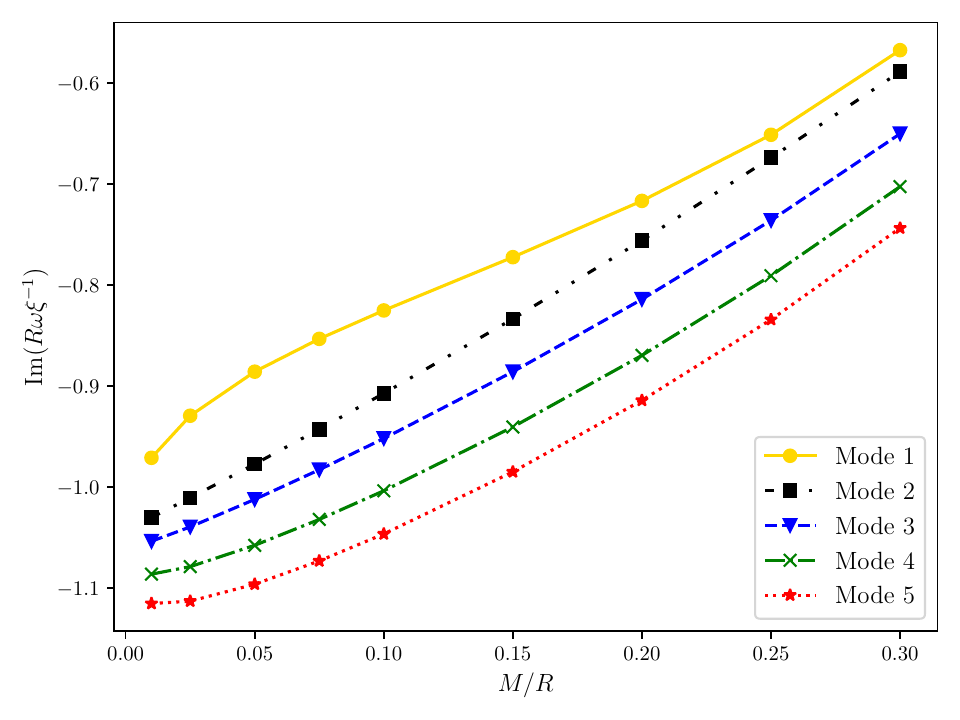} 
\end{center}
\end{minipage}
\vfill
\begin{minipage}[h]{0.47\linewidth}
\begin{center}
\includegraphics[width=1\linewidth]{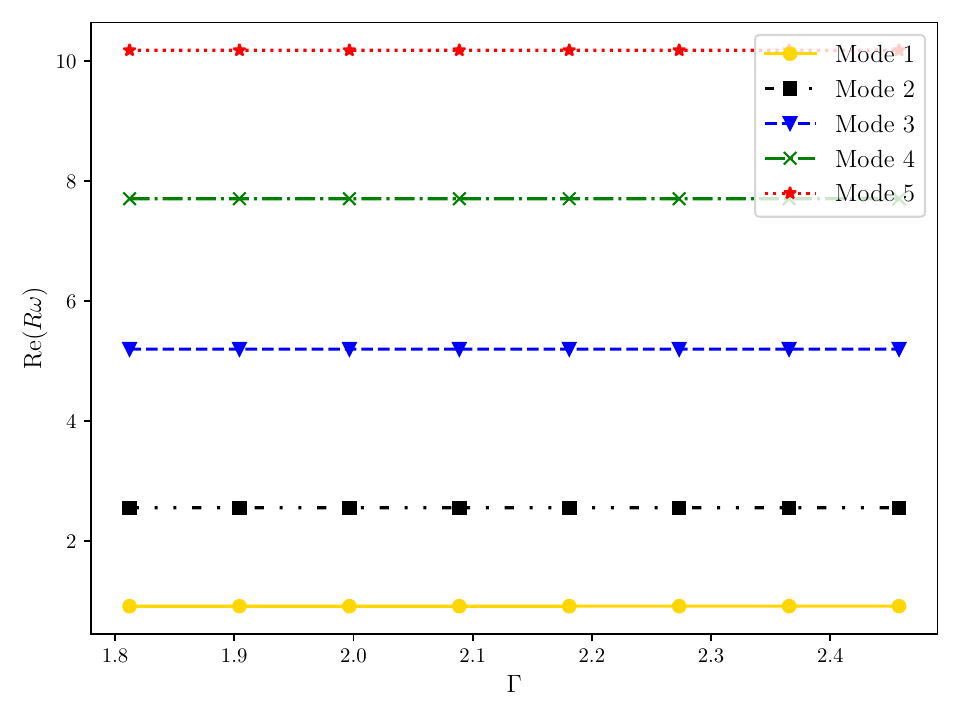} 
\end{center}
\end{minipage}
\hfill
\begin{minipage}[h]{0.47\linewidth}
\begin{center}
\includegraphics[width=1\linewidth]{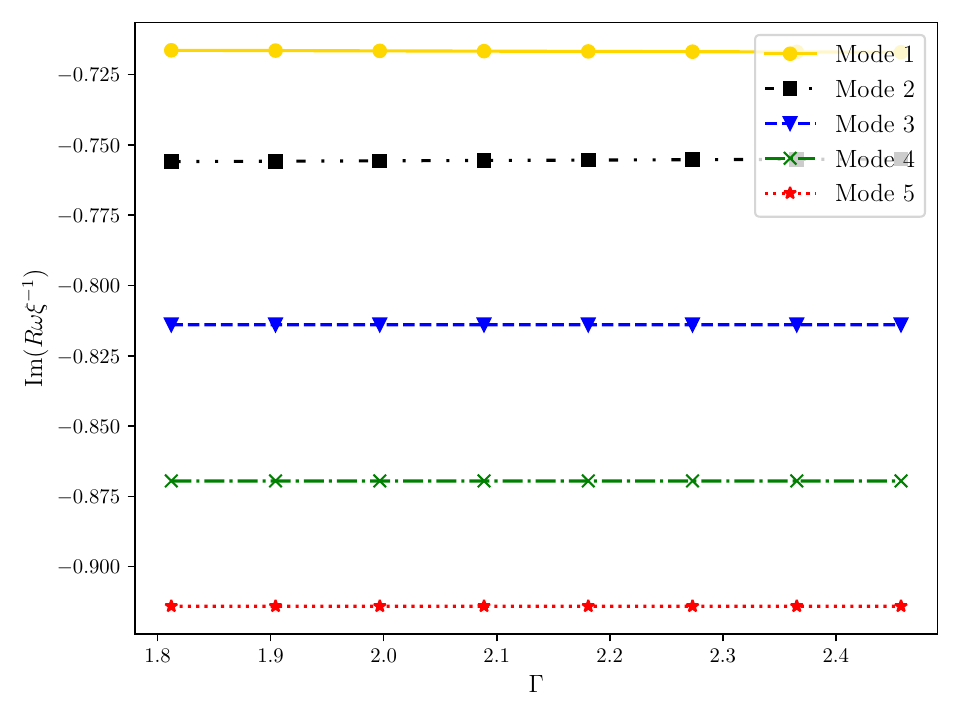} 
\end{center}
\end{minipage}
\caption{Even-parity wave modes: Mode frequencies for $\ell=2$. Upper left: Real part of $R\omega$ in relation to $M/R$. Upper right: Imaginary part of $R\omega/\xi$ in relation to $M/R$, with the factor $\xi := \ln[2\ell(\ell+1)R/M]/2$ introduced in Sec.~\ref{subsec: newt_regime}. The frequencies are computed for $\Gamma=2$. Lower left: Real part of $R\omega$ in relation to $\Gamma$. Lower right: Imaginary part of $R\omega/xi$ in relation to $\Gamma$. The frequencies are computed for $M/R = 0.2$. Yellow dots: fundamental mode. Black dots: first overtone. Blue dots: second overtone. Green dots: third overtone. Red dots: fourth overtone. The dots are linked by guiding lines.}
\label{fig:Rw vs MR, L=2, w-modes even}
\end{figure}

\begin{figure}
\centering
\begin{minipage}{.5\textwidth}
  \centering
  \includegraphics[scale=0.5]{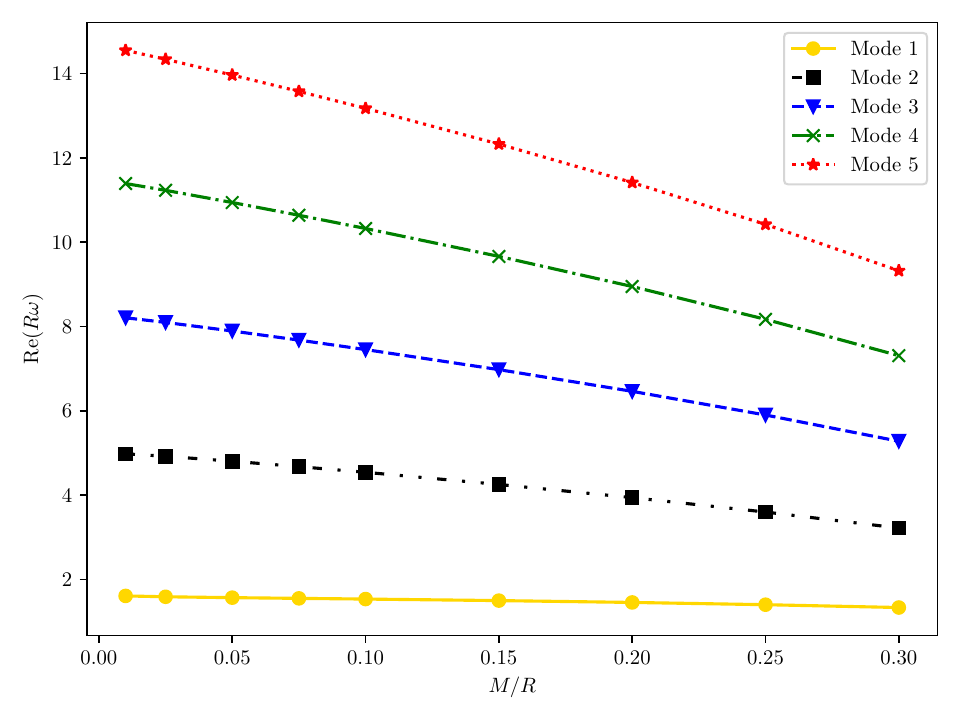}
\end{minipage}%
\begin{minipage}{.5\textwidth}
  \centering
  \includegraphics[scale=0.5]{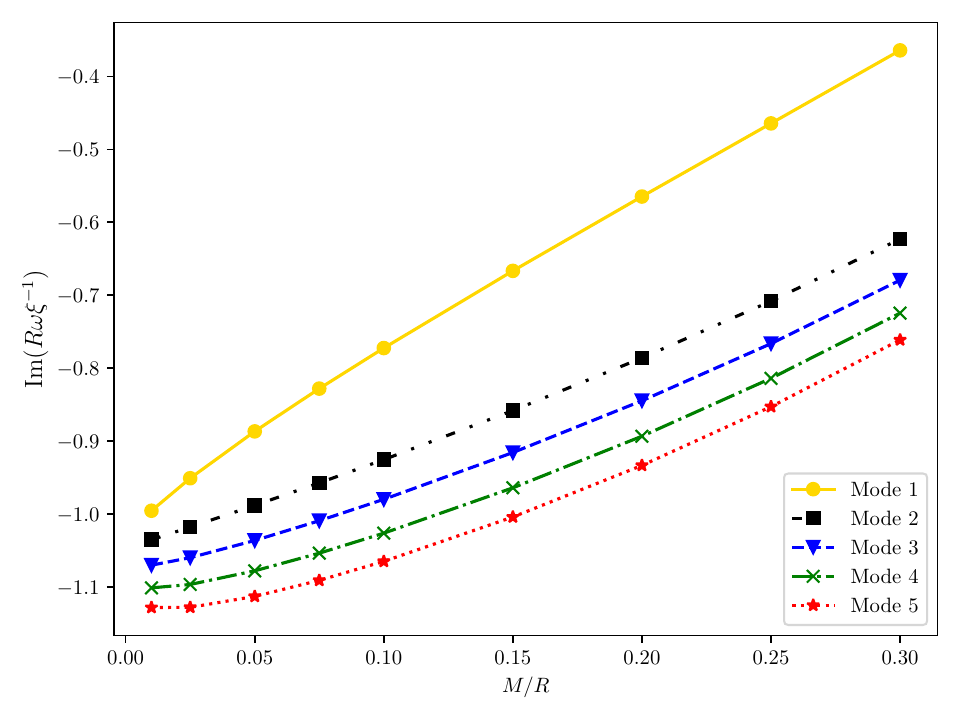}
\end{minipage}
\caption{Odd-parity wave modes: Mode frequencies for $\ell=2$. Left: Real part of $R\omega$ in relation to $M/R$. Right: Imaginary part of $R\omega/xi$ in relation to $M/R$, with $\xi := \ln[2\ell(\ell+1)R/M]/2$. Yellow dots: fundamental mode. Black dots: first overtone. Blue dots: second overtone. Green dots: third overtone. Red dots: fourth overtone. The dots are linked by guiding lines.}
\label{fig:Rw vs Gamma, L=2, w-modes odd}
\end{figure}

In order to obtain the mode frequencies (Sec.~\ref{sec:results}), we integrate the Regge-Wheeler equation numerically for the master variables $\psi_{\rm in}$ and $\psi_{\rm out}$, incorporate the appropriate boundary conditions at $r=0$ and $r=\infty$, and impose all but one of the matching conditions at the shell. This is done for every trial frequency that is sampled by the root-searching algorithm (the samples are made in the complex plane), which aims to impose the last remaining junction condition; a successful search returns one of the mode frequencies. 

We find two classes of modes in the even-parity sector. The first is named {\it matter modes}, and these possess a frequency that scales predominantly as $\omega \sim (M/R^3)^{1/2}$; there are precisely four matter modes for each value of $\ell$, and these describe oscillations of the shell. The second class is termed {\it wave modes}, and these possess a frequency that scales predominantly as $\omega \sim 1/R$; there is an infinity of wave modes, which are essentially oscillations of the spacetime metric. Only wave modes occur in the odd-parity sector of the perturbation. 

For our purposes here, the most interesting mode is a matter mode with a frequency that is purely imaginary and positive. This describes a perturbation that grows exponentially with retarded-time $u$, making the shell dynamically unstable. The mode frequency is shown in Fig.~\ref{fig:matter_imag, L=2, matter-modes even} as a function of the compactness $M/R$ and the adiabatic index $\Gamma$, for the specific case of $\ell=2$. An unstable mode exists also for $\ell=3$, as discussed in Sec.~\ref{sec:results}, and for all other values of $\ell$ that we could sample in our numerical exploration. The instability documented in Ref.~\cite{QNMs_Yang} is therefore not limited to small angular scales, but occurs on all angular scales.

The unstable mode comes with a companion that possesses a frequency that is also purely imaginary, but negative; it is the complex conjugate of the first mode. This mode decays exponentially with advanced-time $u$ and does not participate in the dynamical instability. 

We also identify an additional pair of matter modes, as shown in Fig.~\ref{fig:matter_mixed vs MR, L=2, matter-modes even} for $\ell = 2$. These modes have a complex frequency, with equal and opposite real parts, and a common imaginary part that is negative. The pair describes damped oscillations of the shell, and it also does not participate in the dynamical instability. 

A sample of wave modes is displayed in Figs.~\ref{fig:Rw vs MR, L=2, w-modes even} and \ref{fig:Rw vs Gamma, L=2, w-modes odd}, for the even-parity and odd-parity sectors, respectively. In both cases we have that $\ell = 2$, and the figures feature the fundamental mode and the first four overtones. These modes are all stable. The figures reveal that in the case of even-parity modes, the frequencies are almost independent of the adiabatic index $\Gamma$; there is strictly no dependence on $\Gamma$ in the odd-parity sector. All these modes come in pairs, with frequencies that have identical imaginary parts and opposite real parts.   

\subsection{Tidal constants}
\label{subsec:tidal_const} 

To complete our study of a perturbed thin shell in general relativity, we compute the tidal constants $k_\ell^{\rm even}$ and $k_\ell^{\rm odd}$ (also known as Love numbers) that characterize the deformation of a compact object when it is subjected to tidal forces (Secs.~\ref{sec:tidal_even} and \ref{sec:tidal_odd}). Our main finding is that the even-parity constants are {\it negative} for all values of $\ell$, $\Gamma$, and $M/R$. The connection between a negative tidal constant and the dynamical instability would be interesting to explore further. Such a connection is immediate in Newtonian mechanics, for a three-dimensional fluid body. In this context, the tidal constant can be given a representation as an infinite sum over normal modes, given schematically by (see, for example, Ref.~\cite{pitre-poisson:24})
\begin{equation}
k_\ell = \sum_n \frac{1}{\omega_n^2} \bigl[ (\mbox{overlap})_n \bigr]^2,
\end{equation}
where $n$ is a mode label, $\omega_n$ is the mode frequency, and the overlap is an integral of the tidal force density multiplied by the mode function. The representation makes it clear that a negative $k_\ell$ requires the existence of at least one unstable mode with $\omega_n^2 < 0$. Our results for $k_\ell^{\rm even}$ suggests that a similar connection might exist in general relativity.

\subsection{Newtonian shell}
\label{subsec:newt_shell} 

We bring the story to a close by examining the tidal deformation and dynamical instability of a thin shell in Newtonian mechanics (Sec.~\ref{sec:newton}). We verify the existence of four matter modes for each value of $\ell$, including one with a purely imaginary and positive frequency. The Newtonian shell also is dynamically unstable.    

\subsection{Structure of the paper}

The following sections provide a detailed account of the calculations underlying the results summarized previously. We begin with a description of the matter constituting the thin shell in Sec. \ref{sec:matter}, and we describe the shell's unperturbed state in Sec.~\ref{sec:unperturbed}. In Secs. \ref{sec:even} and \ref{sec:odd} we introduce the even-parity and odd-parity sectors of the perturbation, respectively. In Sec. \ref{sec:RWintegration} we detail the numerical and analytical methods that we employed to solve the Regge-Wheeler equation, which governs the master variable $\psi$ that encodes the metric perturbation in both sectors. Additional results for the matter and wave modes are presented in Sec.~\ref{sec:results}, and we obtain analytical approximations in the regime of small compactness, $M/R \ll 1$. We compute the shell's even-parity and odd-parity tidal constants in Secs.~\ref{sec:tidal_even} and \ref{sec:tidal_odd}, respectively. And finally, in Sec.~\ref{sec:newton} we carry out these computations in a Newtonian setting, and show that this simple model successfully captures the salient points of our relativistic calculations. Some technical developments are relegated to Appendices \ref{app:coefficients} and \ref{sec:asymptotics}.  
\section{Shell matter}
\label{sec:matter}

The thin shell traces out a timelike hypersurface $\Sigma$ in a four-dimensional spacetime. The hypersurface is described by embedding relations $x^\alpha = X^\alpha(y^a)$, in which $y^a$ are coordinates intrinsic to
$\Sigma$. The matter that makes up the shell is taken to be a perfect fluid with areal mass density $\sigma$ (the number density of fluid particles multiplied by the particle's rest mass), areal energy density $\mu$, surface pressure $p$, and velocity field $u^a$. The matter's energy-momentum tensor is
\begin{equation}
T_{\rm shell}^{\alpha\beta} = e^\alpha_a e^\beta_b S^{ab}\, \delta(L),
\end{equation}
where $L$ is the proper distance to $\Sigma$ measured along spacelike geodesics that cross the surface orthogonally, $e^\alpha_a := \partial X^\alpha/\partial y^a$ are vectors tangent to $\Sigma$, and
\begin{equation}
S^{ab} = \mu u^a u^b + p( h^{ab} + u^a u^b )
\label{S_ab}
\end{equation}
describes a perfect fluid. Here $h^{ab}$ is the inverse to the induced metric on the hypersurface,
\begin{equation}
h_{ab} := e^\alpha_a e^\beta_b\, g_{\alpha\beta}.  
\label{h_ab}
\end{equation} 
The fluid is assigned barotropic equations of state of the form
\begin{equation}
\mu = \sigma + \varepsilon(\sigma), \qquad
p = p(\sigma),
\end{equation}
where $\varepsilon$ is the areal density of internal energy. The fluid is assumed to be isentropic, and is subjected to the first law of thermodynamics, which we can write in the form $d(\varepsilon/\sigma) = - p\, d(1/\sigma)$, or $d\varepsilon = (\varepsilon+p)/\sigma\, d\sigma$, or alternatively
\begin{equation}
d\mu = \frac{\mu+p}{\sigma}\, d\sigma.
\label{first_law}
\end{equation}
The adiabatic index $\Gamma$ is defined by 
\begin{equation}
\Gamma := \frac{\sigma}{p} \frac{dp}{d\sigma},
\label{adiab_GR} 
\end{equation}
and in general it is a function of $\sigma$. The first law then implies that
\begin{equation}
dp = \Gamma \frac{p}{\mu+p}\, d\mu. 
\label{dp_vs_dmu}
\end{equation}

The fluid is subjected to two conservation equations,
\begin{equation}
D_a (\sigma u^a) = 0, \qquad
D_b S^{ab} = 0,
\label{conservation}
\end{equation}
where $D_a$ is the covariant-derivative operator compatible with $h_{ab}$. The first statement expresses conservation of mass --- the mass of a fluid element stays the same as it moves on the hypersurface. The second expresses conservation of energy and momentum. The component of $D_b S^{ab} = 0$ in the direction of $u^a$ produces $(\mu+p) u^a D_a \sigma - \sigma u^a D_a \mu = 0$, which is a dynamical version of the first law of Eq.~(\ref{first_law}). The components orthogonal to $u^a$ give rise to
\begin{equation}
(\mu + p) u^b D_b u^a + ( h^{ab} + u^a u^b ) D_b p = 0,
\label{euler}
\end{equation}
the relativistic Euler equation. 

In addition to the fluid equations, the shell is governed by the Israel junction conditions \cite{israel:66}
\begin{subequations}
\label{israel}
\begin{align} 
\bigl[ h_{ab} \bigr] &= 0,
\label{israel_a} \\
\bigl[ K_{ab} \bigr] - \bigl[ h^{cd} K_{cd} \bigr] h_{ab} &= -8\pi S_{ab}, 
\label{israel_b}
\end{align} 
\end{subequations} 
where $[\psi] := \psi_+ - \psi_-$ denotes the jump of a quantity $\psi$ across $\Sigma$. These require the induced metric to be the same on both sides of the hypersurface, and they relate the fluid's energy-momentum tensor to the jump in extrinsic curvature. This is defined by
\begin{equation}
K_{ab} := e^\alpha_a e^\beta_b\, \nabla_\alpha n_\beta,
\label{Kab}
\end{equation}
where $n_\alpha$ is the unit normal to the hypersurface, pointing from the negative side to the positive side, and $\nabla_\alpha$ is the covariant-derivative operator compatible with $g_{\alpha\beta}$. The spacetime metric on both sides of the hypersurface is determined by the Einstein field equations. 

We assume that there is no matter inside and outside the shell, so that $g_{\alpha\beta}$ is a solution to the vacuum field equations. Under these circumstances, it can be shown --- see Eq.~(21) of Ref.~\cite{israel:66} --- that the conservation statement $D_b S^{ab} = 0$ is compatible with the assignment of Eq.~(\ref{israel_b}).

\section{Unperturbed, spherical shell}
\label{sec:unperturbed}

In this section we construct the unperturbed configuration, in which the thin shell is static and spherically symmetric. Spacetime inside the shell is described by the Minkowski metric, while the exterior spacetime is described by the Schwarzschild metric. 

\subsection{Exterior view} 

We give the shell a gravitational mass $M$ and an areal radius $R$. We describe the Schwarzschild metric in terms of the retarded-time coordinate $u := t - r^*$, the areal radius $r$, and the polar angles $\theta$ and $\phi$; $t$ is the familiar time coordinate and $r^* := r + 2M\ln(r/2M-1)$ is the tortoise radius. The metric takes the form  
\begin{equation}
ds^2 = -f\, du^2 - 2\, du dr +r^2 \bigl( d\theta^2 + \sin^2\theta\, d\phi^2 \bigr),
\label{schw} 
\end{equation}
where $f := 1-2M/r$. 

We use $y^a = (u,\vartheta^A)$ as intrinsic coordinates on the hypersurface $\Sigma$, where $u$ is the same retarded-time coordinate as in the bulk spacetime, and $\vartheta^A = (\vartheta,\varphi)$ are {\it Lagrangian coordinates} attached to fluid elements on the hypersurface. When viewed from the exterior spacetime, the embedding relations are $u = u$, $r = R$, $\theta = \vartheta$, and $\phi = \varphi$; in the unperturbed state there is no distinction between the spacetime and intrinsic angles. The tangent vectors are
\begin{equation}
e^\alpha_u = (1,0,0,0), \qquad
e^\alpha_\vartheta = (0,0,1,0), \qquad
e^\alpha_\varphi = (0,0,0,1),
\end{equation}
and the unit normal to the hypersurface is
\begin{equation}
n_\alpha = F^{-1/2} (0, 1, 0, 0), \qquad
F := f(r=R) = 1-2M/R.
\end{equation}
The induced metric is
\begin{equation}
h_{ab}\, dy^a dy^b = -F\, du^2 + R^2\, d\Omega^2,
\label{h_ext} 
\end{equation}
and the nonvanishing components of the extrinsic curvature are
\begin{equation}
K_{uu} = -\frac{M}{R^2} F^{1/2}, \qquad
K_{AB} = R F^{1/2}\, \Omega_{AB},
\end{equation}
where $\Omega_{AB} := \mbox{diag}[1,\sin^2\vartheta]$, so that $\Omega_{AB}\, d\vartheta^A d\vartheta^B = d\vartheta^2 + \sin^2\vartheta\, d\varphi^2$.

\subsection{Interior view} 

The metric inside the shell is written as
\begin{equation}
ds^2 = -F\, du^2 - 2 F^{1/2}\, du dr + r^2 \bigl( d\theta^2 + \sin^2\theta\, d\phi^2 \bigr). 
\label{mink}
\end{equation}
We have rescaled the retarded-time coordinate $t - r$ of the Minkowski spacetime by a factor of $F^{1/2}$ to ensure continuity of the induced metric at the hypersurface. The embedding relations are again given by 
$u = u$, $r = R$, $\theta = \vartheta$, and $\phi = \varphi$; they give rise to the same set of tangent vectors. The unit normal, however, is now given by 
\begin{equation}
n_\alpha = (0, 1, 0, 0). 
\end{equation}
The induced metric is again given by Eq.~(\ref{h_ext}), so that we indeed have continuity across the hypersurface. The nonvanishing components of the extrinsic curvature are now
\begin{equation}
K_{uu} = 0, \qquad
K_{AB} = R\, \Omega_{AB}. 
\end{equation}

\subsection{Surface fluid} 

Computation of $S^{ab}$ from Eq.~(\ref{israel_b}) produces
\begin{equation}
S_{uu} = -\frac{1}{4\pi R} F \bigl( F^{1/2} - 1 \bigr), \qquad
S_{AB} = \frac{R}{8\pi} \biggl( F^{1/2} - 1 + \frac{M}{R} F^{-1/2} \biggr) \Omega_{AB},
\end{equation}
and we observe that these assignments are compatible with the assumed perfect-fluid form of Eq.~(\ref{S_ab}). The fluid can be assigned a velocity vector
\begin{equation}
u^a = F^{-1/2}(1, 0, 0), \qquad
\end{equation}
an areal energy density
\begin{equation}
\mu = \frac{1}{4\pi R} \bigl( 1 - F^{1/2} \bigr),
\label{mu_unpert}
\end{equation}
and a surface pressure
\begin{equation}
p = \frac{1}{8\pi R} \biggl( F^{1/2} - 1 + \frac{M}{R} F^{-1/2} \biggr).
\label{p_unpert}
\end{equation}
It may be verified that in the Newtonian limit $M/R \ll 1$, $\mu$ is approximated by $M/(4\pi R^2)$, while $p$ is approximated by $M^2/(16\pi R^3)$. 

\subsection{Equilibrium sequence} 
\label{subsec: equilibrium sequence}

Equations (\ref{mu_unpert}) and (\ref{p_unpert}) can be inverted \cite{lemaitre-poisson:19} to yield $M$ and $R$ expressed in terms of $\mu$ and $p$. We have
\begin{equation}
M = \frac{4p^2}{\pi \mu^3} \frac{1+2p/\mu}{(1+4p/\mu)^3}, \qquad
R = \frac{p}{\pi\mu^2} \frac{1}{1+4p/\mu}.
\label{MR_vs_mup}
\end{equation}
The equations of state $\mu = \mu(\sigma)$ and $p = p(\sigma)$ imply that Eq.~(\ref{MR_vs_mup}) describes a sequence of equilibrium configurations parameterized by the mass density $\sigma$. The shell becomes compact when $p/\mu$ is allowed to get large. In this regime we have that
\begin{equation}
F = 1 - \frac{2M}{R} = \frac{1}{16} (\mu/p)^2 \biggl[ 1 - \frac{1}{2} (\mu/p) + \frac{3}{16} (\mu/p)^2
+ O\bigl( (\mu/p)^3 \bigr) \biggr],
\end{equation}
and we do see that $F \ll 1$ when $p/\mu \gg 1$. Another way of expressing this is
\begin{equation}
\mu = \frac{1}{8\pi M} \bigl[ 1 + O(F^{1/2}) \bigr], \qquad
p = \frac{1}{16\pi M} F^{-1/2} \bigl[ 1 + O(F^{1/2}) \bigr].
\end{equation}
Because the pressure can never be infinite, the ``black-hole limit'' $F \to 0$ can never be attained.

If we follow $M(\sigma)$ along the equilibrium sequence, taking into account that $dp = \Gamma (p/\sigma)\, d\sigma$ and $d\mu = [(\mu+p)/\sigma]\, d\sigma$, we find that $dM/d\sigma \geq 0$ whenever $\Gamma \geq \Gamma_1$, where\footnote{The criterion appears to be different from the one given in Eq.~(34) of Ref.~\cite{lemaitre-poisson:19}. The reason has to do with inequivalent definitions of the adiabatic index. In Ref.~\cite{lemaitre-poisson:19} it was defined as $d\ln p/d\ln \mu$, whereas it is $d\ln p/d\ln\sigma$ here. To maximize confusion, $\mu$ was denoted $\sigma$ in Ref.~\cite{lemaitre-poisson:19}. The definitions adopted here are more appropriate in view of the thermodynamics of the surface matter.} 
\begin{subequations}
\label{mass_increase}
\begin{align}
\Gamma_1 &:= \frac{3}{2} + 4 (p/\mu) + 4 (p/\mu)^2, \\
&= \frac{1 + 2\sqrt{F} + 3F}{4F}, \\
&= \frac{4 - 6{\cal C} + 2\sqrt{1-2{\cal C}}}{4(1-2{\cal C})},
\end{align}
\end{subequations}
where ${\cal C} := M/R$ is the shell's compactness. When ${\cal C} \ll 1$ the last expression becomes
\begin{equation}
\Gamma_1 = \frac{3}{2} + {\cal C} + \frac{7}{4} {\cal C}^2 + O({\cal C}^3). 
\end{equation}
The point along the sequence at which $dM/d\sigma = 0$ is the configuration of maximum mass for the specified equation of state, and this point also marks the onset of a radial instability --- the shell becomes dynamically unstable against time-dependent, spherically-symmetric perturbations \cite{lemaitre-poisson:19}.

We suppose that the sequence begins at low $\sigma$ with $\Gamma(\sigma) > \Gamma_1$, and that it eventually reaches a configuration of maximum mass when $\Gamma(\sigma) = \Gamma_1$ at some $\sigma$. At this point we have that ${\cal C} = {\cal C}_{\rm max}$ and $F = F_{\rm min}$, where $F_{\rm min} = 1 - 2{\cal C}_{\rm max}$. Inverting Eq.~(\ref{mass_increase}), this is given in terms of $\Gamma_1$ by   
\begin{equation}
F_{\rm min} = \biggl( \frac{1 + \sqrt{4\Gamma_1-2}}{4\Gamma_1-3} \biggr)^2.
\label{Fmin} 
\end{equation}
When $\Gamma_1 \gg 1$ this is approximated by
\begin{equation}
F_{\rm min} = \frac{1}{4\Gamma_1} \biggl[ 1 + \Gamma_1^{-1/2} + \frac{5}{4} \Gamma_1^{-1}
+ \frac{5}{4} \Gamma_1^{-3/2} + O(\Gamma_1^{-2}) \biggr].
\label{Fmin_largeG} 
\end{equation} 
These equations are especially useful when the fluid is polytropic, with $\Gamma$ a specified constant. In this case $\Gamma_1$ must be equal to $\Gamma$, and Eq.~(\ref{Fmin}) can be used to obtain $\mathcal{C}_{\rm max}$ for the given polytropic index. 

\begin{figure}
\includegraphics[width=0.6\linewidth]{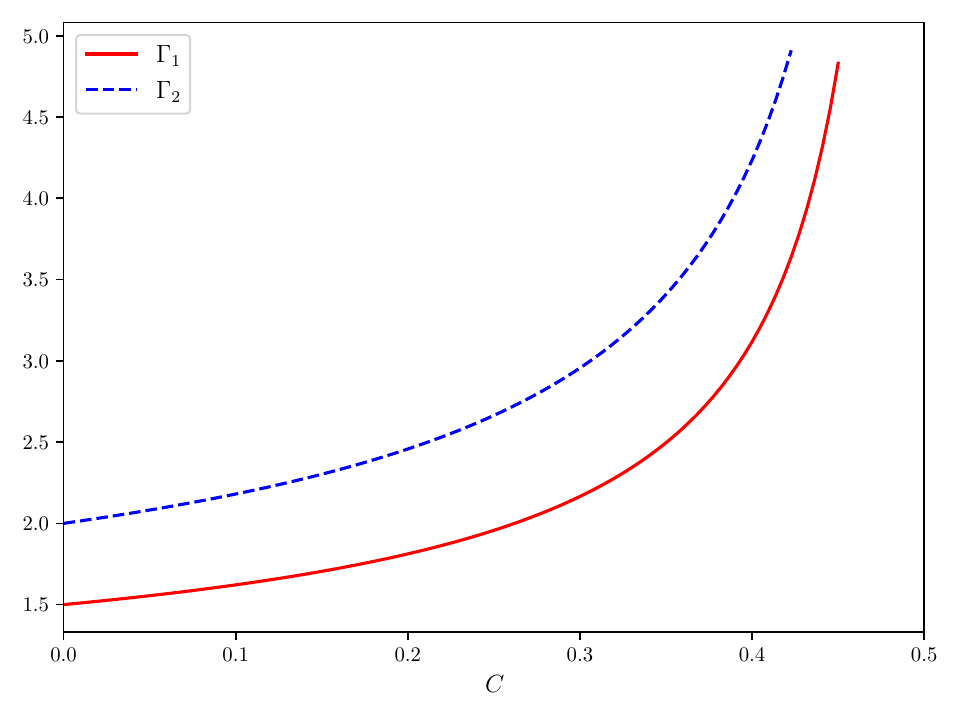}
\caption{Plots of $\Gamma_1$ (red) and $\Gamma_2$ (blue) as functions of ${\cal C} := M/R$. We have that $dM/d\sigma > 0$ when $\Gamma(\sigma)$ is above the $\Gamma_1$ curve, while $dM/d\sigma < 0$ below the curve. On the other hand, we have that $dR/d\sigma < 0$ when $\Gamma(\sigma)$ is below the $\Gamma_2$ curve, while $dR/d\sigma > 0$ above the curve.}  
\label{fig:Gamma} 
\end{figure}

If we now follow $R(\sigma)$ along the equilibrium sequence, we find that $dR/d\sigma \leq 0$ whenever $\Gamma \leq \Gamma_2$, where
\begin{subequations}
\label{radius_decrease}
\begin{align}
\Gamma_2 &:= 2 + 6 (p/\mu) + 4 (p/\mu)^2, \\
&= \frac{1 + 4\sqrt{F} + 3F}{4F}, \\
&= \frac{4 - 6{\cal C} + 4\sqrt{1-2{\cal C}}}{4(1-2{\cal C})}. 
\end{align}
\end{subequations}
When ${\cal C} \ll 1$ this reduces to 
\begin{equation}
\Gamma_2 = 2 + \frac{3}{2} {\cal C} + \frac{5}{2} {\cal C}^2 + O({\cal C}^3). 
\end{equation}
Plots of $\Gamma_1$ and $\Gamma_2$ as functions of ${\cal C}$ are presented in Fig.~\ref{fig:Gamma}.

\subsection{Polytropic sequence}
\label{subsec:polytrope}

To provide a concrete illustration of the foregoing discussion, we take the fluid to possess the polytropic equations of state
\begin{equation}
p = p_0 (\sigma/\sigma_0)^\Gamma, \qquad
\mu = \sigma + \frac{p}{\Gamma-1},
\end{equation}
where $\sigma_0$ and $p_0$ are units of mass density and surface pressure, respectively, and where $\Gamma$ is now a constant; the expression for $\mu$ is a consequence of the first law of Eq.~(\ref{first_law}). We use the notations $x := \sigma/\sigma_0$ and $b := p_0/\sigma_0$.

\begin{figure}
\includegraphics[width=0.6\linewidth]{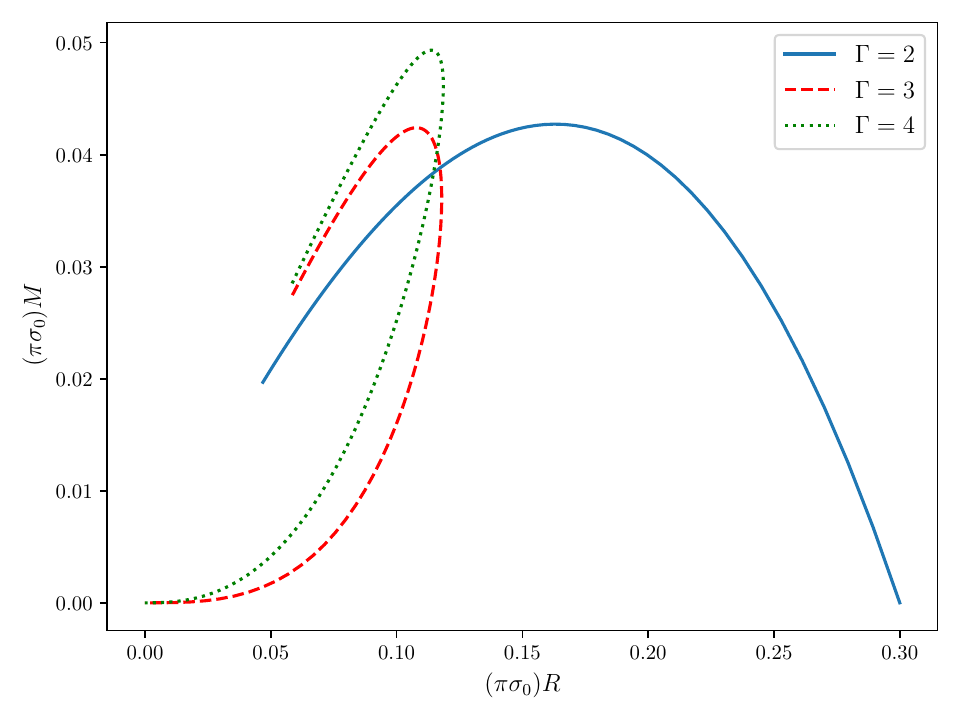}
\caption{Polytropic sequences of equilibria. The figure shows plots of $(\pi \sigma_0) M$ versus $(\pi \sigma_0) R$ for three polytropes. The blue curve represents a sequence with $\Gamma = 2$ and $b = 0.3$, with the parameter $x := \sigma/\sigma_0$ increasing counterclockwise. We see that $R$ decreases everywhere on the sequence, and that $M$ increases toward a maximum of $(\pi \sigma_0) M_{\rm max} \simeq 4.2727 \times 10^{-2}$ at $x \simeq 4.2200 \times 10^{-1}$; the compactness at this maximum is ${\cal C} \simeq 2.6202 \times 10^{-1}$. The red curve represents a sequence with $\Gamma = 3$ and $b = 0.3$, with $x$ again increasing counterclockwise. We see that $R$ first increases along the sequence (while $3 > \Gamma_2$), and that it eventually decreases (when $3 < \Gamma_2$). The maximum mass of $(\pi \sigma_0) M_{\rm max} \simeq 4.2410 \times 10^{-2}$ occurs at $x \simeq  1.0645$, and the corresponding compactness is ${\cal C} \simeq 3.9306 \times 10^{-1}$. For the green curve we have a sequence with $\Gamma = 4$ and $b = 0.3$. We again see that $R$ first increases and then decreases. The maximum mass of $(\pi \sigma_0) M_{\rm max} \simeq 4.9332 \times 10^{-2}$ occurs at $x \simeq  1.1930$, and the corresponding compactness is ${\cal C} \simeq 4.3348 \times 10^{-1}$.}  
\label{fig:polytropes} 
\end{figure}

Making the substitutions in Eqs.~(\ref{MR_vs_mup}), we obtain
\begin{subequations}
\begin{align}
M &= \frac{4b^2 (\Gamma-1)^3}{\pi \sigma_0}
\frac{ (2\Gamma-1) b x^{3\Gamma} + (\Gamma-1) x^{2\Gamma+1} }
{ \bigl[ (4\Gamma-3)b x^\Gamma + (\Gamma-1) x \bigr]^3
  \bigl[ b x^\Gamma + (\Gamma-1) x \bigr] }, \\
R &= \frac{b(\Gamma-1)^2}{\pi \sigma_0}
\frac{x^\Gamma}
{ \bigl[ (4\Gamma-3)b x^\Gamma + (\Gamma-1) x \bigr]
  \bigl[ b x^\Gamma + (\Gamma-1) x \bigr] }.
\end{align}
\end{subequations}
The polytropic sequence is parametrized by $x$, and it depends on two constants, $\Gamma$ and $b$. We display $M$-$R$ diagrams for a few sampled values of these constants in Fig.~\ref{fig:polytropes}.

\section{Quasinormal modes: Even parity}
\label{sec:even}

In this section and the next we perturb the thin-shell spacetime constructed in Sec.~\ref{sec:unperturbed} and derive eigenvalue problems for the complex frequencies of the system's quasinormal modes. These modes are described by a metric perturbation that is required to be regular at $r=0$ and to describe a purely outgoing gravitational wave at $r = \infty$; this is accompanied by a perturbation of the shell's fluid variables. In this section we consider the even-parity sector of these perturbations.

\subsection{Exterior metric}
\label{subsec:even_ext}

We work in the Regge-Wheeler gauge \cite{regge-wheeler:57, martel-poisson:05} and express the perturbed metric outside the shell as
\begin{subequations}
\label{metric_pert_ext}
\begin{align}
g_{uu} &= f \bigl[ -1 + p_{uu}(r)\, Y^{\ell m}(\theta,\phi)\, e^{-i\omega u} \bigr], \\
g_{ur} &= -1 +p_{ur}(r)\, Y^{\ell m}(\theta,\phi)\, e^{-i\omega u}, \\
g_{rr} &= p_{rr}(r)\, Y^{\ell m}(\theta,\phi)\, e^{-i\omega u}, \\
g_{AB} &= r^2 \Omega_{AB} \bigl[ 1 + K(r)\, Y^{\ell m}(\theta,\phi)\, e^{-i\omega u} \bigr], 
\end{align}
\end{subequations}
where $f := 1-2M/r$ and $Y^{\ell m}(\theta,\phi)$ are spherical harmonics. In the context of this equation we use the uppercase latin index $A$ to label the spacetime angles, $\theta^A := (\theta,\phi)$, and we write $\Omega_{AB} := \mbox{diag}[1,\sin^2\theta]$. To declutter the notation we omit the label $\ell m$ on the radial functions $p_{uu}$, $p_{ur}$, $p_{rr}$, and $K$, and we also omit an implicit summation over $\ell$ and $m$. In practice the perturbation equations are independent of $m$, and we shall consider one value of $\ell$ at a time. Our considerations are restricted to $\ell \geq 2$.

Invoking the Einstein field equations linearized on the Schwarzschild background of Eq.~(\ref{schw}), $p_{ur}$ and $p_{rr}$ can be written in terms of $K$ as
\begin{subequations}
\label{p_vs_K_ext}
\begin{align}
p_{rr} &= \biggl[ \frac{\ell(\ell+1)}{2r^2} - \frac{i\omega}{r} \biggr]^{-1}
\Bigl( K'' + \frac{2}{r} K' \Bigr), \\
p_{ur} &= \frac{1}{2} f\, p_{rr}, 
\end{align}
\end{subequations}
where a prime indicates differentiation with respect to $r$. The independent fields are therefore $p_{uu}$ and $K$, and the field equations deliver a system of first-order differential equations for them. We have that
\begin{subequations}
\label{pert_eqns_even_ext}
\begin{align}
r p_{uu}' &= \Bigl\{ 2f \bigl[ \ell(\ell+1) M - 2\omega^2 r^3 \bigr] \Bigr\}^{-1}
\Bigl[ 4i\omega^3 r^4 + 2(\ell^2+\ell-4) \omega^2 r^3 + 28M\omega^2 r^2
- \ell(\ell+1) (\ell^2+\ell-2+2iM\omega) r
\nonumber \\ & \quad \mbox{} 
+ 2\ell(\ell+1)(\ell^2+\ell-4) M + 4\ell(\ell+1) M^2/r \Bigr]\, p_{uu}
\nonumber \\ & \quad \mbox{} 
+ \Bigl\{ 2f^2 \bigl[ \ell(\ell+1) M - 2\omega^2 r^3 \bigr] \Bigr\}^{-1}
\Bigl[ 4\omega^4 r^5 - 4(\ell-1)(\ell+2) \omega^2 r^3
+ 8(\ell^2+\ell-4)M\omega^2 r^2
\nonumber \\ & \quad \mbox{} 
+ (\ell^4+2\ell^3-\ell^2-2\ell + 36M^2\omega^2 )r
- 4(\ell-1)\ell(\ell+1)(\ell+2) M + 4(\ell-1)\ell(\ell+1)(\ell+2)M^2/r \Bigr]\, K, \\
r K' &= \Bigl\{ 2\bigl[ \ell(\ell+1) M - 2\omega^2 r^3 \bigr] \Bigr\}^{-1}
\Bigl[ -4\omega^2 r^3 - (\ell-1)\ell(\ell+1)(\ell+2) r - 4\ell(\ell+1) M \Bigr]\, p_{uu}
\nonumber \\ & \quad \mbox{} 
+ \Bigl\{ 2f \bigl[ \ell(\ell+1) M - 2\omega^2 r^3 \bigr] \Bigr\}^{-1}
\Bigl[ 4i\omega^3 r^4 - 2(\ell-1)(\ell+2) \omega^2 r^3 - 12 M\omega^2 r^2
\nonumber \\ & \quad \mbox{} 
+ \ell(\ell+1)(\ell^2+\ell-2 - 2iM\omega) r
- 2(\ell-1)\ell(\ell+1)(\ell+2) M \Bigr] K.
\end{align}
\end{subequations}
We insert these equations in Eq.~(\ref{p_vs_K_ext}) to express $p_{rr}$ and $p_{ur}$ algebraically in terms of $p_{uu}$ and $K$.

The metric perturbation can be conveniently encoded within the Makkula-Pere\~niguez master variable
\cite{mukkamala-perenigues:25, poisson:25}, defined by
\begin{equation}
\psi := 2i\omega r \Bigl\{ -r f K' + f p_{uu} - 2f p_{ur} + f^2 p_{rr}
+ \frac{1}{2} \bigl[ (\ell-1)(\ell+2) - 2i\omega r \bigr] K \Bigr\}.
\end{equation}
This satisfies the Regge-Wheeler equation \cite{regge-wheeler:57, martel-poisson:05},
\begin{equation}
f \psi'' + 2\biggl( \frac{M}{r^2} + i\omega \biggr) \psi'
- \biggl[ \frac{\ell(\ell+1)}{r^2} - \frac{6M}{r^3} \biggr] \psi = 0,
\label{RW_eqn_ext}
\end{equation}
in spite of the fact that it describes a metric perturbation with even parity. With Eqs.~(\ref{p_vs_K_ext}) and (\ref{pert_eqns_even_ext}), the master variable can be written as
\begin{equation}
\psi = \frac{i\omega r}{\ell(\ell+1) M - 2\omega^2 r^3} \Bigl\{
\ell(\ell+1) f \bigl[ (\ell-1)(\ell+2) r + 6M \bigr]\, p_{uu}
- \bigl[ \tfrac{1}{2} (\ell-1)\ell(\ell+1)(\ell+2)(r-3M) - 6 M\omega^2 r^2 \bigr]\, K \Bigr\}, 
\label{psi_even_ext} 
\end{equation}
algebraically in terms of the independent fields $p_{uu}$ and $K$. The perturbation equations also allow us to express $r\psi'$ in a similar fashion, and we have that $p_{uu}$ and $K$ are in a one-to-one correspondence with $\psi$ and $r\psi'$. All information about the metric perturbation can therefore be obtained by integrating the Regge-Wheeler equation.

\subsection{Interior metric}
\label{subsec:even_in}

The perturbed metric inside the shell is expressed as 
\begin{subequations}
\label{metric_pert_in}
\begin{align}
g_{uu} &= F \bigl[ -1 + p_{uu}(r)\, Y^{\ell m}(\theta,\phi)\, e^{-i\omega u} \bigr], \\
g_{ur} &= F^{1/2} \bigl[ -1 +p_{ur}(r)\, Y^{\ell m}(\theta,\phi)\, e^{-i\omega u} \bigr], \\
g_{rr} &= p_{rr}(r)\, Y^{\ell m}(\theta,\phi)\, e^{-i\omega u}, \\
g_{AB} &= r^2 \Omega_{AB} \bigl[ 1 + K(r)\, Y^{\ell m}(\theta,\phi)\, e^{-i\omega u} \bigr],
\end{align}
\end{subequations}
where $F := 1-2M/R$. The radial functions $p_{uu}$, $p_{ur}$, $p_{rr}$, and $K$ are distinct from those that appear in the metric of Eq.~(\ref{metric_pert_ext}). 

The vacuum Einstein field equations imply that 
\begin{subequations}
\label{p_vs_K_in}
\begin{align}
p_{rr} &= \biggl[ \frac{\ell(\ell+1)}{2r^2} - \frac{iw}{r} \biggr]^{-1}
\Bigl( K'' + \frac{2}{r} K' \Bigr), \\
p_{ur} &= \frac{1}{2}\, p_{rr}, 
\end{align}
\end{subequations}
where $w := F^{-1/2}\, \omega$ is a rescaled frequency and a prime indicates differentiation with respect to $r$. They also produce the system of differential equations
\begin{subequations}
\label{pert_eqns_even_in}
\begin{align}
r p_{uu}' &= -\biggl[ iwr + \frac{1}{2} (\ell^2+\ell-4) 
- \frac{(\ell-1)\ell(\ell+1)(\ell+2)}{4 w^2 r^2} \biggr] p_{uu}
- \biggl[ w^2 r^2 - (\ell-1)(\ell+2) + \frac{(\ell-1)\ell(\ell+1)(\ell+2)}{4 w^2 r^2} \biggr] K, \\
r K' &= \biggl[1 + \frac{(\ell-1)\ell(\ell+1)(\ell+2)}{4 w^2 r^2} \biggr] p_{uu}\,
- \biggl[ iwr - \frac{1}{2} (\ell-1)(\ell+2) + \frac{(\ell-1)\ell(\ell+1)(\ell+2)}{4 w^2 r^2} \biggr] K,
\end{align}
\end{subequations}
for the independent fields $p_{uu}$ and $K$. We insert these equations within Eq.~(\ref{p_vs_K_in}) to express $p_{rr}$ and $p_{ur}$ in terms of $p_{uu}$ and $K$.

The interior perturbation also can be conveniently encoded within a master variable defined by
\begin{equation}
\psi := 2iw r \Bigl\{ -r K' + p_{uu} - 2 p_{ur} + p_{rr}
+ \frac{1}{2} \bigl[ (\ell-1)(\ell+2) - 2iw r \bigr] K \Bigr\}. 
\end{equation}
This satisfies the interior version of the Regge-Wheeler equation,
\begin{equation}
\psi'' + 2iw\, \psi' - \frac{\ell(\ell+1)}{r^2}\, \psi = 0.
\label{RW_eqn_in}
\end{equation}
With the foregoing results we have that $\psi$ can also be expressed as
\begin{equation}
\psi = \frac{(\ell-1)\ell(\ell+1)(\ell+2)}{2iwr} \Bigl( p_{uu} - K \Bigr).
\label{psi_even_in}
\end{equation} 
In a similar way we can write $r\psi'$ in terms of $p_{uu}$ and $K$, and we again find that $p_{uu}$ and $K$ are in a one-to-one correspondence with $\psi$ and $r\psi'$. All information about the metric perturbation can be obtained by integrating the Regge-Wheeler equation.

\subsection{Deformed shell}
\label{subsec:even_shell} 

The shell is now perturbed from its static and spherically symmetric state. We continue to use $y^a = (u,\vartheta^A)$ as intrinsic coordinates on the hypersurface, with the angular coordinates $\vartheta^A = (\vartheta,\varphi)$ assuming double duty as Lagrangian coordinates. By this we mean that a fluid element identified by its coordinates $\vartheta^A$ in the unperturbed configuration keeps these coordinates in the perturbed configuration. The position of this fluid element in the four-dimensional spacetime is altered, however, and this motion is described in terms of a Lagrangian displacement vector, which is decomposed in spherical harmonics.

Either side of the hypersurface (exterior or interior) is described by the embedding relations
\begin{subequations}
\label{embedding_even}
\begin{align}
u &= u, \\
r &= R\bigl[ 1 + \xi_r\, Y^{\ell m}(\vartheta,\varphi)\, e^{-i\omega u} \bigr], \\ 
\theta^A &= \vartheta^A + \xi\, \Omega^{AB} Y^{\ell m}_B(\vartheta,\varphi)\, e^{-i\omega u},
\end{align}
\end{subequations}
where the constants $\xi_r$ and $\xi$ represent the components of the displacement vector. In the context of this equation, and those below, $\Omega^{AB}$ is the inverse to $\Omega_{AB} := \mbox{diag}[1, \sin^2\vartheta]$, and $Y^{\ell m}_A := D_A Y^{\ell m}$ are vector harmonics, with $D_A$ denoting the covariant-derivative operator compatible with $\Omega_{AB}$. We again omit the label $\ell m$ on these quantities, and leave a summation over $\ell$ and $m$ implicit. When the context requires it, we shall be more explicit with our notation and distinguish between the exterior values --- $\xi_r^{\rm out}$ and $\xi^{\rm out}$ --- and the interior values --- $\xi_r^{\rm in}$ and $\xi^{\rm in}$ --- of the Lagrangian displacement.

The tangent vectors $e^\alpha_u$ and $e^\alpha_A$ are obtained by differentiating the embedding relations $x^\alpha = X^\alpha(y^a)$ with respect to the intrinsic coordinates. The normal vector $n_\alpha$ is proportional to the gradient of
\begin{equation} 
r - R \bigl[ 1 + \xi_r\, Y^{\ell m}(\theta,\phi)\, e^{-i\omega u} \bigr],
\end{equation}
with a proportionality factor determined by the normalization condition $g^{\alpha\beta} n_\alpha n_\beta = 1$. Notice that here, the spherical harmonics are expressed in terms of the spacetime angles $(\theta,\phi)$ instead of the intrinsic angles $(\vartheta,\varphi)$; this expression is compatible with Eq.~(\ref{embedding_even}) in the context of first-order perturbation theory. At the end of the computation the normal vector is expressed in terms of $(u,\vartheta^A)$. 

The perfect fluid that makes up the thin shell now possesses a perturbed energy-momentum tensor given by
\begin{equation}
S^{ab} = (\mu + \delta \mu) u^a u^b + (p + \delta p) \bigl( h^{ab} + u^a u^b \bigr),
\label{S_pert_even} 
\end{equation}
in which $\mu$ and $p$ are the unperturbed density and pressure of Eqs.~(\ref{mu_unpert}) and (\ref{p_unpert}), respectively, while $\delta \mu$ and $\delta p$ are their perturbations. We express these as
\begin{subequations}
\label{dmu_dp}
\begin{align}
\delta \mu &= \mu_\ell\, Y^{\ell m}(\vartheta,\varphi)\, e^{-i\omega u}, \\  
\delta p &= p_\ell\, Y^{\ell m}(\vartheta,\varphi)\, e^{-i\omega u},
\end{align}
\end{subequations}
where $\mu_\ell$ and $p_\ell$ are constants. The fluid's equations of state and Eq.~(\ref{dp_vs_dmu}) imply that they are related by
\begin{equation}
p_\ell = \Gamma \frac{p}{\mu+p}\, \mu_\ell,
\label{Gamma_eq}
\end{equation}
where $\Gamma$ is the adiabatic index of Eq.~(\ref{adiab_GR}). The induced metric $h_{ab}$ and velocity field $u^a$ that appear in Eq.~(\ref{S_pert_even}) are perturbed relative to the description given in Sec.~\ref{sec:unperturbed}. The velocity vector has $u^u$ as its only nonvanishing component, which is determined by the normalization condition $h_{ab} u^a u^b = -1$. This follows by virtue of the Lagrangian nature of the angular coordinates $\vartheta^A$: each fluid element moves with a constant value of the intrinsic angles. 

\subsection{Induced metric}
\label{subsec:even_induced} 

The induced metric on the deformed shell is computed from Eq.~(\ref{h_ab}) and the spacetime metrics of Eqs.~(\ref{metric_pert_ext}) and (\ref{metric_pert_in}). The calculation produces the components
\begin{subequations}
\label{induced_metric}
\begin{align}
h_{uu} &= -F + {\cal A}\, Y^{\ell m}(\vartheta,\varphi)\, e^{-i\omega u}, \\
h_{uA} &= -R {\cal B}\, Y_A^{\ell m}(\vartheta,\varphi)\, e^{-i\omega u}, \\
h_{AB} &= R^2 \Omega_{AB} + R^2 \bigl[ {\cal C}\, \Omega_{AB} Y^{\ell m}(\vartheta,\varphi)
+ {\cal D}\, Y^{\ell m}_{AB}(\vartheta,\varphi) \bigr] e^{-i\omega u},
\end{align}
\end{subequations}
where $Y_A^{\ell m} := D_A Y^{\ell m}$ are the vectorial harmonics encountered previously, while
\begin{equation}
Y_{AB}^{\ell m} := \biggl[ D_A D_B + \frac{1}{2} \ell(\ell+1) \Omega_{AB} \biggr] Y^{\ell m}
\end{equation}
are tensorial harmonics, which are tracefree --- $\Omega^{AB} Y_{AB}^{\ell m} = 0$ --- by virtue of the eigenvalue equation for the spherical harmonics. 

When computed from the exterior side, the components of the induced metric are given by
\begin{subequations}
\begin{align}
{\cal A}^{\rm out} &= F p^{\rm out}_{uu}(R) - 2(M/R - i\omega R) \xi^{\rm out}_r, \\
{\cal B}^{\rm out} &= \xi^{\rm out}_r + i\omega R\, \xi^{\rm out}, \\
{\cal C}^{\rm out} &= K^{\rm out}(R) + 2\xi^{\rm out}_r - \ell(\ell+1) \xi^{\rm out}, \\
{\cal D}^{\rm out} &= 2 \xi^{\rm out}.
\end{align}
\end{subequations}
They are
\begin{subequations}
\begin{align}
{\cal A}^{\rm in} &= F p^{\rm in}_{uu}(R) + 2 i\omega R F^{1/2}\, \xi^{\rm in}_r, \\
{\cal B}^{\rm in} &= F^{1/2}\, \xi^{\rm in}_r + i\omega R\, \xi^{\rm in}, \\
{\cal C}^{\rm in} &= K^{\rm in}(R) + 2\xi^{\rm in}_r - \ell(\ell+1) \xi^{\rm in}, \\
{\cal D}^{\rm in} &= 2 \xi^{\rm in}
\end{align}
\end{subequations}
when computed instead from the interior side. Continuity of the induced metric across the shell --- Eq.~(\ref{israel_a}) --- requires
\begin{subequations}
\label{junction1}
\begin{align}
\xi^{\rm in}_r &= F^{-1/2}\, \xi^{\rm out}_r,
\label{j1_a} \\ 
\xi^{\rm in} &= \xi^{\rm out},
\label{j1_b} \\
p^{\rm in}_{uu}(R) &= p^{\rm out}_{uu}(R) - \frac{2M}{R F}\, \xi^{\rm out}_r,
\label{j1_c} \\
K^{\rm in}(R) &= K^{\rm out}(R) + 2(1 - F^{-1/2})\, \xi^{\rm out}_r.
\label{j1_d}
\end{align}
\end{subequations}
We shall henceforth impose these constraints, and it shall no longer be necessary to distinguish between the exterior and interior components of the induced metric.

The conservation equation $D_b S^{ab} = 0$ for the perturbed energy-momentum tensor of Eq.~(\ref{S_pert_even}) implies that $\mu_\ell$ and $p_\ell$ are both determined in terms of the metric perturbation. We have that
\begin{equation} 
\mu_\ell = -(\mu + p) {\cal C}, \qquad
p_\ell = \frac{\mu+p}{2 F} \Bigl( {\cal A} - 2 i \omega R\, {\cal B} \Bigr). 
\label{mup_vs_h}
\end{equation}
Here, $D_a$ stands for the covariant-derivative operator associated with the perturbed metric $h_{ab}$, and $\mu_\ell$ and $p_\ell$ are defined by Eq.~(\ref{dmu_dp}). 

\subsection{Extrinsic curvature}
\label{subsec:even_extrinsic} 

The extrinsic curvature of the deformed shell is calculated from Eq.~(\ref{Kab}) and the spacetime metrics of Eqs.~(\ref{metric_pert_ext}) and (\ref{metric_pert_in}). The exterior computation produces
\begin{subequations}
\label{Kab_out} 
\begin{align}
K^{\rm out}_{uu} &= -\frac{M}{R^2} F^{1/2}
+ R^{-1}\, {\cal G}^{\rm out}\, Y^{\ell m}(\vartheta,\varphi)\, e^{-i\omega u}, \\
K^{\rm out}_{uA} &= {\cal H}^{\rm out}\, Y^{\ell m}_A(\vartheta,\varphi)\, e^{-i\omega u}, \\
K^{\rm out}_{AB} &= R F^{1/2}\, \Omega_{AB}
+ R \bigl[ {\cal I}^{\rm out}\, \Omega_{AB} Y^{\ell m}(\vartheta,\varphi)
+ {\cal J}^{\rm out}\, Y^{\ell m}_{AB}(\vartheta,\varphi) \bigr] e^{-i\omega u},
\end{align}
\end{subequations}
with
\begin{subequations}
\begin{align}
{\cal G}^{\rm out} &= \frac{1}{2} F^{3/2} R\, p'_{uu}(R)
+ \frac{1}{2} F^{1/2} (3M/R - i\omega R)\, p_{uu}(R)
- F^{1/2} (M/R - i\omega R)\, p_{ur}(R) + F^{3/2} \frac{M}{2R}\, p_{rr}(R)
\nonumber \\ & \quad \mbox{}
+ F^{-1/2} \bigl( \omega^2 R^2 + 2 i M\omega + 2M/R - 5 M^2/R^2 \bigr) \xi_r, \\
{\cal H}^{\rm out} &= \frac{1}{2} F^{1/2}\, p_{uu}(R) - \frac{1}{2} F^{1/2}\, p_{ur}(R)
+ F^{-1/2} (i\omega R - M/R)\, \xi_r - i\omega R F^{1/2}\, \xi, \\
{\cal I}^{\rm out} &= \frac{1}{2} F^{1/2} R\, K'(R) - \frac{1}{2} F^{1/2}\, p_{uu}(R)
+ F^{1/2}\, p_{ur}(R) - \frac{1}{2} F^{3/2}\, p_{rr}(R)
+ \frac{1}{2} F^{-1/2} (i\omega R + 2 - 4M/R)\, K(R)
\nonumber \\ & \quad \mbox{}
+ \frac{1}{2} F^{-1/2} \bigl[ \ell(\ell+1) + 2 - 2M/R \bigr]\, \xi_r
- \ell(\ell+1) F^{1/2}\, \xi, \\
{\cal J}^{\rm out} &= -F^{-1/2}\, \xi_r + 2 F^{1/2}\, \xi,
\end{align}
\end{subequations}
where we omit the label ``out'' on all the quantities that occur on the right-hand side. The interior calculation yields
\begin{subequations}
\label{Kab_in} 
\begin{align}
K^{\rm in}_{uu} &= R^{-1}\, {\cal G}^{\rm in}\, Y^{\ell m}(\vartheta,\varphi)\, e^{-i\omega u}, \\
K^{\rm in}_{uA} &= {\cal H}^{\rm in}\, Y^{\ell m}_A(\vartheta,\varphi)\, e^{-i\omega u}, \\
K^{\rm in}_{AB} &= R\, \Omega_{AB}
+ R \bigl[ {\cal I}^{\rm in}\, \Omega_{AB} Y^{\ell m}(\vartheta,\varphi)
+ {\cal J}^{\rm in}\, Y^{\ell m}_{AB}(\vartheta,\varphi) \bigr] e^{-i\omega u},
\end{align}
\end{subequations}
with
\begin{subequations}
\begin{align}
{\cal G}^{\rm in} &= \frac{1}{2} F R\, p'_{uu}(R) - \frac{1}{2} i\omega R F^{1/2}\, p_{uu}(R)
+ i\omega R F^{1/2}\, p_{ur}(R) + \omega^2 R^2\, \xi_r, \\
{\cal H}^{\rm in} &= \frac{1}{2} F^{1/2}\, p_{uu}(R) - \frac{1}{2} F^{1/2}\, p_{ur}(R)
+ i\omega R\, \xi_r - i\omega R\, \xi, \\
{\cal I}^{\rm in} &= \frac{1}{2} R\, K'(R) - \frac{1}{2}\, p_{uu}(R)
+ p_{ur}(R) - \frac{1}{2}\, p_{rr}(R)
+ \frac{1}{2} F^{-1/2} (i\omega R + 2F^{1/2} )\, K(R)
\nonumber \\ & \quad \mbox{}
+ \frac{1}{2} \bigl[ \ell(\ell+1) + 2 \bigr]\, \xi_r
- \ell(\ell+1)\, \xi, \\
{\cal J}^{\rm out} &= -\xi_r + 2 \xi,
\end{align}
\end{subequations}
where we omit the label ``in'' on all the quantities that occur on the right-hand side. By virtue of Eqs.~(\ref{p_vs_K_ext}), (\ref{pert_eqns_even_ext}), (\ref{p_vs_K_in}), (\ref{pert_eqns_even_in}), and (\ref{junction1}), the components of the extrinsic curvature (on both the outside and inside faces of the shell) can be expressed entirely in terms of the set
\begin{equation}
\Bigl\{ p_{uu}^{\rm out}(R), K^{\rm out}(R), \xi_r^{\rm out}, \xi^{\rm out} \Bigr\}
\label{variables_even} 
\end{equation}
of perturbation variables. 

\subsection{Matching equations and eigenvalue problem}
\label{subsec:even_eigenvalue} 

The Israel junction conditions imply that $S_{ab}$, the surface tensor of Eq.~(\ref{S_pert_even}), must be related to $[K_{ab}]$, the jump in extrinsic curvature across the shell, by Eq.~(\ref{israel_b}). This requirement returns a single algebraic equation that implicates the set of variables listed in Eq.~(\ref{variables_even}). On the other hand, the perturbations in density and pressure obtained in Eq.~(\ref{mup_vs_h}) must be related by the fluid's equation of state, which implies Eq.~(\ref{Gamma_eq}). This requirement returns a second algebraic equation involving the variables of Eq.~(\ref{variables_even}). These two equations are then solved to return $\xi_r^{\rm out}$ and $\xi^{\rm out}$ in terms of $p_{uu}^{\rm out}(R)$ and $K^{\rm out}(R)$.

Next we return to the matching equations (\ref{j1_c}) and (\ref{j1_d}), which relate the inside and outside values of the perturbation fields $p_{uu}$ and $K$. As was observed below Eqs.~(\ref{psi_even_ext}) and (\ref{psi_even_in}), these can all be expressed in terms of the (interior and exterior) master function $\psi$ and its derivative. Combining all this, we have that the matching conditions take the form of
\begin{equation}
\bigl[ r \psi' \bigr] = P_1\, R \psi'_{\rm out}(R) + P_2\, \psi_{\rm out}, \qquad
\bigl[ \psi \bigr] = P_3\, R \psi'_{\rm out}(R) + P_4\, \psi_{\rm out},
\label{match1_even} 
\end{equation}
in which the coefficients $P_1$, $P_2$, $P_3$, $P_4$ are functions of $\ell$, $\Gamma$, $M/R$, and $i\omega R$.

To see how Eq.~(\ref{match1_even}) can be turned into an eigenvalue equation for the quasinormal mode frequencies $\omega$, we imagine that the Regge-Wheeler equation (\ref{RW_eqn_in}) is integrated with a trial (complex) frequency $\omega$ inside the shell, with the boundary condition that $\psi$ be nonsingular at $r=0$; the integration delivers a solution $\hat{\psi}_{\rm in}(r, \omega)$ and its  derivative. We also imagine that Eq~(\ref{RW_eqn_ext}) is integrated outside the shell, with the boundary condition that $\psi$ be regular at $r=\infty$ (so that the master variable describes a gravitational wave that is purely outgoing at future null infinity); this returns a solution $\hat{\psi}_{\rm out}(r, \omega)$ and its derivative. The interior and exterior solutions are both defined up to an overall multiplicative constant, and we write
\begin{equation}
\psi_{\rm in} = N_{\rm in}\, \hat{\psi}_{\rm in}, \qquad
\psi_{\rm out} = N_{\rm out}\, \hat{\psi}_{\rm out}
\end{equation}
for the correct solution to the global problem, where $N_{\rm in}$ and $N_{\rm out}$ are constants, and where the normalization of the hatted solutions is chosen arbitrarily. The global solution to the Regge-Wheeler equation must satisfy Eq.~(\ref{match1_even}). The second of these equations determines the ratio $N_{\rm out}/N_{\rm in}$ --- the overall normalization of the master function is arbitrary. The first equation becomes
\begin{equation}
C_1\, \eta_{\rm out} + C_2\, \eta_{\rm in} + C_3\, \eta_{\rm in} \eta_{\rm out} - C_4 = 0,
\label{match2_even}
\end{equation}
in which
\begin{equation}
\eta_{\rm in} := \frac{ R \hat{\psi}'_{\rm in}(R, \omega) }{ \hat{\psi}_{\rm in}(R, \omega) }, \qquad
\eta_{\rm out} := \frac{ R \hat{\psi}'_{\rm out}(R, \omega) }{ \hat{\psi}_{\rm out}(R, \omega) },
\label{eta_def} 
\end{equation}
while $C_1 := (1-P_1)$, $C_2 := (P_4-1)$, $C_3 := P_3$, and $C_4 :=P_2$. Equation (\ref{match2_even}) is an eigenvalue equation for the mode frequencies.  

The quantities $C_n$ that appear in Eq.~(\ref{match2_even}) are functions $\ell$, $\Gamma$, $M/R$, and $i\omega R$. After eliminating a common factor to simplify the expressions, we find that they possess the structure
\begin{subequations}
\label{Cn_structure} 
\begin{align}
C_1 &=c_1^0 + c_1^1\, (i\omega R) + c_1^2\, (i\omega R)^2 + c_1^3\, (i\omega R)^3
+ c_1^4\, (i\omega R)^4 + c_1^5\, (i\omega R)^5, \\
C_2 &= (1-2M/R)^{-1/2} \bigl[ c_2^0 + c_2^1\, (i\omega R) + c_2^2\, (i\omega R)^2
+ c_2^3\, (i\omega R)^3 + c_2^4\, (i\omega R)^4 + c_2^5\, (i\omega R)^5 \bigr], \\
C_3 &=c_3^0 + c_3^2\, (i\omega R)^2 + c_3^4\, (i\omega R)^4, \\
C_4 &= \frac{1}{4} (1-2M/R)^{-1/2} \bigl[ c_4^0 + c_4^1\, (i\omega R) + c_4^2\, (i\omega R)^2
+ c_4^3\, (i\omega R)^3 + c_4^4\, (i\omega R)^4 + c_4^5\, (i\omega R)^5 \bigr],
\end{align}
\end{subequations}
with coefficients $c_n^p$ that depend on $\ell$, $\Gamma$, and $M/R$. These are listed in Appendix~\ref{app:coefficients}.

\subsection{Symmetry}

It is easy to verify that the perturbation equations (\ref{pert_eqns_even_ext}) and (\ref{pert_eqns_even_in}) are such that the variables $p^*_{uu}$ and $K^*$ are solutions with frequency $-\omega^*$ when $p_{uu}$ and $K$ are solutions with frequency $\omega$; here the asterisk denotes complex conjugation. The same observation applies to the master variable $\psi$, and we conclude that the eigenvalue problem of Eq.~(\ref{match2_even}) is satisfied with the frequency $-\omega^*$ when it is solved with the frequency $\omega$. Writing $\omega = \omega_{\Re} + i \omega_{\Im}$ with $\omega_{\Re}$ and $\omega_{\Im}$ real, the symmetry implies that quasinormal modes come in frequency pairs
\begin{equation}
\omega_{\Re} + i \omega_{\Im}, \qquad -\omega_{\Re} + i \omega_{\Im},
\label{mode_pairs}
\end{equation}
so that the spectrum is reflection-symmetric about the imaginary axis.

\section{Quasinormal modes: Odd parity}
\label{sec:odd}

We continue our study of the quasinormal modes of the thin-shell spacetime constructed in Sec.~\ref{sec:unperturbed}. Here we turn to the odd-parity sector of the gravitational and fluid perturbations. 

\subsection{Exterior metric}
\label{subsec:odd_ext} 

We again work in the Regge-Wheeler gauge and express the perturbed metric outside the shell as $g_{uu} = -f$ (with $f := 1-2M/r$), $g_{ur} = -1$, $g_{AB} = r^2 \Omega_{AB}$, and
\begin{subequations}
\label{metric_odd_ext}
\begin{align}
g_{uA} &= r p_u(r)\, X^{\ell m}_A(\theta,\phi)\, e^{-i\omega u}, \\
g_{rA} &= r p_r(r)\, X^{\ell m}_A(\theta,\phi)\, e^{-i\omega u},
\end{align}
\end{subequations}
where $X^{\ell m}_A := -\varepsilon_A^{\ B} D_B Y^{\ell m}$ are odd-parity vector harmonics. In the context of this equation, the unit two-sphere is charted with the spacetime angles $\theta^A = (\theta, \phi)$, and $\varepsilon_{AB}$ (with independent component $\varepsilon_{\theta\phi} = \sin\theta$) is the Levi-Civita tensor on the two-sphere. We again omit the label $\ell m$ on the radial functions $p_{u}$ and $p_{r}$, as well as an implicit summation over these labels. Our considerations are again restricted to $\ell \geq 2$. 

The Einstein field equations, linearized about the Schwarzschild solution of Eq.~(\ref{schw}), produce the system of first-order differential equations 
\begin{subequations}
\label{pert_eqns_odd_ext}
\begin{align}
r p_{u}' &= \biggl[ 1 + \frac{(\ell-1)(\ell+2)}{i\omega r} \biggr]\, p_u
- \biggl[ i\omega r + \frac{(\ell-1)(\ell+2) f}{i\omega r} \biggr]\, p_r, \\
r f p_{r}' &= \biggl[ 2 + \frac{(\ell-1)(\ell+2)}{i\omega r} \biggr]\, p_u
- \biggl[ 1 + 2i\omega r + \frac{(\ell-1)(\ell+2) f}{i\omega r} \biggr]\, p_r
\end{align}
\end{subequations}
for the perturbation variables. The metric perturbation can be encoded within the Cunningham-Price-Moncrief master function \cite{cunningham-etal:78, martel-poisson:05}, 
\begin{equation}
\psi := \frac{i\omega r}{(\ell-1)(\ell+2)} \Bigl( r p'_u - p_u + i\omega r\, p_r \Bigr),
\end{equation}
which satisfies the Regge-Wheeler equation (\ref{RW_eqn_ext}). With Eqs.~(\ref{pert_eqns_odd_ext}), we find that the master variable and its derivative can be expressed as 
\begin{equation}
\psi = p_u - f\, p_r, \qquad
r\psi' = -p_u + (i\omega r + f)\, p_r,
\label{psi_odd_ext} 
\end{equation}
so that $p_{u}$ and $p_r$ are in a one-to-one correspondence with $\psi$ and $r\psi'$.

\subsection{Interior metric}
\label{subsec:odd_in} 

The perturbed metric inside the shell is written as $g_{uu} = -F$ (with $F := 1-2M/R$), $g_{ur} = -F^{1/2}$, $g_{AB} = r^2 \Omega_{AB}$, and
\begin{subequations}
\label{metric_odd_in}
\begin{align}
g_{uA} &= F^{1/2}\, r p_u(r)\, X^{\ell m}_A(\theta,\phi)\, e^{-i\omega u}, \\
g_{rA} &= r p_r(r)\, X^{\ell m}_A(\theta,\phi)\, e^{-i\omega u}.
\end{align}
\end{subequations}
The radial functions $p_{u}$ and $p_{r}$ are distinct from those that appear in the metric of Eq.~(\ref{metric_odd_ext}). 

The vacuum Einstein field equations produce the system of differential equations
\begin{subequations}
\label{pert_eqns_odd_in}
\begin{align}
r p_{u}' &= \biggl[ 1 + \frac{(\ell-1)(\ell+2)}{iw r} \biggr]\, p_u
- \biggl[ iw r + \frac{(\ell-1)(\ell+2)}{iw r} \biggr]\, p_r, \\
r p_{r}' &= \biggl[ 2 + \frac{(\ell-1)(\ell+2)}{iw r} \biggr]\, p_u
- \biggl[ 1 + 2iw r + \frac{(\ell-1)(\ell+2)}{iw r} \biggr]\, p_r
\end{align}
\end{subequations}
for the perturbation variables, where $w := F^{-1/2}\, \omega$ is a rescaled frequency. The interior version of the master variable is defined by
\begin{equation}
\psi := \frac{iw r}{(\ell-1)(\ell+2)} \Bigl( r p'_u - p_u + iw r\, p_r \Bigr),
\end{equation}
and it satisfies Eq.~(\ref{RW_eqn_ext}). With Eqs.~(\ref{pert_eqns_odd_in}), we have that 
\begin{equation}
\psi = p_u - p_r, \qquad
r\psi' = -p_u + (iw r + 1)\, p_r,
\label{psi_odd_in} 
\end{equation}
so that $p_{u}$ and $p_r$ are again in a one-to-one correspondence with $\psi$ and $r\psi'$.

\subsection{Deformed shell}
\label{subsec:odd_shell} 

Next we examine the perturbation of the shell. We again use $(u,\vartheta^A)$ as intrinsic coordinates on the hypersurface, with the angular coordinates $\vartheta^A = (\vartheta,\varphi)$ playing the role of Lagrangian labels for fluid elements. The shell's embedding relations are now 
\begin{subequations}
\label{embedding_odd}
\begin{align}
u &= u, \\
r &= R, \\ 
\theta^A &= \vartheta^A + \xi\, \Omega^{AB} X^{\ell m}_B(\vartheta,\varphi)\, e^{-i\omega u},
\end{align}
\end{subequations}
where $\xi$ is a constant and the vector harmonics are now expressed in terms of the intrinsic angles. In the context of this equation and those below, $\Omega_{AB} := \mbox{diag}[1,\sin^2\vartheta]$. The tangent vectors $e^\alpha_u$ and $e^\alpha_A$ are again obtained by differentiating the embedding relations with respect to the intrinsic coordinates. The normal vector $n_\alpha$ is now proportional to the gradient of $r-R$; on the exterior face of the shell its only nonvanishing component is $n_r = F^{-1/2}$, while it is $n_r = 1$ on the interior face. 

The physics of the shell's matter was described in Sec.~\ref{sec:matter}. Because the energy density $\mu$ and pressure $p$ are scalars, they are not altered by an odd-parity perturbation. We also have that the velocity field $u^a$ is unchanged; this follows because (i) $u^u$ cannot acquire an odd-parity perturbation, and (ii) $u^A = 0$ by virtue of the Lagrangian nature of the intrinsic angles $\vartheta^A$. The surface energy-momentum tensor Eq.~(\ref{S_ab}) is perturbed only because the induced metric $h_{ab}$ is perturbed.    

\subsection{Induced metric}
\label{subsec:odd_induced} 

The induced metric $h_{ab}$ on the deformed shell is computed from Eq.~(\ref{h_ab}) and the spacetime metrics of Eqs.~(\ref{metric_odd_ext}) and (\ref{metric_odd_in}). We obtain 
\begin{subequations}
\label{induced_metric_odd}
\begin{align}
h_{uu} &= -F, \\
h_{uA} &= R {\scrpt A}\, X_A^{\ell m}(\vartheta,\varphi)\, e^{-i\omega u}, \\
h_{AB} &= R^2 \Omega_{AB} + R^2 {\scrpt B}\, X^{\ell m}_{AB}(\vartheta,\varphi)\, e^{-i\omega u},
\end{align}
\end{subequations}
where $X_{AB}^{\ell m} := \frac{1}{2}( D_A X_B^{\ell m} + D_B X_A^{\ell m} )$ are odd-parity tensor harmonics; we recall that $D_A$ is the covariant-derivative operator compatible with $\Omega_{AB}$. The components of the induced metric are given by
\begin{subequations}
\begin{align}
{\scrpt A}^{\rm out} &= p^{\rm out}_u(R) - i\omega R\, \xi^{\rm out}, \\ 
{\scrpt B}^{\rm out} &= 2\xi^{\rm out}
\end{align}
\end{subequations}
when it is computed on the exterior side of the shell. On the interior side we have instead 
\begin{subequations}
\begin{align}
{\scrpt A}^{\rm in} &= F^{1/2}\, p^{\rm in}_u(R) - i\omega R\, \xi^{\rm in}, \\ 
{\scrpt B}^{\rm in} &= 2\xi^{\rm in}.
\end{align}
\end{subequations}
Continuity of the induced metric across the shell implies that 
\begin{equation}
\xi^{\rm in} = \xi^{\rm out}, \qquad
p^{\rm in}_u(R) = F^{-1/2}\, p^{\rm out}_u(R). 
\label{junction1_odd}
\end{equation} 
We shall enforce these conditions, and no longer distinguish between the exterior and interior components of the induced metric.

The conservation equation $D_b S^{ab} = 0$ implies that ${\scrpt A} = 0$, so that 
\begin{equation}
i\omega R\, \xi^{\rm out} = p^{\rm out}_u(R).
\label{xi_sol_odd}
\end{equation}
The Lagrangian displacement is therefore determined in terms of the metric perturbation. 

\subsection{Extrinsic curvature}
\label{subsec:odd_extrinsic} 

The extrinsic curvature of the deformed shell is computed from Eq.~(\ref{Kab}) and the spacetime metrics of  Eqs.~(\ref{metric_odd_ext}) and (\ref{metric_odd_in}). We obtain 
\begin{subequations}
\label{Kab_odd_out} 
\begin{align}
K^{\rm out}_{uu} &= -\frac{M}{R^2} F^{1/2}, \\
K^{\rm out}_{uA} &= {\scrpt C}^{\rm out}\, X^{\ell m}_A(\vartheta,\varphi)\, e^{-i\omega u}, \\
K^{\rm out}_{AB} &= R F^{1/2}\, \Omega_{AB}
+ R {\scrpt D}^{\rm out}\, X^{\ell m}_{AB}(\vartheta,\varphi)\, e^{-i\omega u},
\end{align}
\end{subequations}
for the shell's exterior face, and 
\begin{subequations}
\label{Kab_odd_in} 
\begin{align}
K^{\rm in}_{uu} &= 0, \\
K^{\rm in}_{uA} &= {\scrpt C}^{\rm in}\, X^{\ell m}_A(\vartheta,\varphi)\, e^{-i\omega u}, \\
K^{\rm in}_{AB} &= R\, \Omega_{AB}
+ R {\scrpt D}^{\rm in}\, X^{\ell m}_{AB}(\vartheta,\varphi)\, e^{-i\omega u}
\end{align}
\end{subequations}
for the interior face. We have introduced
\begin{subequations}
\begin{align}
{\scrpt C}^{\rm out} &:= \frac{1}{2} F^{1/2} \Bigl( R p'_u(R) + p_u(R) + i \omega R\, p_r(R) - 2i\omega R\, \xi \Bigr), \\
{\scrpt D}^{\rm out} &:= F^{1/2} \Bigl( F^{-1}\, p_u(R) - p_r(R) + 2\xi \Bigr),
\end{align}
\end{subequations}
and
\begin{subequations}
\begin{align}
{\scrpt C}^{\rm in} &:= \frac{1}{2} F^{1/2} \Bigl( R p'_u(R) + p_u(R) + i w R\, p_r(R) - 2iw R\, \xi \Bigr), \\
{\scrpt D}^{\rm in} &:= p_u(R) - p_r(R) + 2\xi, 
\end{align}
\end{subequations}
where we omit the labels ``out'' or ``in'' on the quantities that occur on the right-hand sides. Equations (\ref{pert_eqns_odd_ext}),  (\ref{pert_eqns_odd_in}), (\ref{junction1_odd}), and (\ref{xi_sol_odd}) imply that the components of the extrinsic curvature can be expressed algebraically in terms of the external and internal values of $p_u(R)$ and $p_r(R)$.   

\subsection{Matching equations and eigenvalue problem}

The shell's energy-momentum tensor, given by Eq.~(\ref{S_ab}), must be related to the jump in extrinsic curvature by the Israel condition of Eq.~(\ref{israel_b}). This requirement produces
\begin{equation}
p^{\rm in}_r(R) = F^{1/2}\, p^{\rm out}_r(R).
\label{junction2_odd}
\end{equation}
This matching condition, combined with Eq.~(\ref{junction1_odd}),
\begin{equation}
p^{\rm in}_u(R) = F^{-1/2}\, p^{\rm out}_u(R),
\label{junction1_odd_repeat}
\end{equation}
gives rise to an eigenvalue problem for the mode frequency $\omega$.

We use Eqs.~(\ref{psi_odd_ext}) and (\ref{psi_odd_in}) to express $p_u$ and $p_r$ in terms of the master function $\psi$ and its derivative, and rewrite Eqs.~(\ref{junction2_odd}) and (\ref{junction1_odd_repeat}) as
\begin{equation}
\bigl[ r\psi' \bigr] = -\bigl(1 - F^{-1/2} \bigr) \psi_{\rm out}, \qquad
\bigl[ \psi \bigr] = \bigl(1 - F^{-1/2} \bigr) \psi_{\rm out}.
\label{match1_odd} 
\end{equation}
The strategy is now identical to what was described in Sec.~\ref{subsec:even_eigenvalue}. Integration of the interior Regge-Wheeler equation (\ref{RW_eqn_in}) returns a solution $\hat{\psi}_{\rm in}(r, \omega)$ and its derivative, while integration of the exterior equation (\ref{RW_eqn_ext}) delivers $\hat{\psi}_{\rm out}(r, \omega)$ and its derivative. The global fields are 
\begin{equation}
\psi_{\rm in} = N_{\rm in}\, \hat{\psi}_{\rm in}, \qquad
\psi_{\rm out} = N_{\rm out}\, \hat{\psi}_{\rm out}, 
\end{equation}
where $N_{\rm in}$ and $N_{\rm out}$ are normalization constants. The equation for $[\psi]$ determines the ratio $N_{\rm out}/N_{\rm in}$, and the equation for $[r\psi']$ becomes
\begin{equation}
F^{1/2}\, \eta_{\rm out} - \eta_{\rm in} + F^{1/2} - 1 = 0, 
\label{match2_odd}
\end{equation}
in which
\begin{equation}
\eta_{\rm in} := \frac{ R \hat{\psi}'_{\rm in}(R, \omega) }{ \hat{\psi}_{\rm in}(R, \omega) }, \qquad
\eta_{\rm out} := \frac{ R \hat{\psi}'_{\rm out}(R, \omega) }{ \hat{\psi}_{\rm out}(R, \omega) }. 
\end{equation}
Equation (\ref{match2_odd}) is an eigenvalue equation for the mode frequencies. In this odd-parity case, the dependence on $\omega$ is contained entirely within $\eta_{\rm in}$ and $\eta_{\rm out}$; compare this with Eq.~(\ref{match2_even}), in which the coefficients $C_n$ are quintic polynomials in $i\omega R$. 

\subsection{Symmetry}

It is also true of the odd-parity quasinormal modes that they come in pairs 
\begin{equation}
\omega_{\Re} + i \omega_{\Im}, \qquad -\omega_{\Re} + i \omega_{\Im},
\end{equation}
as was first expressed by Eq.~(\ref{mode_pairs}) in the even-parity case. Here the conclusion follows from the perturbation equations (\ref{pert_eqns_odd_ext}) and (\ref{pert_eqns_odd_in}), which reveal that the variables $p^*_u$ and $p^*_r$ are solutions with frequency $-\omega^*$ when $p_u$ and $p_r$ are solutions with frequency $\omega$. The same observation applies to the master variable $\psi$, and we see that Eq.~(\ref{match2_odd}) is satisfied with the frequency $-\omega^*$ when it holds for the frequency $\omega$.

\section{Integration of the Regge-Wheeler equation}
\label{sec:RWintegration}

In this section we describe the numerical and analytical methods we employed to integrate the Regge-Wheeler equation for the metric perturbation both inside and outside the shell. 

\subsection{Interior function}

The interior version of the Regge-Wheeler equation, given by Eq.~(\ref{RW_eqn_in}), admits an analytic solution in terms of spherical Bessel functions,
\begin{equation}
\psi_{\rm in}(r,\omega) = e^{-iwr}\, wr\, j_\ell(wr),
\label{psi_in_sln} 
\end{equation}
where we recall that $w := F^{-1/2}\omega$ with $F := 1-2M/R$. This choice of solution ensures that the master function is regular at $r=0$. For low values of $\ell$ the spherical Bessel function can be written explicitly in terms of trigonometric functions and polynomials; the numerical evaluation of Eq.~(\ref{psi_in_sln}) and its derivative is entirely straightforward.

For the analytic work to be presented below it is useful to record the representation
\begin{equation}
\psi_{\rm in} = \frac{1}{2} (-i)^{\ell+1} \Biggl[
\sum_{k=0}^\ell (i)^k \frac{(\ell+k)!}{2^k k! (\ell-k)!} \frac{1}{(wr)^k}
+ (-1)^{\ell+1} e^{-2iwr} \sum_{k=0}^\ell (-i)^k \frac{(\ell+k)!}{2^k k! (\ell-k)!} \frac{1}{(wr)^k} \Biggr],
\label{psi_in_sum}
\end{equation}
which follows from Eqs.~(10.49.6) and (10.49.7) of Ref.~\cite{NIST:10}. 

\subsection{Exterior function: numerical methods} 

The exterior version of the Regge-Wheeler equation, displayed in Eq.~(\ref{RW_eqn_ext}), must be integrated numerically with the condition that its solution be regular at $r=\infty$. In view of the implicit factor $e^{-i\omega u}$ attached to the master variable, the regular solution describes a wave that is purely outgoing at future null infinity. An incoming wave from past null infinity would instead be proportional to
\begin{equation}
e^{-i\omega v} = e^{-2i\omega r^*} e^{-i\omega u},
\end{equation} 
where $v := t + r^*$; this solution to the Regge-Wheeler equation is singular at infinity.

The regular solution to the Regge-Wheeler equation admits the asymptotic expansion
\begin{equation}
\psi_{\rm out}(r,\omega) = 1 - \frac{1}{2} \ell(\ell+1)\, \frac{1}{i\omega r}
+ \biggl[ \frac{1}{8}(\ell-1)\ell(\ell+1)(\ell+2) + \frac{3}{2} iM\omega \biggr]\, \frac{1}{(i\omega r)^2}
+ \cdots 
\label{psi_out_asymp}
\end{equation}
when $i\omega r \gg 1$, in which the ellipses denote terms of higher order in $(i\omega r)^{-1}$. This justifies the boundary conditions
\begin{equation}
\lim_{r \to \infty} \psi_{\rm out} = 1, \qquad
\lim_{r \to \infty} \varpi_{\rm out} = 0 
\end{equation} 
for the numerical integration, in which $\varpi_{\rm out} := r \psi'_{\rm out}$. We explored several ways of performing these integrations, and eventually settled on a collocation method based on a simultaneous expansion of $\psi_{\rm out}$ and $\varpi_{\rm out}$ in Chebyshev polynomials. For this we rewrite the system of first-order differential equations for the two functions in terms of the independent variable $y := 1-2M/r$, which we then subject to a linear transformation so that the final variable $x$ lies in the interval $-1 \leq x < 1$ when $R \leq r < \infty$. The collocation method allows us to implement the boundary conditions directly at $r=\infty$, and it produces accurate results (up to 8 significant digits) when the expansion is carried out with 50 terms. The method struggles, however, when $\epsilon := 2M\omega$ is small, because the master function grows large when $r$ becomes comparable to $3M$, which takes it inside the potential barrier of the Regge-Wheeler equation. (This difficulty is associated with a transmission coefficient proportional to $\epsilon^{-(\ell+1)}$ and is therefore inherent to the differential equation itself; it is not specific to this collocation method.) Fortunately, $\epsilon$ turns out to be not too small for most of the modes computed in Sec.~\ref{sec:results}, and our techniques perform well for these modes. For other modes, those for which $\epsilon$ is indeed small, we can turn to the analytical methods described below.

Because the numerics are delicate, we made sure to validate them with an entirely independent method of calculation. Following Mano, Suzuki, and Takasugi \cite{mano-suzuki-takasugi:96b}, we represent the master variable as an infinite sum of hypergeometric functions (away from $r=\infty$), or as a sum of Coulomb wave functions (away from $r=2M$); each set of basis functions depends on a parameter $\nu$ that is determined so that the expansions converge uniformly. From a comparison with the MST representation we could establish the reliability of our collocation method through 8 significant digits. For an additional check, we also compared our answers to those returned by the Regge-Wheeler solver supplied by the Black Hole Perturbation Toolkit \cite{BHPToolkit}; here also we get agreement through 8 significant digits.  

\subsection{Exterior function: analytical methods}
\label{subsec:approx}

The Regge-Wheeler equation (\ref{RW_eqn_ext}) can be integrated approximately when $\epsilon := 2M\omega$ is small. This can be achieved through the MST expansions described previously, or alternatively and equivalently, through the results of Poisson and Sasaki in Ref.~\cite{poisson-sasaki:95}. We write
\begin{equation}
\psi_{\rm out} = z e^{-iz} \bigl[ Y_0(z) + \epsilon Y_1(z) + O(\epsilon^2) \bigr],
\label{psi_vs_Y}
\end{equation}
in which $z := \omega r$, insert this within Eq.~(\ref{RW_eqn_ext}), and expand in powers of $\epsilon$. At zeroth order we find that the differential equation reduces to a spherical Bessel equation for $Y_0$, and we adopt the solution
\begin{equation}
Y_0 = (i)^{\ell+1} h_\ell^{(1)}(z),
\label{Y0} 
\end{equation}
where $h_\ell^{(1)}(z)$ is a spherical Hankel function that describes an outgoing wave at infinity; the numerical factor is chosen so that $\psi_{\rm out} \to 1$ when $z \to \infty$. At first order in $\epsilon$ we obtain a spherical Bessel equation for $Y_1$, with a source term constructed from $Y_0$. The solution is
\begin{equation}
Y_1 = (i)^{\ell+1} \biggl\{ c_\ell\, h_\ell^{(1)}
+ \Bigl[ \frac{\pi}{2} - \mbox{Si}(2z) + i \mbox{Ci}(2z) \Bigr] h_\ell^{(2)}
- A_\ell\, h_\ell^{(1)} + B_\ell\, h_\ell^{(2)}
- \frac{(\ell-2)(\ell+2)}{2\ell(2\ell+1)}\, h_{\ell-1}^{(1)}
+ \frac{(\ell-1)(\ell+3)}{2(\ell+1)(2\ell+1)}\, h_{\ell+1}^{(1)} \biggr\},
\label{Y1} 
\end{equation}
in which $\mbox{Si}(x) := \int_0^x t^{-1} \sin t\, dt$ and $\mbox{Ci}(x) := -\int_x^\infty t^{-1} \cos t\, dt$ are the sine and cosine integrals, respectively, and 
\begin{subequations}
\begin{align}
A_\ell &:= \frac{i}{2} \biggl[ 1
+ z^2 \sum_{p=1}^{\ell-1} \biggl( \frac{1}{p} + \frac{1}{p+1} \biggr) h_p^{(1)} h_p^{(2)} \biggr], \\
B_\ell &:= \frac{i}{2} \biggl[ -e^{2iz} 
+ z^2 \sum_{p=1}^{\ell-1} \biggl( \frac{1}{p} + \frac{1}{p+1} \biggr) h_p^{(1)} h_p^{(1)} \biggr].
\end{align}
\end{subequations}
The constant $c_\ell$ is arbitrary, and it corresponds to the freedom to renormalize the master function; in practice it can be chosen so that the asymptotic behavior $\psi_{\rm out} \to 1$ as $z \to \infty$ is respected through order $\epsilon$. 

For our purposes below we shall be interested in the behavior of $\psi_{\rm out}$ when $z \ll 1$. To extract this from Eqs.~(\ref{psi_vs_Y}), (\ref{Y0}), and (\ref{Y1}) we rely on Eqs.~(6.6.5), (6.6.7), (10.49.6), and (10.49.7) of Ref.~\cite{NIST:10}, together with the fact that
\begin{equation} 
-A_\ell\, h_\ell^{(1)} + B_\ell\, h_\ell^{(2)} = O(z^{-\ell}),
\end{equation} 
a statement that summarizes a large sample of special cases and is therefore likely to be correct in general (we were not able to devise a formal proof). With all this we arrive at
\begin{align}
\psi_{\rm out} &= (i)^\ell (2\ell-1)!!\, \frac{1}{z^\ell} \Biggl\{
(1 + \epsilon c_\ell) \biggl[ 1 - i z - \frac{\ell-1}{2\ell-1} z^2 + i \frac{\ell-2}{3(2\ell-1)} z^3
+ O(z^4) \biggr]
\nonumber \\ & \quad \mbox{} 
+ \epsilon \biggl[ \frac{(\ell-1)(\ell+3)}{2(\ell+1)} \biggl( \frac{1}{z} - i \biggr)
- \biggl( \frac{\pi}{2} + i\gamma + i \ln(2z) \biggr) + O(z) \biggr] \Biggr\},
\label{psi_out_approx}
\end{align}
where $\gamma \simeq 0.57721$ is the Euler-Mascheroni constant. In Eq.~(\ref{psi_out_approx}) we kept all terms that suit our purposes later, and discarded higher-order terms. From this we get
\begin{equation}
\eta_{\rm out} = -\ell - iZ + \frac{1}{2\ell-1} Z^2 + O(Z^4)
- \epsilon \biggl[ \frac{(\ell-1)(\ell+3)}{2(\ell+1)} \frac{1}{Z} + i + O(Z) \biggr],
\label{eta_out_approx}
\end{equation}
in which $Z := \omega R$ and $\eta_{\rm out}$ is defined by Eq.~(\ref{eta_def}).

For completeness we note that Eq.~(\ref{psi_in_sln}) leads to
\begin{equation}
\eta_{\rm in} = \ell + 1 - i \bar{Z} - \frac{1}{2\ell+3} \bar{Z}^2 + O(\bar{Z}^4),
\label{eta_in_approx}
\end{equation}
in which $\bar{Z} := (1-2M/R)^{-1/2}\, \omega R$; the factor of $(1-2M/R)^{-1/2}$ accounts for the ratio of frequencies $w/\omega$ between the interior and exterior versions of the Regge-Wheeler equation.

\section{Results: Quasinormal modes of a thin shell}
\label{sec:results} 

\subsection{Matter and wave modes}

It is useful to classify the spectrum of quasinormal modes according to the dominant scaling of $\epsilon := 2M\omega$ with $M/R$ when $M/R$ is small. We call a mode a {\it matter mode} when $\epsilon$ scales as $(M/R)^{3/2}$, so that $\omega$ scales as $(M/R^3)^{1/2}$; this is the behavior that emerges in the Newtonian limit, as we show in Sec.~\ref{sec:newton}. The family of matter modes can be followed in a continuous manner as $M/R$ increases, and the label is therefore unambiguous for any value of the shell's compactness. We call a mode a {\it wave mode} when $\epsilon$ scales instead as $M/R$, so that $\omega$ scales as $1/R$; these modes do not have a have Newtonian analogue. The family of wave modes is also  continuous as $M/R$ varies, and the label stays meaningful for any value of $M/R$. Matter modes exist only for the even-parity sector of the perturbation; wave modes exist for both sectors.  

We shall see that for each value of $\ell$ and for given values of $\Gamma$ and $M/R$, there are precisely four matter modes. As far as we can gather from our numerical results, two of these modes are purely imaginary, with $\omega_{\Re} = 0$, where we write the complex frequency as
\begin{equation}
\omega = \omega_{\Re} + i \omega_{\Im},
\end{equation}
with $\omega_{\Re}$ and $\omega_{\Im}$ both real. One of the modes has a positive $\omega_{\Im}$, while the other comes with a negative $\omega_{\Im}$; these values are equal in magnitude. In view of the exponential factor in front of all perturbation variables,
\begin{equation}
e^{-i\omega u} = e^{\omega_{\Im} u}\, e^{-i\omega_{\Re} u},
\end{equation} 
the first mode signals an {\it instability} of the system for any value of $\ell$ --- the perturbation grows exponentially with retarded-time $u$. The remaining two modes are complex, with $\omega_{\Im}$ small and negative; these describe a {\it stable} perturbation. The real components are equal in magnitude and opposite in sign. In view of the pairing behavior identified in Eq.~(\ref{mode_pairs}), each imaginary mode constitutes its own (degenerate) pair, while the two complex modes form a third pair with frequencies $\omega_{\Re} + i\omega_{\Im}$ and $-\omega_{\Re} + i\omega_{\Im}$.  

In both sectors of the perturbation (even-parity and odd-parity) there appears to be an infinite sequence of wave modes for each value of $\ell$ and given values of $\Gamma$ and $M/R$. These modes also come in pairs, with $\omega_{\Im} < 0$; these modes describe a {\it stable} perturbation of the system.  

We have already presented our results for $\ell=2$ in Sec.~\ref{sec: intro qnms}. Below we examine the case $\ell = 3$ in some detail. We have also computed quasinormal modes for $\ell = 4$ and observed that they are qualitatively similar to those for $\ell = 2$ and $\ell = 3$; there is no need to describe them here.   

\subsection{Even-parity matter modes: numerical results}

We find the same unstable matter mode for $\ell = 3$ as we did for $\ell = 2$. This is displayed in Fig.~\ref{fig:matter_imag, L=3, matter-modes even}, and we see that the frequency is purely imaginary, with a positive sign for all values of $M/R$ and $\Gamma$. This mode comes with a stable companion with a negative imaginary part. 

\begin{figure}[h]
\centering
\begin{minipage}{.5\textwidth}
  \centering
  \includegraphics[scale=0.5]{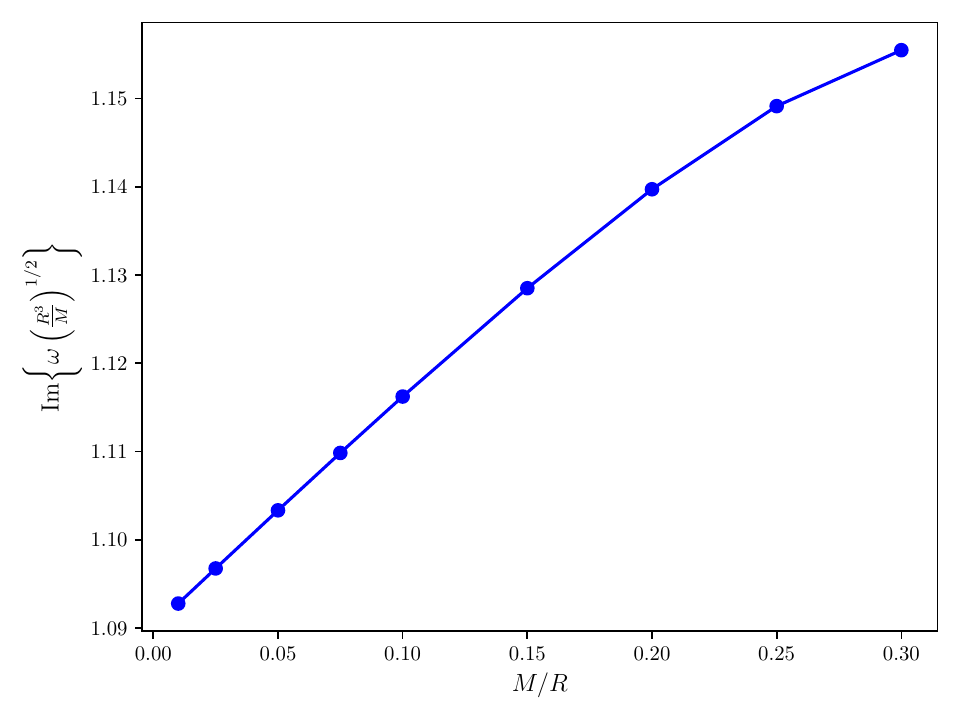}
\end{minipage}%
\begin{minipage}{.5\textwidth}
  \centering
  \includegraphics[scale=0.5]{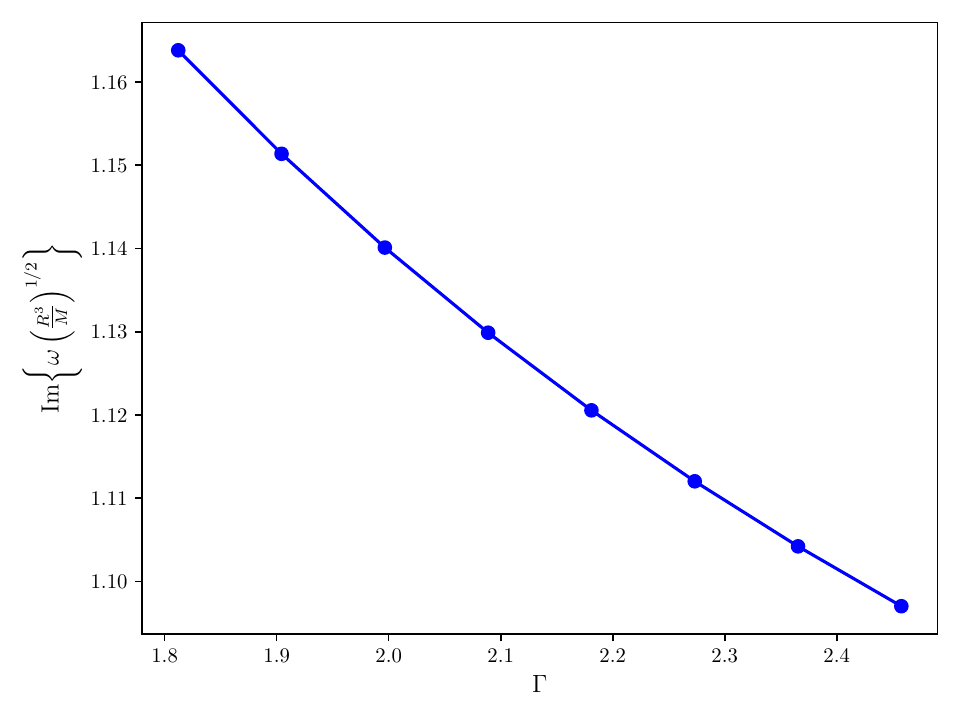}
\end{minipage}
\caption{Even-parity, unstable matter mode for $\ell = 3$. Left: Imaginary part of $\omega(R^3/M)^{1/2}$ in relation to $M/R$ for $\Gamma = 2$. Right: Imaginary part of the frequency in relation to $\Gamma$ for $M/R = 0.2$. The real part of the frequency vanishes. The dots are linked by guiding lines.}
\label{fig:matter_imag, L=3, matter-modes even}
\end{figure}

The matter modes for $\ell = 3$ also include a pair with a complex frequency, with a real part that can be of either sign and an imaginary part that is negative; these modes are stable. The results are displayed in Figure \ref{fig:matter_mixed vs MR, L=3, matter-modes even}.  

\begin{figure}[h]
\begin{minipage}[h]{0.47\linewidth}
\begin{center}
\includegraphics[width=1\linewidth]{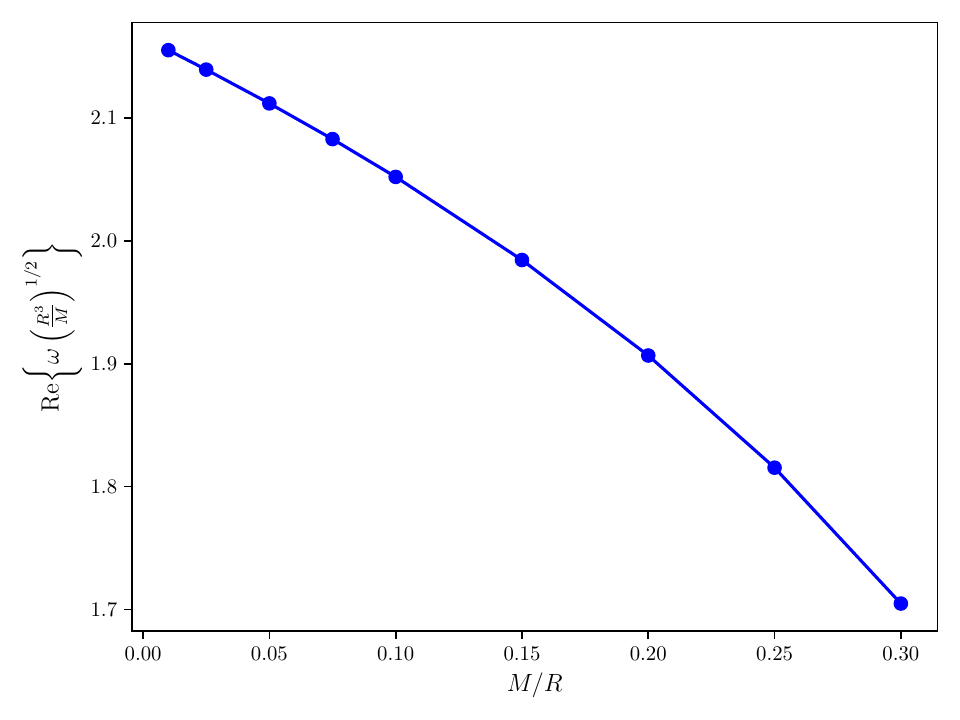} 
\end{center} 
\end{minipage}
\hfill
\begin{minipage}[h]{0.47\linewidth}
\begin{center}
\includegraphics[width=1\linewidth]{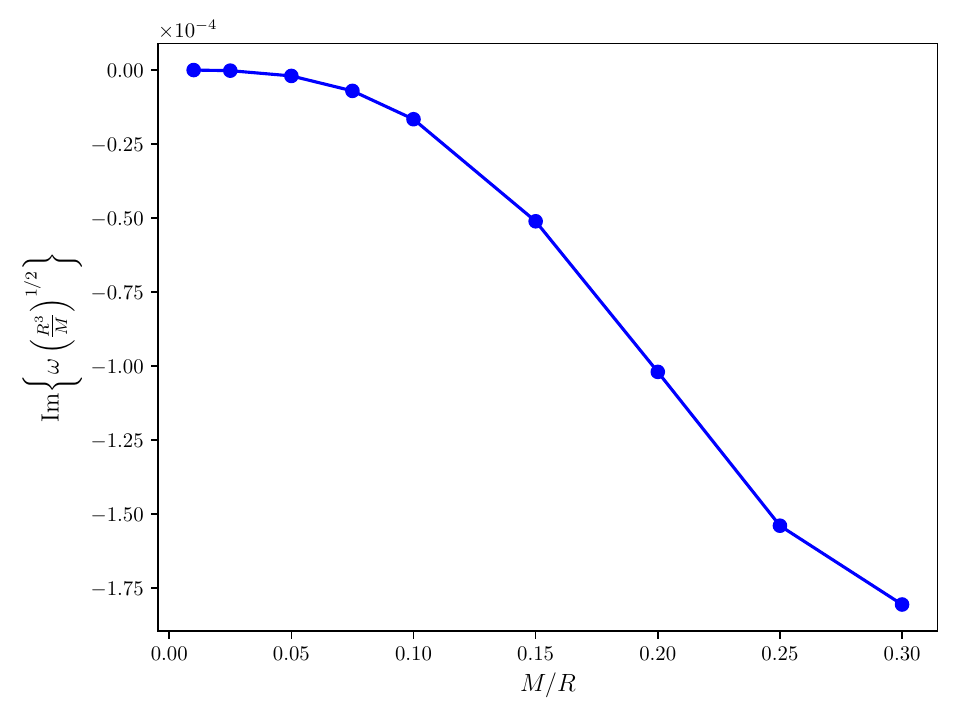} 
\end{center}
\end{minipage}
\vfill
\begin{minipage}[h]{0.47\linewidth}
\begin{center}
\includegraphics[width=1\linewidth]{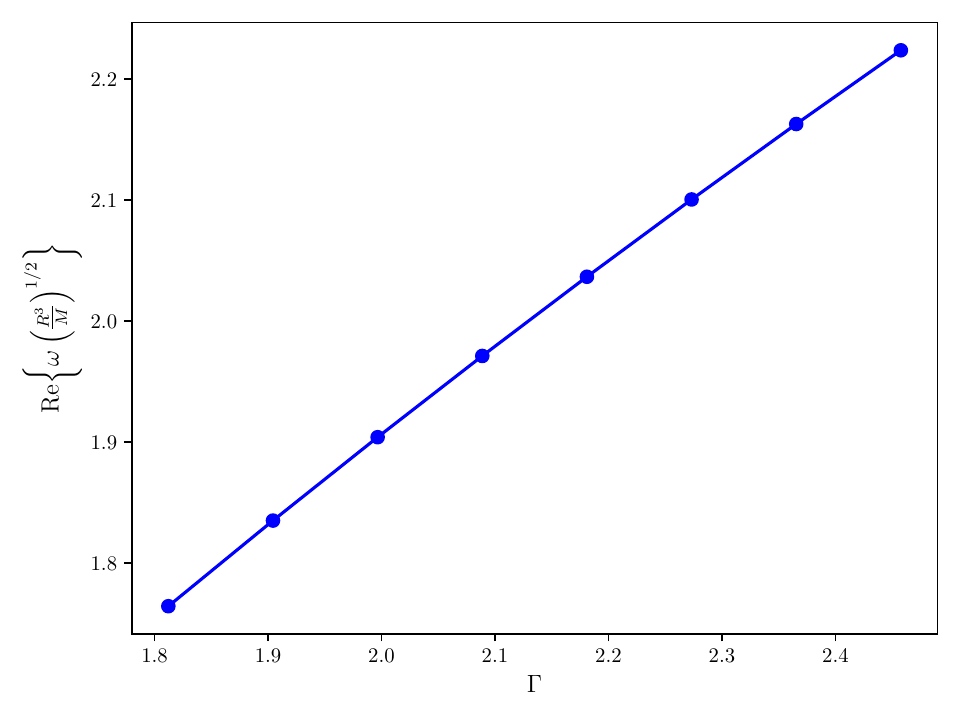} 
\end{center}
\end{minipage}
\hfill
\begin{minipage}[h]{0.47\linewidth}
\begin{center}
\includegraphics[width=1\linewidth]{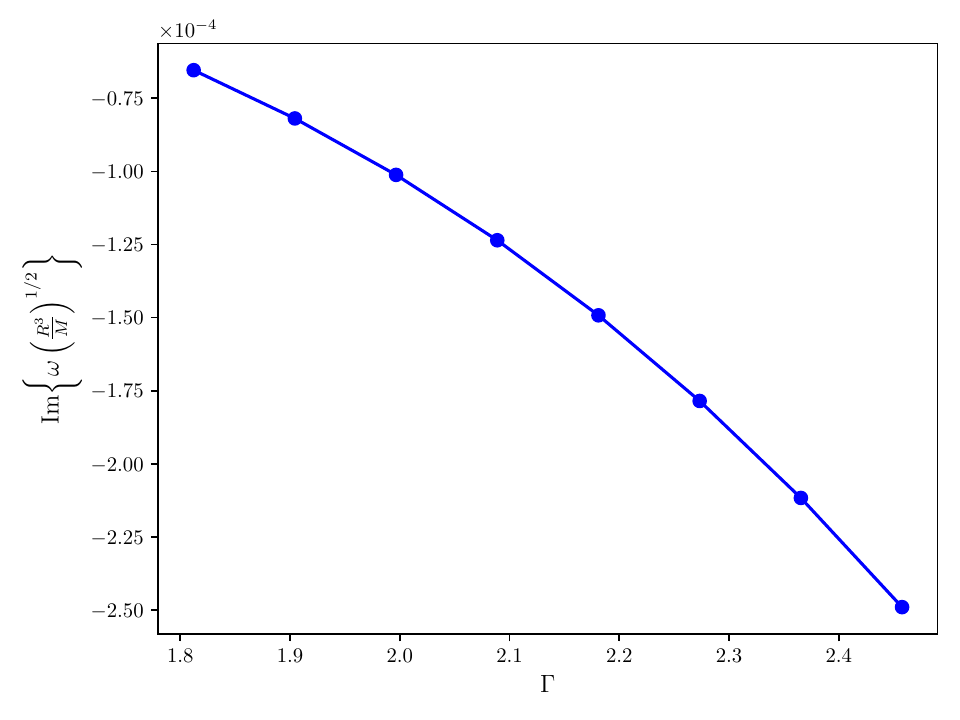} 
\end{center}
\end{minipage}
\caption{Even-parity, stable matter modes for $\ell=3$. Upper left: Real part of $\omega(R^3/M)^{1/2}$ in relation to $M/R$. Upper right: Imaginary part of the frequency in relation to $M/R$. Lower left: Real part of the frequency in relation to $\Gamma$. Lower right: Imaginary part of the frequency in relation to $\Gamma$. Upper plots: $\Gamma = 2$. Lower plots: $M/R = 0.2$. The dots are linked by guiding lines.}
\label{fig:matter_mixed vs MR, L=3, matter-modes even}
\end{figure}

\subsection{Even-parity matter modes: post-Newtonian regime}

We consider the spectrum of quasinormal modes when the shell compactness is small, $M/R \ll 1$, and examine those modes that come with a frequency parameter $\epsilon := 2M\omega$ that is also small. Our numerical results indicate that this requirement is met by matter modes only, which exist in the even-parity sector of the perturbation. We are therefore looking for eigenvalues for $\epsilon$ that scale as $(M/R)^{3/2}$ when $M/R$ is small, and here we find them with analytical methods. 

It is helpful to introduce a scaling parameter $v$, defined so that $v^2 := M/R$. With this we have that $\epsilon$ scales as $v^3$, and we write this as
\begin{equation}
\epsilon = 2\varsigma v^3,
\end{equation}
in which $\varsigma$ is a substitute for the frequency parameter, which we expand as
\begin{equation}
\varsigma = \varsigma_0 + \varsigma_1\, v + \varsigma_2\, v^2 + \varsigma_3\, v^3 + O(v^4). 
\end{equation}
We are working in a regime in which both $\epsilon$ and $Z := \omega R = \varsigma v$ are small, and we can make use of the analytical expressions for $\eta_{\rm out}$ and $\eta_{\rm in}$ obtained in Sec.~\ref{subsec:approx}. We insert the scaling relations within Eqs.~(\ref{eta_out_approx}) and (\ref{eta_in_approx}), to find that 
\begin{subequations}
\begin{align} 
\eta_{\rm out} &= -\ell - i\varsigma_0\, v - \biggl[ \frac{(\ell-1)(\ell+3)}{\ell+1} + \frac{\varsigma_0^2}{2\ell-1}
- i\varsigma_1 \biggr]\, v^2 + \biggl[ \frac{2\varsigma_0 \varsigma_1}{2\ell-1} - 2i\varsigma_0 - i\varsigma_2 \biggr]\, v^3
+ O(v^4), \\
\eta_{\rm in} &= \ell + 1 - i\varsigma_0\, v - \biggl( \frac{\varsigma_0^2}{2\ell+3} - i \varsigma_1 \biggr)\, v^2
- \biggl( \frac{2\varsigma_0 \varsigma_1}{2\ell+3} + i\varsigma_0 + i\varsigma_2 \biggr)\, v^3
+ O(v^4). 
\end{align}
\end{subequations} 
The coefficients $C_n$ that appear in the eigenvalue problem of Eq.~(\ref{match2_even}) can also be expanded in powers of $v$. We find that each coefficient begins at order $v^4$, and we keep additional terms of relative order $v$, $v^2$, and $v^3$.

With all this we find that Eq.~(\ref{match2_even}) becomes a sequence of algebraic equations, one at each order in $v$, for the expansion coefficients of the frequency parameter $\varsigma$. At order $v^0$ we get an equation for $\varsigma_0$,
\begin{equation} 
0 = \varsigma_0^4 - \frac{1}{4} \bigl[ (\ell^2+\ell+4)\Gamma - (\ell^2+\ell+6) \bigr]\, \varsigma_0^2 
- \frac{(\ell-1)\ell(\ell+1)(\ell+2)}{16(2\ell+1)} \bigl[ (2\ell-3) \Gamma - 4 \bigr].
\label{sigma0_eq} 
\end{equation}
At order $v$ we find that $\varsigma_1 = 0$. At order $v^2$ we obtain an equation for $\varsigma_2$,
\begin{align}
0 &= \biggl[ 4 \varsigma_0^2 - \frac{1}{2} (\ell^2+\ell+4) \Gamma
+ \frac{1}{2} (\ell^2+\ell+6) \biggr] \varsigma_0\, \varsigma_2
\nonumber \\ & \quad \mbox{} 
+ \frac{1}{16(2\ell-1)(2\ell+1)(2\ell+3)} \Bigl[ (24\ell^5+68\ell^4+238\ell^3+265\ell^2-73\ell-72) \Gamma
\nonumber \\ & \quad \mbox{} 
- 2(8\ell^5+16\ell^4+114\ell^3+131\ell^2-59\ell-30) \Bigr] \varsigma_0^2
\nonumber \\ & \quad \mbox{} 
+ \frac{(\ell-1)(\ell+1)\ell}{64(2\ell-1)(2\ell+1)^2(2\ell+3)}
\Bigl[ (2\ell+5)(40\ell^4-68\ell^3-98\ell^2+81\ell+90)\Gamma
\nonumber \\ & \quad \mbox{} 
- 4(\ell-2)(56\ell^3-12\ell^2-100\ell-69) \Bigr]. 
\label{sigma2_eq} 
\end{align}
And at order $v^3$ we discover that $\varsigma_3 = 0$. 

The solutions to Eq.~(\ref{sigma0_eq}) are
\begin{align}
\varsigma^2_{0\pm} &= \frac{1}{8} \bigl[ (\ell^2+\ell+4)\Gamma - (\ell^2+\ell+6) \bigr]
\pm \frac{1}{8} \biggl[ (\ell^2+\ell+4)^2 \Gamma^2
\nonumber \\ & \quad \mbox{} 
+ 2 \frac{2\ell^5-3\ell^4-40\ell^3-33\ell^2-46\ell-24}{2\ell+1} \Gamma
+ \frac{2\ell^5-11\ell^4+60\ell^3+53\ell^2+52\ell+36}{2\ell+1} \biggr]^{1/2}.
\label{sigma0_slns} 
\end{align}
For $\ell \geq 2$ and $\Gamma \geq 3/2$, which ensures that the shell is radially stable, we find that
\begin{equation} 
\varsigma^2_{0+} > 0, \qquad \varsigma^2_{0-} < 0.
\end{equation} 
Equation (\ref{sigma0_slns}) simplifies to
\begin{subequations}
\begin{align}
\varsigma^2_{0+} &= \frac{1}{4} (\ell^2+\ell+4) \Gamma + O(1), \\
\varsigma^2_{0-} &= -\frac{(\ell-1)\ell(\ell+1)(\ell+2)(2\ell-3)}{4(2\ell+1)(\ell^2+\ell+4)}
+ O(\Gamma^{-1}) 
\end{align}
\end{subequations}
when $\Gamma$ is large. 

\begin{figure}
\includegraphics[width=0.49\linewidth]{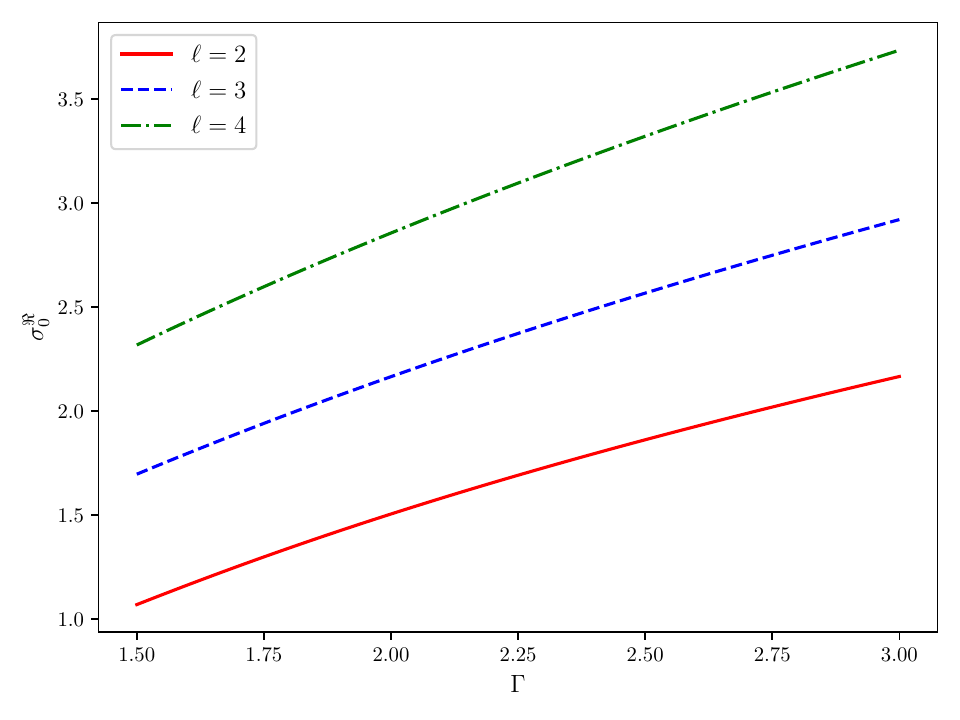}
\includegraphics[width=0.49\linewidth]{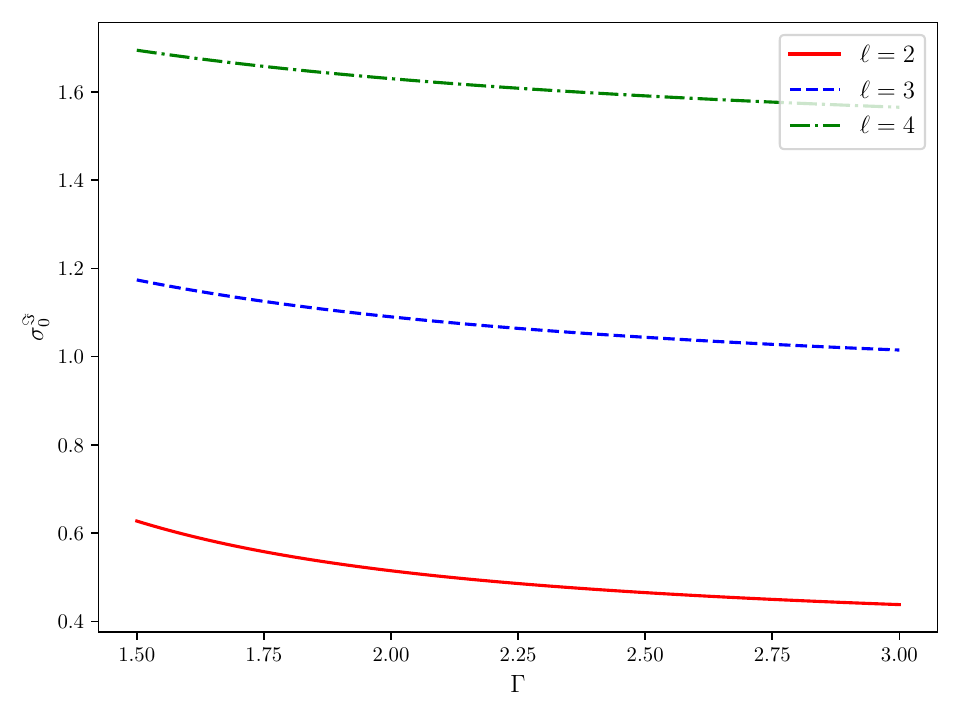}
\caption{Solutions to Eq.~(\ref{sigma0_eq}) for selected values of $\ell$, plotted as functions of $\Gamma$. Left panel: $\varsigma_0^{\Re}$. Right panel: $\varsigma_0^{\Im}$. Red curves: $\ell = 2$. Blue curves: $\ell = 3$. Green curves: $\ell = 4$.} 
\label{fig:sigma0} 
\end{figure}

We have a total of four solutions for $\varsigma_0$, two real values issued from $\varsigma^2_{0+}$, equal in magnitude and opposite in sign, and two imaginary values issued from $\varsigma^2_{0-}$, also equal and opposite. We use the notation
\begin{subequations}
\label{sigma0_list} 
\begin{align}
+\varsigma_0^{\Re} &: \mbox{real, positive} \\
-\varsigma_0^{\Re} &: \mbox{real, negative} \\
+i\varsigma_0^{\Im} &: \mbox{imaginary, positive} \\
-i\varsigma_0^{\Im} &: \mbox{imaginary, negative}
\end{align}
\end{subequations} 
to describe these solutions. The solutions $\varsigma_0^{\Re}$ and $\varsigma_0^{\Im}$ are plotted as functions of $\Gamma$ in Fig.~\ref{fig:sigma0}, for selected values of $\ell$.

\begin{figure}
\includegraphics[width=0.49\linewidth]{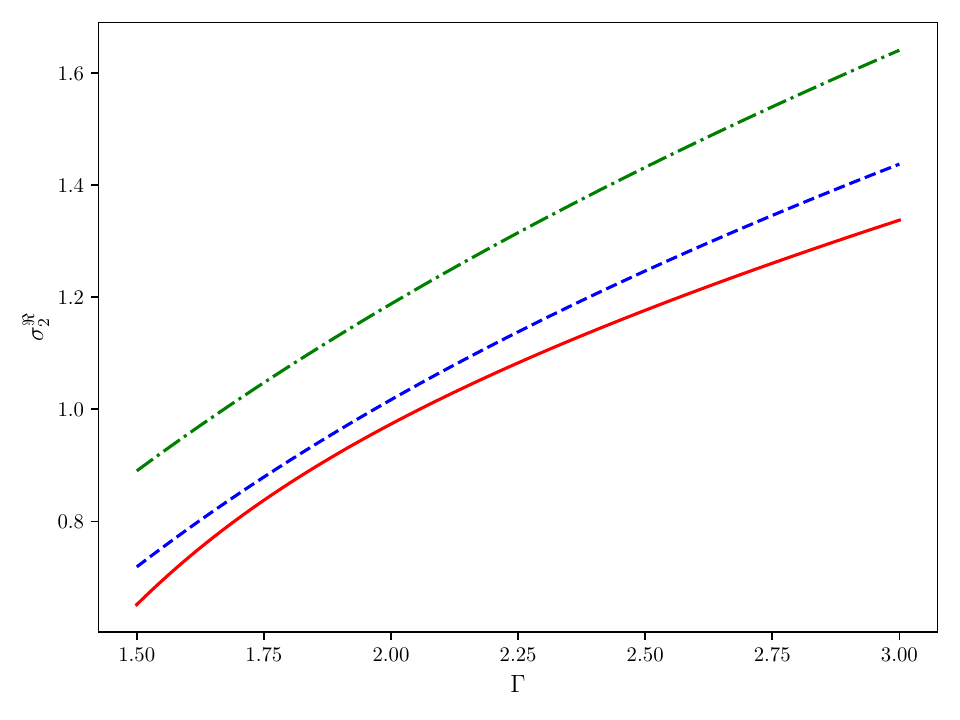}
\includegraphics[width=0.49\linewidth]{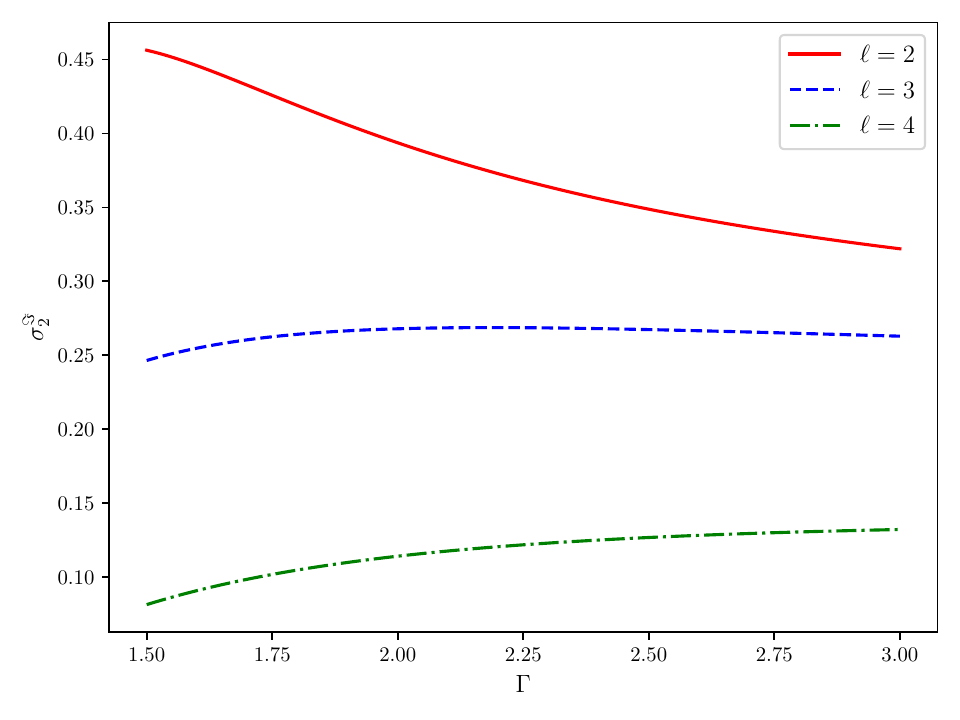}
\caption{Solutions to Eq.~(\ref{sigma2_eq}) for selected values of $\ell$, plotted as functions of $\Gamma$. Left panel: $\varsigma_2^{\Re}$. Right panel: $\varsigma_2^{\Im}$. Red curves: $\ell = 2$. Blue curves: $\ell = 3$. Green curves: $\ell = 4$.} 
\label{fig:sigma2} 
\end{figure}

With $\varsigma_0$ determined, Eq.~(\ref{sigma2_eq}) provides a unique assignment for $\varsigma_2$, one for each of the four solutions for $\varsigma_0$. The structure of Eq.~(\ref{sigma2_eq}) is such that
\begin{subequations}
\label{sigma2_list} 
\begin{align}
+\varsigma_0^{\Re} &: \quad \varsigma_2 = -\varsigma_2^{\Re} \\
-\varsigma_0^{\Re} &: \quad \varsigma_2 = +\varsigma_2^{\Re} \\
+i\varsigma_0^{\Im} &: \quad \varsigma_2 = +i\varsigma_2^{\Im} \\
-i\varsigma_0^{\Im} &: \quad \varsigma_2 = -i\varsigma_2^{\Im}, 
\end{align}
\end{subequations} 
where $\varsigma_2^{\Re}$ and $\varsigma_2^{\Im}$ are defined to be real and positive. In words, in the case of the real solutions for $\varsigma_0$, the positive value is corrected by a negative $\varsigma_2$ while the negative value is corrected by a positive $\varsigma_2$; in the case of the imaginary solutions, $\varsigma_2$ comes with the same sign as $\varsigma_0$. The solutions $\varsigma_2^{\Re}$ and $\varsigma_2^{\Im}$ are plotted as functions of $\Gamma$ in Fig.~\ref{fig:sigma2}, for selected values of $\ell$.

We summarize our findings by stating that in the post-Newtonian regime considered here, the matter-mode frequencies can be written as
\begin{equation}
\omega = (M/R^3)^{1/2} \bigl[ \varsigma_0 + \varsigma_2 (M/R) + O(M^2/R^2) \bigr],
\label{omega_PN}
\end{equation}
with the four solutions for $\varsigma_0$ listed in Eq.~(\ref{sigma0_list}) and determined from Eq.~(\ref{sigma0_slns}), and the corresponding four solutions for $\varsigma_2$ listed in Eq.~(\ref{sigma2_list}) and determined from Eq.~(\ref{sigma2_eq}). Two of these frequencies are real (with values equal in magnitude and opposite in sign), and the remaining two are imaginary (also with equal and opposite values). The presence of a mode with $\omega_{\Im} > 0$ for each value of $\ell$ implies that the shell is necessarily unstable to nonspherical perturbations. Our numerical results indicate that the conclusion stays valid beyond the post-Newtonian regime considered here, and holds for shells with a large compactness. 

\subsection{Even-parity wave modes: numerical results}

The even-parity wave-mode frequencies for $\ell = 3$ are displayed in Figure \ref{fig:Rw vs Gamma, L=3, w-modes even}. All frequencies have a negative imaginary part and are thus stable. 

In our numerical exploration of the eigenvalue problem we noticed what seemed to be a third class of modes, which did not behave like matter modes or wave modes. One such mode seemed to appear suddenly near $M/R = 0.1$ as we scanned the shell's compactness, and more modes were revealed as we kept increasing it ---  we could count as many as four modes at $M/R = 0.4$. We cannot claim that these new modes actually exist, as our numerical techniques (both the collocation method and the MST representation) were unable to return reliable solutions for the mode frequencies. We mention them here as a topic of further study in future work.  

\begin{figure}[h]
\centering
\begin{minipage}{.5\textwidth}
  \centering
  \includegraphics[scale=0.5]{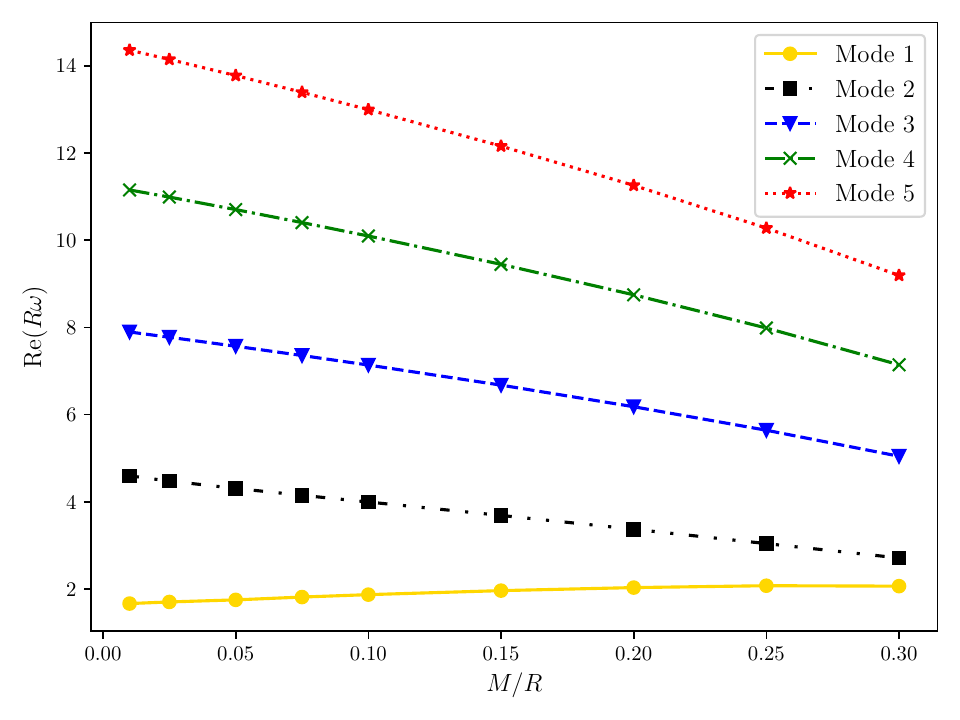}
\end{minipage}%
\begin{minipage}{.5\textwidth}
  \centering
  \includegraphics[scale=0.5]{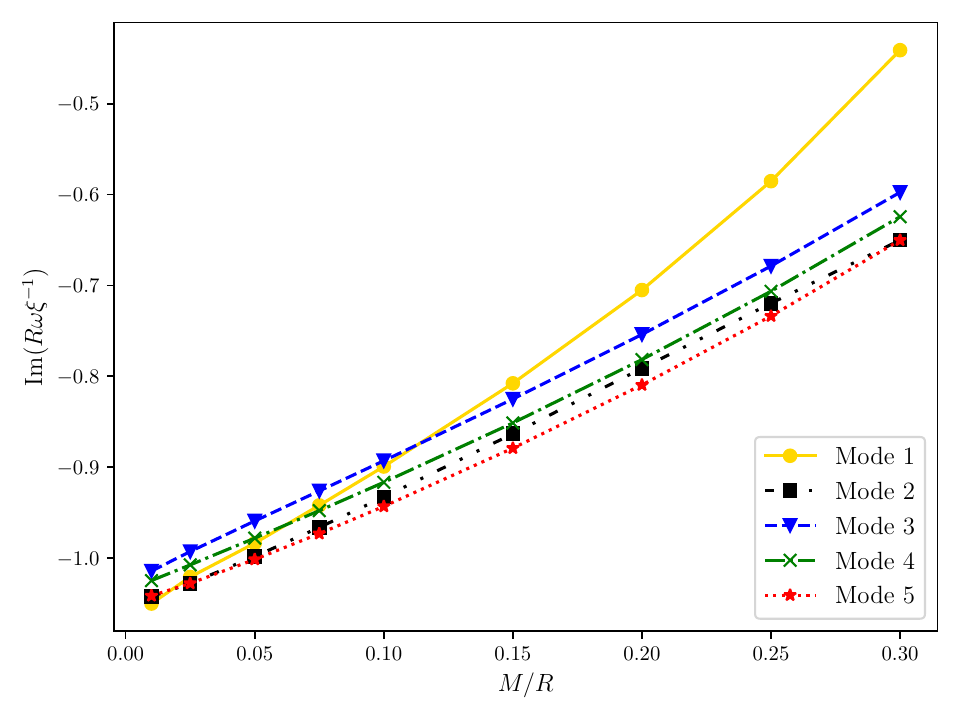}
\end{minipage}
\caption{Odd-parity wave modes for $\ell = 3$ and $\Gamma = 2$. Left: Real part of $R\omega$ in relation to $M/R$. Right: Imaginary part of $R\omega/\xi$ in relation to $M/R$, in which $\xi := \ln[2\ell(\ell+1)R/M]/2$ is introduced in Sec.~\ref{subsec: newt_regime}. Yellow dots: fundamental mode. Black dots: first overtone. Blue dots: second overtone.
Green dots: third overtone. Red dots: fourth overtone. The dots are linked by guiding lines.}
\label{fig:Rw vs Gamma, L=3, w-modes even}
\end{figure}

\subsection{Odd-parity wave modes: numerical results}

The frequencies for the $\ell = 3$, odd-parity wave modes are shown in Figure \ref{fig:Rw vs MR, L=3, w-modes odd}. All these modes have a frequency with a negative imaginary part and are therefore stable.

The ``curious'' modes mentioned previously were also encountered in the odd-parity case. Once more we cannot be sure about their actual existence, given the poor accuracy of our numerical solutions. 

\begin{figure}[h]
\centering
\begin{minipage}{.5\textwidth}
  \centering
  \includegraphics[scale=0.5]{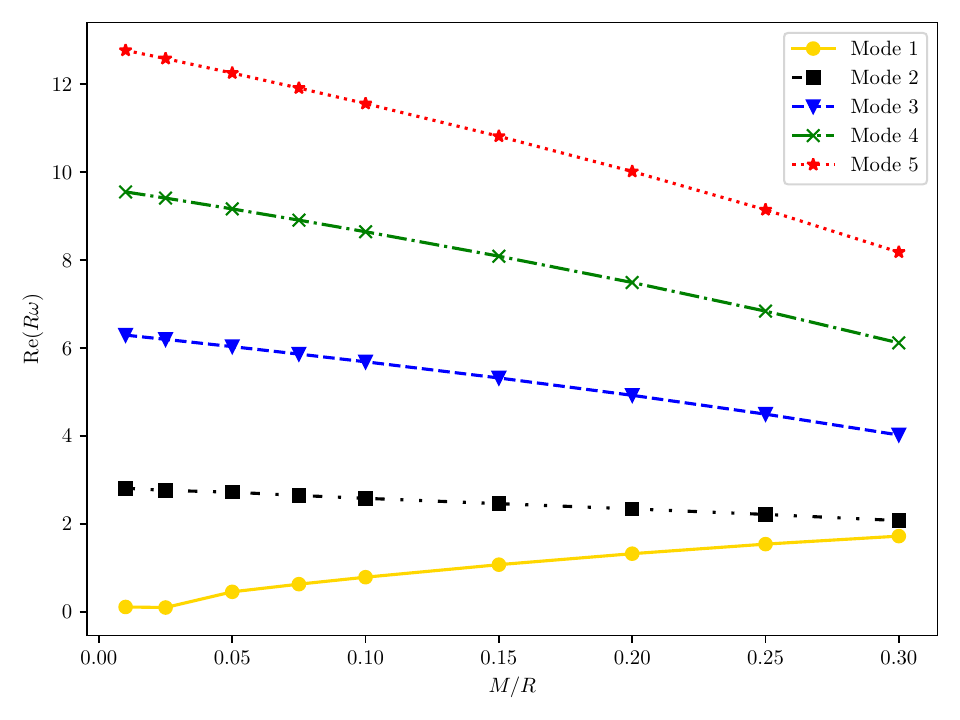}
\end{minipage}%
\begin{minipage}{.5\textwidth}
  \centering
  \includegraphics[scale=0.5]{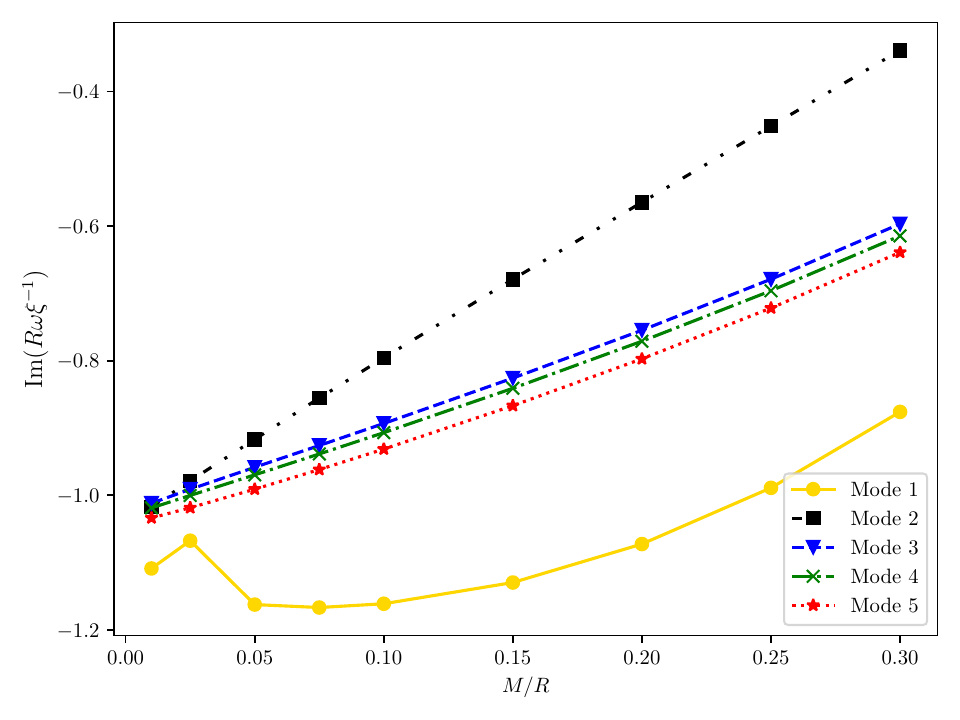}
\end{minipage}
\caption{Odd-parity wave modes for $\ell=3$. Left: Real part of $R\omega$ in relation to $M/R$. Right: Imaginary part of $R\omega/\xi$ in relation to $M/R$, in which $\xi := \ln[2\ell(\ell+1)R/M]/2$. Yellow dots: fundamental mode. Black dots: first overtone. Blue dots: second overtone. Green dots: third overtone. Red dots: fourth overtone. The dots are linked by guiding lines.}
\label{fig:Rw vs MR, L=3, w-modes odd}
\end{figure}

\subsection{Wave modes: Newtonian regime}
\label{subsec: newt_regime}

We again consider the spectrum of quasinormal modes when the shell compactness is small, $M/R \ll 1$, but this time we focus our attention on wave modes. Our considerations apply to both even-parity and odd-parity perturbations. We take advantage of the fact that according to our numerical results, $\omega R$ is larger than unity; we use this observation to derive an analytic solution to the eigenvalue equation. Our developments in this subsection are rather crude ($\omega R$ is actually not that large), but they do capture the main behavior of our numerical results, and they do reveal the dominant scaling of the wave modes when $M/R \ll 1$.

The solution to the interior Regge-Wheeler equation can be written as in Eq.~(\ref{psi_in_sum}), which we rewrite as
\begin{align}
\psi_{\rm in} &= \frac{1}{2} (-i)^{\ell+1} \Biggl\{
1 - \frac{1}{2} \ell(\ell+1) \frac{1}{i w r} + \frac{1}{8} (\ell-1)\ell(\ell+1)(\ell+2) \frac{1}{(i w r)^2} + \cdots
\nonumber \\ & \quad \mbox{} 
+ (-1)^{\ell+1} e^{-2iwr} \biggl[ 1 + \frac{1}{2} \ell(\ell+1) \frac{1}{i w r}
  + \frac{1}{8} (\ell-1)\ell(\ell+1)(\ell+2) \frac{1}{(i w r)^2} + \cdots \biggr] \Biggr\}, 
\end{align} 
where the ellipses denote terms of higher order in $(iwr)^{-1}$. We anticipate that the eigenvalues $w$ will come with a negative imaginary part, and therefore treat the exponential factor $e^{-2iwr}$ as a small quantity. Under these circumstances, a calculation of $\eta_{\rm in} := R\psi'_{\rm in}(R)/\psi_{\rm in}(R)$ returns
\begin{equation}
\eta_{\rm in} = \frac{\ell(\ell+1)}{2iwR} \Biggl\{ 1 + \frac{1}{iwR} + \cdots
- 2 (-1)^{\ell+1} e^{-2iwR} \biggl[ 1 + \frac{\ell(\ell+1)}{iwR} + \cdots \biggr] \Biggr\}.
\label{eta_in_wave}
\end{equation}
On the other hand, Eq.~(\ref{psi_out_asymp}) produces
\begin{align}
\eta_{\rm out} &= \frac{\ell(\ell+1)}{2i\omega R} \Biggl\{ 1 + \frac{1}{i\omega R} + \cdots
- \frac{M}{R} \biggl[ \frac{6}{\ell(\ell+1)} + \cdots \biggr] \Biggr\}
\nonumber \\ 
&= \frac{\ell(\ell+1)}{2iw R} \Biggl\{ 1 + \frac{1}{iw R} + \cdots
+ \frac{M}{R} \biggl[ \frac{(\ell-2)(\ell+3)}{\ell(\ell+1)} + \frac{2}{iwR} + \cdots \biggr] \Biggr\};
\label{eta_out_wave}
\end{align}
in the second line we re-expressed the exterior frequency $\omega$ in terms of the interior frequency $w$, and expanded through first order in $M/R$; recall that $\omega = (1-2M/R)^{1/2} w$.

In the even-parity sector of the perturbation, the eigenvalue problem for the mode frequencies is given by Eq.~(\ref{match2_even}), with coefficients $C_n$ listed in Eq.~(\ref{Cn_structure}). We subject these coefficients to a simultaneous expansion in powers of $(iwR)^{-1}$ and $M/R$, and obtain
\begin{subequations}
\begin{align}
C_1 &= k\Biggl\{ 1
- \frac{M}{R} \biggl[ \frac{12}{(\ell-1)\ell(\ell+1)(\ell+2)} iwR + \frac{3}{4} + \cdots \biggr]\Biggr\}, \\
C_2 &= k\Biggl\{ -1
- \frac{M}{R} \biggl[ \frac{12}{(\ell-1)\ell(\ell+1)(\ell+2)} iwR + \frac{1}{4} + \cdots \biggr] \Biggr\}, \\
C_3 &= - k \frac{M}{R} \biggl[ \frac{12}{(\ell-1)\ell(\ell+1)(\ell+2)} + \cdots \biggr], \\   
C_4 &= - k \frac{M}{R} \biggl[ \frac{\ell^2+\ell+10}{(\ell-1)(\ell+2)} + \cdots \biggr], 
\end{align}
\end{subequations}
in which $k := \frac{1}{8}(\ell-1)\ell(\ell+1)(\ell+2) (iwR)^4$. Note that $\Gamma$ does not occur in these asymptotic expansions. We insert all these results within the eigenvalue equation, retain only the leading-order terms, and obtain
\begin{equation}
(-1)^{\ell+1} \ell(\ell+1) \frac{e^{-2iwR}}{iwR}= -\frac{M}{R},
\label{Xeq_even}
\end{equation}
a transcendental equation for $iwR$.

The eigenvalue problem for the odd-parity sector of the perturbation is given by Eq.~(\ref{match2_odd}), which we also expand in powers of $M/R$. We insert our previous expressions for $\eta_{\rm in}$ and $\eta_{\rm out}$, retain only leading-order terms, and get
\begin{equation}
(-1)^{\ell+1} \ell(\ell+1) \frac{e^{-2iwR}}{iwR} = +\frac{M}{R},
\label{Xeq_odd}
\end{equation}
which is the same as Eq.~(\ref{Xeq_even}), except for the sign on the right-hand side.

We combine Eqs.~(\ref{Xeq_even}) and (\ref{Xeq_odd}) into the single equation
\begin{equation}
2iwR\, e^{2iwR} =\iota \frac{2\ell(\ell+1)}{M/R},
\label{Xeq_comb}
\end{equation}
in which $\iota$ is a sign parameter given by
\begin{equation}
\iota := \left\{
\begin{array}{ll}
  +1 & \quad \mbox{even parity, $\ell$ even; odd parity, $\ell$ odd} \\
  -1 & \quad \mbox{even parity, $\ell$ odd; odd parity, $\ell$ even}.
\end{array}
\right. 
\end{equation} 
Equation (\ref{Xeq_comb}) is solved by the Lambert-W function \cite{corless-etal:96}, $2iw R = W(\chi)$, with $\chi$ denoting the right-hand side of the equation. Because $M/R \ll 1$ we may appeal to the function's asymptotic behavior, $W(\chi) = \mbox{Ln}\,\chi + \mbox{Ln}(\mbox{Ln}\,\chi) + \cdots$. Keeping only the leading term, we have that
\begin{equation}
2iw R = \mbox{Ln}\biggl[ \iota \frac{2\ell(\ell+1)}{M/R}\biggr]
= \mbox{Ln}\, \iota + \ln \frac{2\ell(\ell+1)}{M/R},
\end{equation}
in which we distinguish between the multivalued logarithm ($\mbox{Ln}$) and its principal branch ($\ln$). When $\iota = +1$ we have that $\mbox{Ln}\, \iota = 2ni\pi$, while $\mbox{Ln}\, \iota = (2n+1)i\pi$ when $\iota = -1$; in both cases we have that $n$ runs over all integers (positive, negative, and zero). 

Putting these ingredients together and converting from $w$ to $\omega$ (ignoring the correction of order $M/R$), we finally obtain that the spectrum of wave modes in the Newtonian regime $M/R \ll 1$ is given by
\begin{equation}
\omega R = n\pi - i\, \xi
\label{wavemode_1}
\end{equation}
when $\iota = +1$ (even parity, $\ell$ even; odd parity, $\ell$ odd), while
\begin{equation}
\omega R = \biggl(n + \frac{1}{2} \biggr) \pi - i\, \xi 
\label{wavemode_2}
\end{equation}
when $\iota = -1$ (even parity, $\ell$ odd; odd parity, $\ell$ even). Here,
\begin{equation}
\xi := \frac{1}{2} \ln \frac{2\ell(\ell+1)}{M/R}
\end{equation}
is the logarithmic factor that was introduced in the various figures that display the frequencies of wave modes. The expressions of Eqs.~(\ref{wavemode_1}) and (\ref{wavemode_2}) are in good agreement with our numerical results when $M/R \ll 1$. These results bear a striking resemblance to those obtained by Andersson \cite{andersson:96} on the basis of simple toy problems.

\section{Tidal deformation of a thin shell: Even parity}
\label{sec:tidal_even}

In this section we examine the deformation of a thin shell when it is placed in a static tidal environment of a given multipole order $\ell$. The main goal is to compute the metric tidal constants $k^{\rm even}_\ell$ associated with a tidal deformation of even parity. The case of an odd-parity deformation will be considered next in Sec.~\ref{sec:tidal_odd}. 

\subsection{Metric perturbation}

The perturbed metric outside the deformed shell is expressed as in Eq.~(\ref{metric_pert_ext}), in which we now set $\omega = 0$. The solution to the perturbation equations listed in Sec.~\ref{subsec:even_ext} is given by \cite{binnington-poisson:09}
\begin{subequations}
\label{pert_even_static_ext}
\begin{align}
p^{\rm out}_{uu} &= -\frac{2}{(\ell-1)\ell} f \biggl( r^\ell\, A_1
+ 2 k^{\rm even}_\ell \frac{R^{2\ell+1}}{r^{\ell+1}} B_1 \biggr)\, {\cal E}^{\ell m}, \\
K^{\rm out} &= -\frac{2}{(\ell-1)\ell} \biggl( r^\ell\, A_2
+ 2 k^{\rm even}_\ell \frac{R^{2\ell+1}}{r^{\ell+1}} B_2 \biggr)\, {\cal E}^{\ell m},
\end{align}
\end{subequations}
where the constant ${\cal E}^{\ell m}$ denotes a tidal multipole moment, which provides a characterization of the tidal environment, $k^{\rm even}_\ell$ is the metric tidal constant, which encapsulates the shell's response to the applied tidal forces, and
\begin{subequations}
\label{AB_even}
\begin{align}
A_1 &:= \mbox{}_2 F_1(-\ell,-\ell+2;-2\ell;2M/r), \\
B_1 &:= \mbox{}_2 F_1(\ell+1,\ell+3;2\ell+2;2M/r), \\
A_2 &:= \frac{\ell+1}{\ell-1}\, \mbox{}_2 F_1(-\ell,-\ell;-2\ell;2M/r)
- \frac{2}{\ell-1}\, \mbox{}_2 F_1(-\ell-1,-\ell;-2\ell;2M/r), \\
B_2 &:= \frac{\ell}{\ell+2}\, \mbox{}_2 F_1(\ell+1,\ell+1;2\ell+2;2M/r)
+ \frac{2}{\ell+2}\, \mbox{}_2 F_1(\ell,\ell+1;2\ell+2;2M/r),
\end{align}
\end{subequations}
in which $\mbox{}_2 F_1(a,b;c;z)$ is the hypergeometric function. The functions $A_1$ and $A_2$ are polynomials in $2M/r$, while $B_1$ and $B_2$ can be written in terms of polynomials and $\ln(1-2M/r)$, which diverges in the (unrealized) limit $r \to 2M$. All instances of the radial functions admit an expansion of the form $1 + O(2M/r)$ when $2M/r \ll 1$.

The metric inside the shell is written as in Eq.~(\ref{metric_pert_in}), still with $\omega = 0$. In this case the field equations deliver
\begin{equation}
p^{\rm in}_{uu} = K^{\rm in} = -\frac{2}{(\ell-1)\ell}\, \lambda_\ell\, r^\ell\, {\cal E}^{\ell m},
\label{pert_even_static_in}
\end{equation}
where $\lambda_\ell$ is a constant to be determined.

\subsection{Deformed shell and junction conditions} 

The deformed shell is described as in Sec.~\ref{subsec:even_shell}, and the induced metric and extrinsic curvature are computed as in Secs.~\ref{subsec:even_induced} and \ref{subsec:even_extrinsic}. In this static case the conservation equation $D_b S^{ab} = 0$ delivers a relation between $p_\ell$ and the induced metric, as in Eq.~(\ref{mup_vs_h}), but it provides no information about $\mu_\ell$. To determine this we need the equation of state, implemented as in Eq.~(\ref{Gamma_eq}) --- the perturbations in the density and pressure are related by $\Gamma$, the fluid's adiabatic index.

The junction condition $[h_{ab}] = 0$ again produces Eq.~(\ref{junction1}), and the Israel condition of Eq.~(\ref{israel_b}) delivers a single piece of information, an expression for $\xi_r^{\rm out} = F^{1/2}\, \xi_r^{\rm in}$ in terms of $p_{uu}(R)$, $K(R)$, and $K'(R)$ evaluated on each face of the shell. In this static case, the angular component of the Lagrangian displacement vector, denoted $\xi$, is left undetermined.

\subsection{Tidal constant} 

The constants $k^{\rm even}_\ell$ and $\lambda_\ell$ are determined by the matching conditions
\begin{equation}
p^{\rm in}_{uu}(R) = p^{\rm out}_{uu}(R) - \frac{2M}{R F}\, \xi^{\rm out}_r, \qquad 
K^{\rm in}(R) = K^{\rm out}(R) + 2(1 - F^{-1/2})\, \xi^{\rm out}_r, 
\end{equation}
in which we insert the previously mentioned expression for $\xi^{\rm out}_r$ and the solutions of Eqs.~(\ref{pert_even_static_ext}) and (\ref{pert_even_static_in}). We shall not write down our result for $\lambda_\ell$, which is uninteresting. Our expression for the tidal constant is
\begin{equation}
k^{\rm even}_\ell = \frac{(\ell+1)(\ell+2)}{(\ell-1)\ell}
\frac{ (\ell+1) {\cal C}\, \PP_1 \FF_1 + \ell\, \PP_2 \FF_2 }
{ \ell {\cal C}\, \PP_3 \FF_3 + (\ell+1)\, \PP_4 \FF_4 },
\label{kL_even}
\end{equation}
where ${\cal C} := M/R$ is the shell's compactness,
\begin{subequations}
\label{F1234} 
\begin{align} 
\FF_1 &:= \mbox{}_2 F_1(-\ell,-\ell;-2\ell+1;2 {\cal C}), \\
\FF_2 &:= \mbox{}_2 F_1(-\ell-1,-\ell-1;-2\ell;2 {\cal C}), \\
\FF_3 &:= \mbox{}_2 F_1(\ell+1,\ell+1;2\ell+3;2 {\cal C}), \\
\FF_4 &:= \mbox{}_2 F_1(\ell,\ell;2\ell+2;2 {\cal C}),
\end{align}
\end{subequations}
and
\begin{subequations}
\label{P1234}
\begin{align}
\PP_1 &:= \bigl[ -6(4\Gamma - 3) {\cal C} + 2(\ell+5) \Gamma - 13 \bigr] \sqrt{1-2{\cal C}}
+ \bigl[ 4(\ell^2-\ell-3)\Gamma + 21 \bigr] {\cal C} 
- 2(\ell^2-\ell-3) \Gamma - 11, \\
\PP_2 &:= \bigl[ -6(4\Gamma - 3) {\cal C} + 2(\ell^2+5) \Gamma - (3\ell+13) \bigr] \sqrt{1-2{\cal C}} 
\nonumber \\ & \quad \mbox{} 
+ \bigl[ 4(\ell^2-4\ell-3)\Gamma + 3(3\ell+7) \bigr] {\cal C}
- 2(\ell^2-4\ell-3) \Gamma - (5\ell+11), \\
\PP_3 &:= \bigl[ -12(4\Gamma - 3) {\cal C} + 4(\ell+5) \Gamma - 26 \bigr] \sqrt{1-2{\cal C}}
+ \bigl[ 8(\ell^2-\ell-3)\Gamma + 42 \bigr] {\cal C}
- 4(\ell^2-\ell-3) \Gamma - 22, \\
\PP_4 &:= \bigl[ -12(4\Gamma - 3) {\cal C}
- 4(\ell^2-\ell-6) \Gamma + 2(3\ell-10) \bigr] \sqrt{1-2{\cal C}}
\nonumber \\ & \quad \mbox{} 
+ \bigl[ 8\ell(\ell+2)\Gamma - 6(3\ell-4) \bigr] {\cal C}
- 4\ell(\ell+2) \Gamma + 2(5\ell-6).
\end{align}
\end{subequations}
We observe that for each value of $\ell$, the tidal constant depends on the adiabatic index $\Gamma$ and the shell's compactness ${\cal C}$.

\begin{figure}
\includegraphics[width=0.49\linewidth]{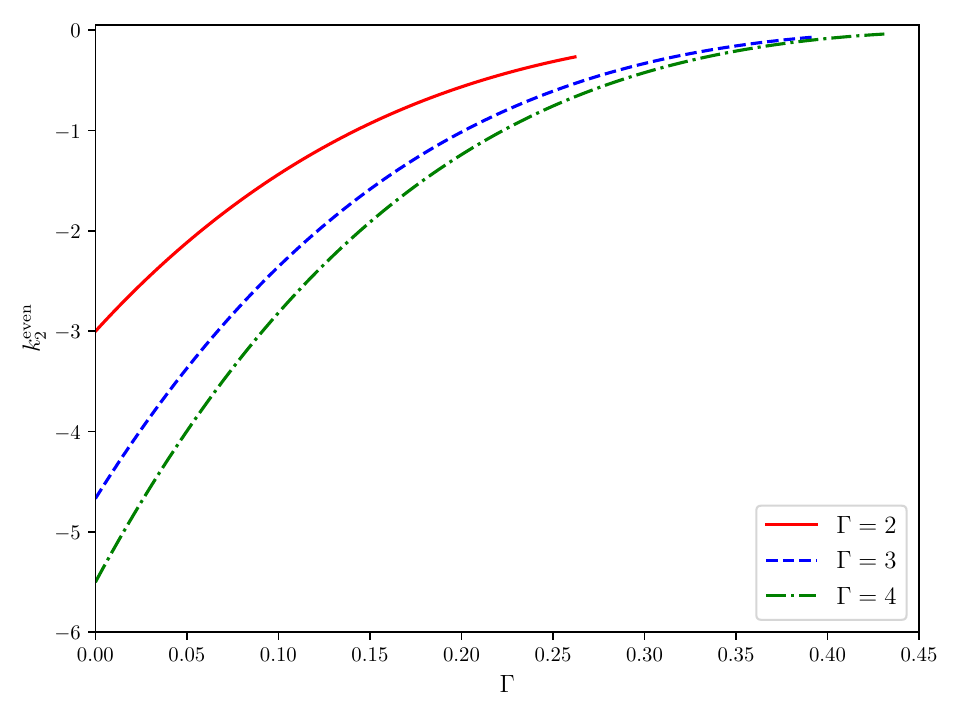}
\includegraphics[width=0.49\linewidth]{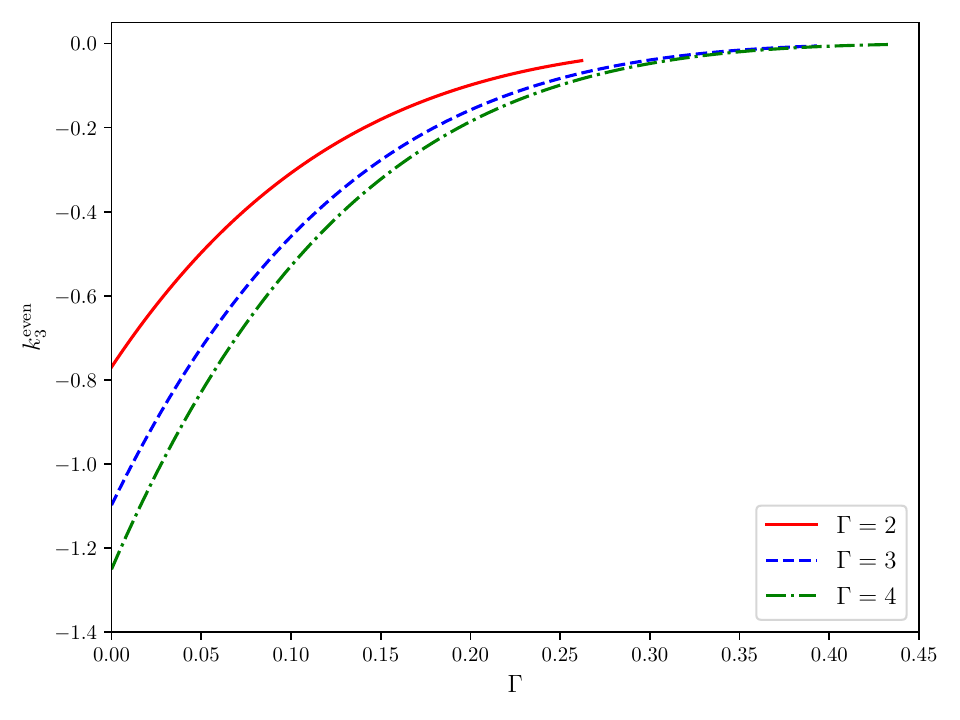}
\caption{Even-parity tidal constant $k^{\rm even}_\ell$ of a polytropic thin shell, as a function of the compactness ${\cal C} := M/R$. Left: $\ell = 2$. Right: $\ell = 3$. Red curves: $\Gamma = 2$. Blue curves: $\Gamma = 3$. Green curves: $\Gamma = 4$. The tidal constants are plotted in the interval $0 < C < C_{\rm max}$, with $C_{\rm max}$ corresponding to the configuration of maximum mass, at which the sequence becomes radially unstable.} 
\label{fig:love1_even} 
\end{figure} 

\begin{figure}
\includegraphics[width=0.49\linewidth]{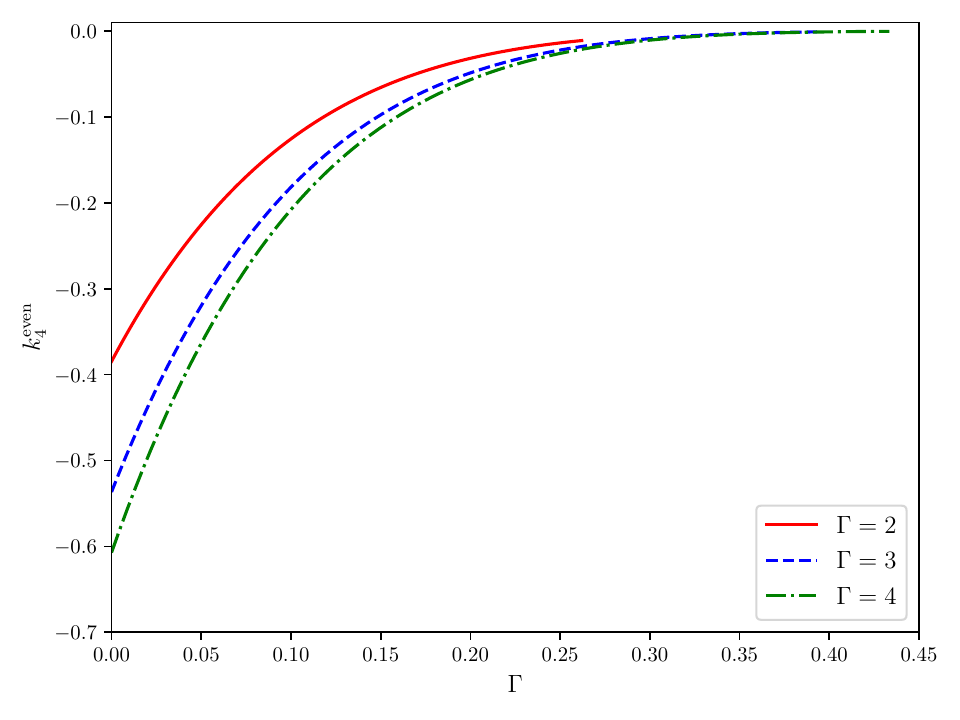}
\includegraphics[width=0.49\linewidth]{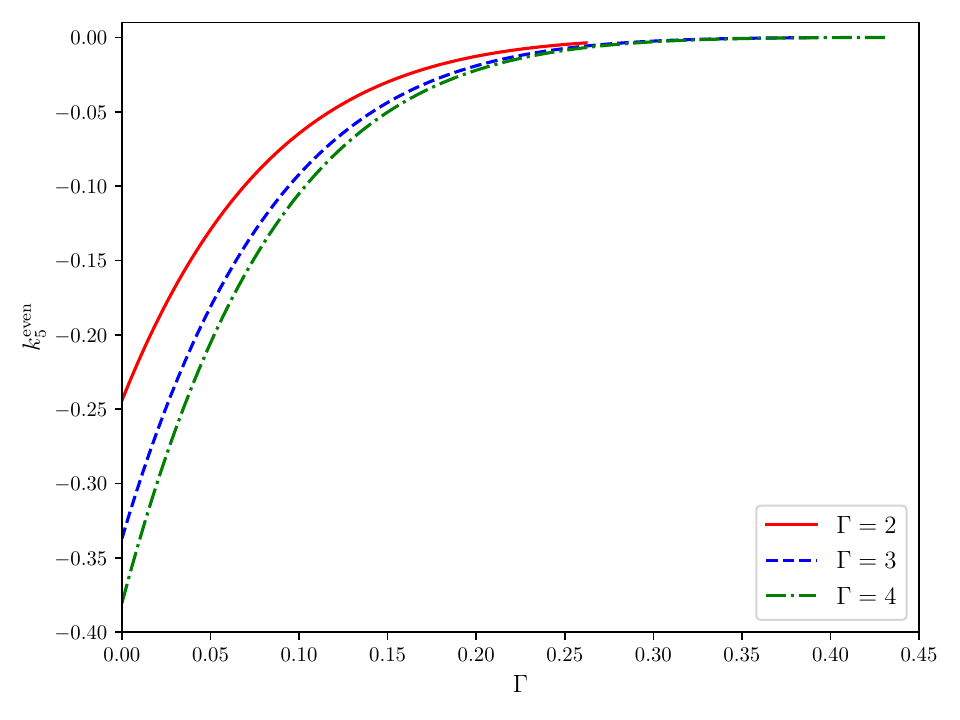}
\caption{Even-parity tidal constant $k^{\rm even}_\ell$ of a polytropic thin shell, as a function of the compactness ${\cal C} := M/R$. Left: $\ell = 4$. Right: $\ell = 5$. Red curves: $\Gamma = 2$. Blue curves: $\Gamma = 3$. Green curves: $\Gamma = 4$.} 
\label{fig:love2_even} 
\end{figure} 

\begin{table} 
\caption{\label{tab:love_N} Even-parity tidal constant at ${\cal C}=0$, the Newtonian limit.} 
\begin{ruledtabular}
\begin{tabular}{ccccc}
$\Gamma$ & $k^{\rm even}_2$ & $k^{\rm even}_3$ & $k^{\rm even}_4$ & $k^{\rm even}_5$ \\ 
\hline 
 2 & $-3$ & $-10/3$ & $-5/13$ & $-21/86$ \\  
 3 & $-14/3$ & $-45/41$ & $-22/41$ & $-91/270$ \\
 4 & $-11/2$ & $-5/4$ & $-17/28$ & $-35/92$  
\end{tabular} 
\end{ruledtabular} 
\end{table} 

\begin{table} 
\caption{\label{tab:love_max} Even-parity tidal constant at ${\cal C}={\cal C}_{\rm max}$, the configuration of maximum mass.} 
\begin{ruledtabular}
\begin{tabular}{cccccc}
$\Gamma$ & $C_{\rm max}$ & $k^{\rm even}_2$ & $k^{\rm even}_3$ & $k^{\rm even}_4$ & $k^{\rm even}_5$ \\ 
\hline 
  2 & $2.6202 \times 10^{-1}$  & $-2.6509 \times 10^{-1}$ & $-4.0708 \times 10^{-2}$
& $-1.1025 \times 10^{-2}$  & $-3.6902 \times 10^{-3}$ \\  
  3 & $3.9306 \times 10^{-1}$  & $-6.6199 \times 10^{-2}$ & $-5.7757 \times 10^{-3}$
& $-8.7979 \times 10^{-4} $ & $-1.6541 \times 10^{-4}$ \\
  4 & $4.3348 \times 10^{-1}$  & $-3.4132 \times 10^{-2}$ & $-2.3172 \times 10^{-3}$
& $-2.6884 \times 10^{-4} $ & $-3.8349 \times 10^{-5}$
\end{tabular} 
\end{ruledtabular} 
\end{table} 

Plots of $k^{\rm even}_\ell$ for $\ell = \{ 2, 3, 4, 5 \}$ and for polytropic shells with $\Gamma = \{2, 3, 4\}$ are presented in Figs.~\ref{fig:love1_even} and \ref{fig:love2_even}. The figures reveal that for the entire range of ${\cal C}$ that produces a radially stable shell, {\it the even-parity tidal constants are negative}. They decrease in magnitude as ${\cal C}$ increases, approaching (but never reaching) zero as ${\cal C} \to {\cal C}_{\rm max}$, which corresponds to the configuration of maximum mass. The values of $k^{\rm even}_\ell$ at the Newtonian limit of ${\cal C}=0$ are listed in Table~\ref{tab:love_N}. The values at ${\cal C} = {\cal C}_{\rm max}$ are shown in Table~\ref{tab:love_max}.

\subsection{Limiting cases} 

The exact expression of Eq.~(\ref{kL_even}) is not terribly illuminating, but it can be manipulated to deliver useful information in two interesting limiting situations. The first is the post-Newtonian regime corresponding to ${\cal C} \ll 1$. A straightforward expansion of Eq.~(\ref{kL_even}) in powers of ${\cal C}$ delivers
\begin{equation}
k^{\rm even}_\ell = k_\ell^{\rm N} + k_\ell^{\rm PN} {\cal C} + O({\cal C}^2),
\end{equation}
where
\begin{equation}
k^{\rm N}_\ell = -\frac{2(\ell+2)\bigl[ (\ell+2)\Gamma - (\ell+3) \bigr]}
{(\ell-1) \bigl[ (\ell+2)(2\ell-3) \Gamma - 4(\ell-2)\bigr]} 
\label{kL_Newton}
\end{equation}
is the Newtonian limit, and 
\begin{equation}
k^{\rm PN}_\ell = \frac{(2\ell+1)(\ell+2) \bigl[ (\ell+2)(8\ell^2 + 7\ell - 27) \Gamma^2
  - 2(\ell+2)(4\ell^2 + 15\ell - 35) \Gamma + 16(\ell-2)(\ell+3) \bigr]}
{2(\ell-1) \bigl[(\ell+2)(2\ell-3) \Gamma - 4(\ell-2)\bigr]^2}
\end{equation}
is the factor in front of ${\cal C}$ in the first post-Newtonian correction. The Newtonian expression for the tidal constant is reproduced in Sec.~\ref{sec:newton} in a purely Newtonian calculation. 

The second limiting situation is the highly compact regime in which ${\cal C}$ is set equal to ${\cal C}_{\rm max}$ and $F_{\rm min} = 1 - 2{\cal C}_{\rm max}$ is taken to be small; in view of Eq.~(\ref{Fmin_largeG}) this corresponds to $\Gamma_1 \gg 1$, where $\Gamma_1$ is the adiabatic index at the configuration of maximum mass. To compute the tidal constant in this regime we rely on the asymptotic relations
\begin{subequations}
\label{F1234_asymp}
\begin{align}
\FF_1 &= \frac{\ell!^2}{(2\ell-1)!}\, F + O(F^2),
\label{F1_asymp} \\
\FF_2 &= \frac{(\ell+1)!^2}{2(2\ell)!}\, F^2 + O(F^3),
\label{F2_asymp} \\
\FF_3 &= \frac{(2\ell+2)!}{(\ell+1)!^2} \Bigl\{ 1 + (\ell+1)^2 F \bigl[ \ln F
+ 2\bigl( \psi(\ell+2) + \gamma \bigr) - 1 \bigr] + O(F^2) \Bigr\},
\label{F3_asymp} \\
\FF_4 &= \frac{(2\ell+1)!}{(\ell+1)!^2} \Bigl\{ 1 - \ell^2 F - \tfrac{1}{2} \ell^2 (\ell+1)^2 F^2
\bigl[ \ln F + 2\bigl( \psi(\ell+2) + \gamma \bigr) - \tfrac{3}{2} \bigr] + O(F^3) \Bigr\},
\label{F4_asymp} 
\end{align}
\end{subequations}
where $\psi(z) := d\ln\Gamma(z)/dz$ is the digamma function, and $\gamma$ is the Euler-Mascheroni constant. These relations are derived in Appendix~\ref{sec:asymptotics}, and we note that the combination $\psi(\ell+2) + \gamma$ is actually rational when $\ell$ is an integer. Making the substitutions in Eq.~(\ref{kL_even}) and simplifying, we arrive at
\begin{align}
k^{\rm even}_\ell(C_{\rm max})
&= -\frac{(\ell-2)!(\ell-1)!(\ell+1)!(\ell+2)!}{4 (2\ell-1)! (2\ell+1)!} F_{\rm min}^{1/2} 
\Biggl\{ 1
\nonumber \\ & \quad \mbox{} 
+ (\ell+1) F_{\rm min}^{1/2} \biggl[ \ln F_{\rm min} + 2 \bigl( \psi(\ell+2) + \gamma \bigr)
- \frac{(\ell-2)(\ell+2)}{(\ell-1)\ell} \biggr] + O(F_{\rm min}) \Biggr\}.
\end{align}
An alternative expression is obtained by inserting Eq.~(\ref{Fmin_largeG}),
\begin{align}
k^{\rm even}_\ell(C_{\rm max})
&= -\frac{(\ell-2)!(\ell-1)!(\ell+1)!(\ell+2)!}{8 (2\ell-1)! (2\ell+1)!} \Gamma_1^{-1/2} \Biggl\{ 1
\nonumber \\ & \quad \mbox{} 
- \Gamma_1^{-1/2} \biggl[ \frac{1}{2} (\ell+1) \ln (4\Gamma_1)
- (\ell+1) \bigl( \psi(\ell+2) + \gamma \bigr)
+ \frac{\ell^3-3\ell-4}{2(\ell-1)\ell} \biggr] + O(\Gamma_1^{-1}) \Biggr\}.
\end{align}
This version is especially useful when the fluid is polytropic, with $\Gamma_1 = \Gamma = \mbox{constant}$. Either expression reveals that the tidal constant at maximum mass becomes very small (in magnitude) when $F_{\rm min} \ll 1$ and $\Gamma_1 \gg 1$.

\section{Tidal deformation of a thin shell: Odd parity}
\label{sec:tidal_odd}

In this section we examine the odd-parity sector of the tidal deformation of a thin shell, and compute the metric tidal constant $k^{\rm odd}_\ell$.

\subsection{Metric perturbation}

The metric outside the shell is written as in Eq.~(\ref{metric_odd_ext}) with $\omega = 0$. The perturbation equations (\ref{pert_eqns_odd_ext}) do not have a limit when $\omega \to 0$ and therefore cannot be applied directly to the static case. The Einstein field equations, however, can be manipulated into a second-order differential equation for $p_u$ and an algebraic equation for $p_r$. The solutions are \cite{binnington-poisson:09}
\begin{equation}
p^{\rm out}_u = \frac{2}{3(\ell-1)\ell} \biggl( r^\ell\, A_3
- \frac{2(\ell+1)}{\ell} k_\ell^{\rm odd} \frac{R^{2\ell+1}}{r^{\ell+1}} B_3 \biggr)\, {\cal B}^{\ell m},
\label{pert_odd_static_ext}
\end{equation}
and $p^{\rm out}_r = f^{-1} p^{\rm out}_u$. Here, the constant ${\cal B}^{\ell m}$ represents a tidal multipole moment, $k^{\rm odd}_\ell$ is the tidal metric constant, and
\begin{subequations}
\label{AB_odd}
\begin{align}
A_3 &:= \mbox{}_2 F_1(-\ell-2,-\ell+1;-2\ell;2M/r), \\
B_3 &:= \mbox{}_2 F_1(\ell-1,\ell+2;2\ell+2;2M/r).
\end{align}
\end{subequations}
The function $A_3$ is a polynomial in $2M/r$, while $B_3$ also involves $\ln(1-2M/r)$; in both cases the functions admit an expansion of the form $1 + O(2M/r)$ when $2M/r \ll 1$.

The metric inside the shell is expressed as in Eq.~(\ref{metric_odd_in}) with $\omega = 0$. In this case the solutions to the perturbation equations are 
\begin{equation}
p^{\rm in}_{u} = p^{\rm in}_{r} = \frac{2}{3(\ell-1)\ell}\, \chi_\ell\, r^\ell\, {\cal B}^{\ell m},
\label{pert_odd_static_in}
\end{equation}
where $\chi_\ell$ is a constant to be determined.

\subsection{Deformed shell and junction conditions} 

We adapt the discussion of Sec.~\ref{subsec:odd_shell} to the static case, and the induced metric and extrinsic curvature are calculated as in Secs.~\ref{subsec:odd_induced} and \ref{subsec:odd_extrinsic}. In this case the conservation equation $D_b S^{ab} = 0$ produces no useful information, and the Lagrangian displacement vector is left undetermined. 

Continuity of the induced metric across the shell continues to produce Eq.~(\ref{junction1}), and Eq.~(\ref{israel_b}) gives rise to two equations,
\begin{equation}
RF \bigl[ p_u'(R) \bigr] + \bigl(1 - 4M/R - F^{1/2} \bigr) p^{\rm out}_u(R) = 0
\label{match1_odd_static}
\end{equation}
and
\begin{equation}
p^{\rm in}_r(R) = F^{1/2} p^{\rm out}_r(R).
\label{match2_odd_static}
\end{equation}
This last equation is identical in content to Eq.~(\ref{junction1}), by virtue of the relationships between $p_u$ and $p_r$. 

\subsection{Tidal constant} 

\begin{figure}
\includegraphics[width=0.6\linewidth]{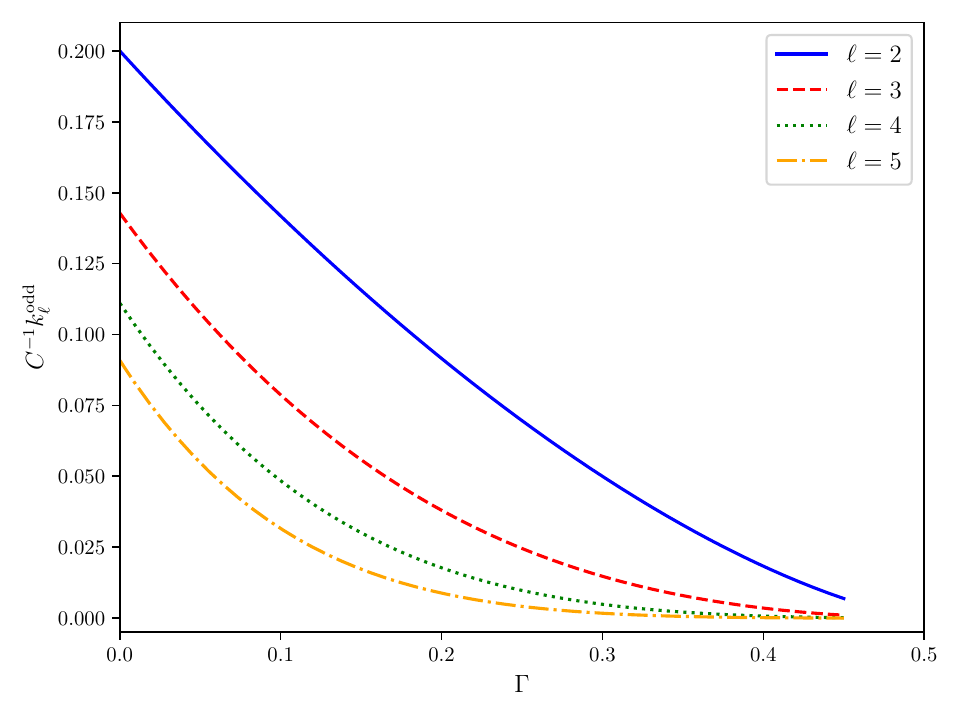}
\caption{Odd-parity tidal constant of a thin shell: ${\cal C}^{-1}\, k^{\rm odd}_\ell$ is plotted as a function of the shell's compactness ${\cal C} := M/R$. Blue curve: $\ell = 2$. Red curve: $\ell = 3$. Green curve: $\ell = 4$. Orange curve: $\ell = 5$. The curves are terminated arbitrarily at ${\cal C} = 0.45$; the correct termination point should be at the configuration of maximum mass, which depends on the shell's equation of state. Except for the termination point, the curves are universal and apply to a shell with any equation of state.} 
\label{fig:love_odd} 
\end{figure} 

The constants $k^{\rm odd}_\ell$ and $\chi_\ell$ are determined by the matching conditions of Eqs.~(\ref{match1_odd_static}) and (\ref{match2_odd_static}), in which we insert the solutions of Eqs.~(\ref{pert_odd_static_ext}) and (\ref{pert_odd_static_in}). We arrive at 
\begin{equation}
k^{\rm odd}_\ell = \frac{1}{2}
\frac{ (\ell-1)(\ell+2) {\cal C}\, \PP_5 \FF_5 + \ell\, \PP_6 \FF_6 }
{ -(\ell-1)(\ell+2) {\cal C}\, \PP_7 \FF_7 + (\ell+1)\, \PP_8 \FF_8 }
\label{kL_odd}
\end{equation}
for the tidal constant, with ${\cal C} := M/R$ denoting the shell's compactness,
\begin{subequations}
\label{F5678} 
\begin{align} 
\FF_5 &:= \mbox{}_2 F_1(-\ell-1,-\ell+2;-2\ell+1;2 {\cal C}), \\
\FF_6 &:= \mbox{}_2 F_1(-\ell-2,-\ell+1;-2\ell;2 {\cal C}), \\
\FF_7 &:= \mbox{}_2 F_1(\ell,\ell+3;2\ell+3;2 {\cal C}), \\
\FF_8 &:= \mbox{}_2 F_1(\ell-1,\ell+2;2\ell+2;2 {\cal C}),
\end{align}
\end{subequations}
and
\begin{subequations}
\label{P5678}
\begin{align}
\PP_5 &:= (1-2{\cal C})^{3/2}, \\
\PP_6 &:= \bigl[ -2(\ell+2) {\cal C} + \ell + 1 \bigr] \sqrt{1-2{\cal C}}
+ 2(\ell+1) {\cal C} - \ell - 1, \\
\PP_7 &:= (1-2{\cal C})^{3/2}, \\
\PP_8 &:= \bigl[ 2(\ell-1) {\cal C} - \ell \bigr] \sqrt{1-2{\cal C}}
+ 2(\ell+1) {\cal C} - \ell - 1. 
\end{align}
\end{subequations}
In the odd-parity case, the tidal constant depends on ${\cal C}$ only, and is independent of the fluid's adiabatic index $\Gamma$. 

Plots of ${\cal C}^{-1}\, k^{\rm odd}_\ell$ for $\ell = \{ 2, 3, 4, 5 \}$ are shown in Fig.~\ref{fig:love_odd};
the tidal constant is divided by the compactness so that the limit when ${\cal C} \to 0$ is a nonvanishing constant. The figure reveals that {\it the odd-parity tidal constants are positive}. We observe also that 
${\cal C}^{-1}\, k^{\rm odd}_\ell$ decreases in magnitude as ${\cal C}$ increases, approaching zero in the formal (and unrealized) limit ${\cal C} \to 1/2$. 

An expansion of Eq.~(\ref{kL_odd}) in powers of ${\cal C}$ gives rise to
\begin{equation}
k^{\rm odd}_\ell = \frac{{\cal C}}{2\ell+1} \biggl[
1 - \frac{\ell(32\ell^2+2\ell-39)}{4(2\ell-1)(2\ell+1)}\, {\cal C} + O({\cal C}^2) \biggr].
\end{equation}
This expression confirms that ${\cal C}^{-1}\, k^{\rm odd}_\ell$ approaches a nonvanishing constant when ${\cal C} \to 0$. 

Because the odd-parity tidal constant does not depend on the fluid's adiabatic index, its dependence on the compactness ${\cal C}$ is universal; it is the same for any surface fluid with any equation of state. In this context, there is no particular need to examine $k^{\rm odd}_\ell$ at the configuration of maximum mass for any choice of equation of state, as we did in Sec.~\ref{sec:tidal_even} for the even-parity sector.

\section{Newtonian thin shell}
\label{sec:newton}

In this final section we calculate the normal modes of vibration and tidal constants of a thin spherical shell of matter in Newtonian gravity. It is advantageous to provide a unified treatment by taking the shell to be immersed within a time-dependent tidal field. In the special case in which this field vanishes, we obtain the shell's normal modes. In the special case in which the tidal field is static, we recover the shell's tidal constants. 

\subsection{Governing equations}
\label{subsec:governing}

We consider a thin shell of fluid matter in Newtonian gravity. The shell is idealized as infinitely thin, and it occupies a closed, two-dimensional surface $\SSS(t)$ described by embedding relations $x^a = X^a(t, \vartheta^A)$, where $x^a$ are arbitrary coordinates in three-dimensional Euclidean space, and $\vartheta^A$ are intrinsic coordinates on the surface. We take these to be {\it Lagrangian coordinates}, so that fluid elements on the surface move with constant values of $\vartheta^A$. The embedding relations, therefore, also describe the motion of fluid elements in three-dimensional space. The surface comes with a unit normal vector $n^a$ and a set of tangent vectors $e^a_A := \partial X^a/\partial \vartheta^A$. 

The shell possesses a areal mass density $\sigma(t, \vartheta^A)$ and a surface pressure $p(t, \vartheta^A)$. It creates a gravitational field described by the Newtonian potential $U(t, x^a)$. The velocity of a fluid element is given by $v^a = \partial_t X^a(t,\vartheta^A)$, and it can be decomposed as
\begin{equation}
v^a = v_n\, n^a + v^A\, e^a_A,
\label{v_decomposed}
\end{equation}
in terms of normal and tangential components. The surface comes with an induced metric
\begin{equation}
h_{AB} := e^a_A e^b_B\, g_{ab}, 
\label{induced_metric_N}
\end{equation}
where $g_{ab}$ is the metric of three-dimensional space in coordinates $x^a$, and an extrinsic curvature
\begin{equation}
K_{AB} := e^a_A e^b_B\, \nabla_a n_b.
\label{extrinsic_curvature}
\end{equation}
We let $K := h^{AB} K_{AB}$ denote the trace of the extrinsic curvature, with $h^{AB}$ denoting the matrix inverse to $h_{AB}$. The induced metric is used to lower indices on tangent tensors such as $v^A$, and its inverse is used to raise indices. We continue to use the notation $\Omega_{AB} := \mbox{diag}(1,\sin^2\vartheta)$ for the metric on the unit two-sphere.

The surface fluid and its gravitational field are governed by a number of equations (see Sec.~IV of Ref.~\cite{cadogan-poisson:24b}). The statement of mass conservation is
\begin{equation}
\partial_t \sigma + \sigma \bigl( v_n K + D_A v^A ) = 0,
\label{mass_cons}
\end{equation}
where $D_A$ is the covariant-derivative operator compatible with $h_{AB}$. Conservation of momentum is expressed by
\begin{subequations}
\label{fluid_eqns}
\begin{align}
0 &= \sigma\bigl( \partial_t v_n + v^A \partial_A v_n - K_{AB} v^A v^B
- \langle g \rangle \bigr) - p K,
\label{fluid_a} \\
0 &= \sigma \bigl( \partial_t v_A - v_n \partial_A v_n - v_B D_A v^B - \partial_A U \bigr) + \partial_A p,
\label{fluid_b} 
\end{align}
\end{subequations} 
where $g := n^a \partial_a U$ is the normal component of the gravitational field evaluated on each side of the surface, and $\langle g \rangle$ is its arithmetic average over the two sides. Inside and outside of $\SSS(t)$ the gravitational potential is a solution to Laplace's equation,
\begin{equation} 
\nabla^2 U = 0, 
\label{Laplace}
\end{equation}
and it comes with the junction conditions
\begin{equation}
[ U ] = 0, \qquad
[ g ] = -4\pi G \sigma,
\label{gravity_eqns}
\end{equation}
where $[\psi]$ is the jump of a quantity $\psi$ across the shell --- the value on the positive side minus the value on the negative side, with the convention that $n^a$ points toward the positive side. These equations follow directly from Poisson's equation $\nabla^2 U = -4\pi G \rho$ when the volume mass density is given by $\rho = \sigma\, \delta(L)$, where $L$ is the orthogonal distance to $\SSS(t)$ and $\delta$ is the Dirac distribution. Together with an equation of state of the form $p = p(\sigma, s)$, where $s$ is the fluid's specific entropy (entropy per unit mass), Eqs.~(\ref{mass_cons}), (\ref{fluid_eqns}), (\ref{Laplace}), and (\ref{gravity_eqns}) form a complete set of dynamical equations for the surface fluid and its gravitational field.

In a static situation Eq.~(\ref{mass_cons}) becomes trivial, Eqs.~(\ref{fluid_eqns}) reduce to 
\begin{equation}
\sigma \langle g \rangle + pK = 0, \qquad
\sigma \partial_A U - \partial_A p = 0,
\label{fluid_eqns_static} 
\end{equation}
and Eqs.~(\ref{Laplace}), (\ref{gravity_eqns}) stay unchanged. A generalized version of the first of Eqs.~(\ref{fluid_eqns_static}) reads $\sigma \langle g \rangle + p K = [P]$, and it expresses the balance of normal forces across a fluid boundary. On the right-hand side we have the jump of the bulk pressure $P$ across the two faces of the boundary; this vanishes in our case because there is no fluid beyond the surface. The first term on the left-hand side is the average of the gravitational force per unit area acting on each face of the boundary, and the second term is the contribution from the surface pressure. In the absence of gravity the equation reduces to $pK = [P]$, the statement of the Young-Laplace law (which is usually written in terms of the surface tension $-p$). The second of Eqs.~(\ref{fluid_eqns_static}) expresses the balance of tangential forces on the shell.

\subsection{Static and spherical shell}
\label{subsec:spherical}

The shell is static and spherically symmetric in its unperturbed state. Using spherical coordinates $x^a = (r,\theta,\phi)$ for the ambiant space, and selecting $\vartheta^A = (\vartheta,\varphi)$ for the intrinsic coordinates, we have that the embedding relations for the surface $\SSS$ are $r = R$, $\theta = \vartheta$, and $\phi = \varphi$, where $R$ is the shell's radius. The vectors $e^a_A = \partial_A X^a$, given explicitly by
\begin{equation}
e^a_\vartheta = (0,1,0), \qquad
e^a_\varphi = (0,0,1),
\label{tangent_unpert} 
\end{equation}
are tangent to $\SSS$, and
\begin{equation}
n_a = (1,0,0)
\label{normal_unpert} 
\end{equation}
is the unit normal. The induced metric on $\SSS$ is given by
\begin{equation}
h_{AB}\, d\vartheta^A d\vartheta^B
= R^2 \bigl( d\vartheta^2 + \sin^2\vartheta\, d\varphi^2 \bigr)
= R^2\, \Omega_{AB}\, d\vartheta^A d\vartheta^B.
\label{induced_unpert} 
\end{equation}
The surface's extrinsic curvature is given by $K_{AB} = R\, \Omega_{AB}$, and its trace is $K = 2/R$. 

The gravitational potentials inside and outside the shell are given by
\begin{equation}
U_{\rm in} = \frac{GM}{R}, \qquad
U_{\rm out} = \frac{GM}{r},  
\end{equation}
where $M$ denotes the shell's mass. The normal components of the gravitational fields at the shell are
\begin{equation} 
g_{\rm in} = 0, \qquad
g_{\rm out} = -\frac{GM}{R^2},
\end{equation}
so that $\langle g \rangle = -GM/(2R^2)$ and $[g] = -GM/R^2$.

The shell's mass density is calculated with the help of Eq.~(\ref{gravity_eqns}), and we get the expected 
\begin{equation}
\sigma = \frac{M}{4\pi R^2}.
\label{sigma_unpert}
\end{equation}
Equation (\ref{fluid_eqns_static}) gives us the shell's surface pressure, and we obtain 
\begin{equation}
p = \frac{GM^2}{16\pi R^3}.
\label{p_unpert_N}
\end{equation}
These equations can be inverted to give $M$ and $R$ in terms of $\sigma$ and $p$:
\begin{equation}
M = \frac{4}{\pi G^2} \frac{p^2}{\sigma^3}, \qquad
R = \frac{1}{\pi G} \frac{p}{\sigma^2}.
\end{equation}
When $p$ is related to $\sigma$ by an equation of state of the form $p = p(\sigma, s)$, these equations describe a family of equilibrium configurations parametrized by $\sigma$ and $s$. When the equation of state is of the barotropic form $p = p(\sigma)$, the family becomes a one-dimensional sequence parametrized by $\sigma$.

It is known \cite{lemaitre-poisson:19} that these equilibrium configurations are radially stable whenever $\Gamma \geq 3/2$, where
\begin{equation}
\Gamma := \biggl( \frac{\partial \ln p}{\partial \ln \sigma} \biggr)_s
\label{Gamma_N} 
\end{equation} 
is the adiabatic index associated with the equation of state (also a function of $\sigma$ and $s$). By ``radially stable'' we mean that a time-dependent, spherically symmetric perturbation of the shell will produce a bounded oscillation about the equilibrium point; perturbing an unstable shell would produce instead a catastrophic implosion or explosion. 

\subsection{Geometry of a deformed shell} 
\label{subsec:geometry}

We now introduce a small deformation of the formerly spherical shell, described by the new embedding relations
\begin{subequations}
\label{embedding_deformed} 
\begin{align}
r &= R\bigl[ 1 + h^{\ell m}(t)\, Y^{\ell m}(\vartheta,\varphi) \bigr], \\
\theta^A &= \vartheta^A + \Omega^{AB} \bigl[ 
j^{\ell m}(t)\, Y_B^{\ell m}(\vartheta,\varphi)
+ f^{\ell m}(t)\, X_B^{\ell m}(\vartheta,\varphi) \bigr],
\end{align}
\end{subequations}
where $h^{\ell m}(t)$, $j^{\ell m}(t)$, and $f^{\ell m}(t)$ are dimensionless functions of time to be determined, taken to be much smaller than unity. The deformation comes with a specific multipole order $\ell$, and it is further characterized by an azimuthal integer $m$; a general deformation is obtained by summing over $\ell$ and $m$. For completeness, and to make contact with the relativistic treatment of the previous sections, the angular displacement in Eq.~(\ref{embedding_deformed}) includes both even-parity and odd-parity components. As we shall see, the odd-parity piece proportional to $f^{\ell m}$ does not produce interesting consequences in the Newtonian context. 

Differentiation of Eqs.~(\ref{embedding_deformed}) with respect to $\vartheta^A$ provides us with the tangent vectors $e^a_A$, and to first order in the deformation, the induced metric of Eq.~(\ref{induced_metric}) is
\begin{equation}
h_{AB} = R^2 \biggl\{ \Omega_{AB}
+ \bigl[ 2 h^{\ell m} - \ell(\ell+1) j^{\ell m} \bigr] \Omega_{AB} Y^{\ell m}
+ 2 j^{\ell m}\, Y^{\ell m}_{AB} + 2 f^{\ell m}\, X^{\ell m}_{AB} \biggr\}.
\end{equation}
A computation of the extrinsic curvature requires an expression for the unit normal $n_a$. To obtain this we infer from Eq.~(\ref{embedding_deformed}) that to first order in the deformation, a description of the surface is provided by
\begin{equation}
0 = \Phi(x^a) := r - R \bigl[ 1 + h^{\ell m}\, Y^{\ell m}(\theta,\phi) \bigr],
\end{equation}
in which the spherical harmonics are now expressed in terms of the polar angles $(\theta,\phi)$ associated with the coordinates $x^a$. We have that $n_a = \partial_a \Phi$ --- it is easy to verify that this is properly normalized --- and after evaluation on $\SSS(t)$, we find that
\begin{equation}
n_r = 1, \qquad
n_A = -h^{\ell m} R\, Y^{\ell m}_A(\vartheta,\varphi),
\end{equation}
where the spherical harmonics are once more expressed in terms of the intrinsic coordinates $(\vartheta,\varphi)$. Notice the abuse of notation: On the left-hand side, the index $A$ on $n_A$ refers to the ambiant coordinates $\theta^A$, while on the right-hand side they refer to the intrinsic angles $\vartheta^A$.

The extrinsic curvature is calculated from Eq.~(\ref{extrinsic_curvature}), and we find that
\begin{equation}
K_{AB} = R \biggl\{ \Omega_{AB}
+ \Bigl[ \tfrac{1}{2} (\ell^2+\ell+2) h^{\ell m} - \ell(\ell+1) j^{\ell m} \Bigr]
\Omega_{AB} Y^{\ell m}
+ \bigl( 2 j^{\ell m}  - h^{\ell m} \bigr) Y^{\ell m}_{AB}
+ 2 f^{\ell m}\, X^{\ell m}_{AB} \biggr\}.
\end{equation}
Its trace is
\begin{equation}
K = \frac{2}{R} \Bigl[ 1 + \tfrac{1}{2} (\ell-1)(\ell+2) h^{\ell m}\, Y^{\ell m} \Bigr],
\end{equation}
and we see that it is independent of $j^{\ell m}$. 

\subsection{Gravitational potential and field}
\label{subsec:gravi}

The shell is immersed in a tidal field of multipole order $\ell$, described by
\begin{equation}
U_{\rm tidal} = -\frac{1}{(\ell-1)\ell} \frac{GM}{R}\, \Bigl[ e^{\ell m}(t)\, (r/R)^\ell \Bigr]\,
Y^{\ell m}(\theta,\phi),
\label{U_tidal} 
\end{equation}
where $e^{\ell m}(t)$ are dimensionless tidal moments. The tidal forces cause the mass distribution on the shell to change, and the shell acquires dimensionless mass multipole moments $q^{\ell m}(t)$. The shell's tidal response is described by
\begin{equation}
U_{\rm resp} = -\frac{1}{(\ell-1)\ell} \frac{GM}{R}\, \Bigl[ 2 q^{\ell m}(t)\, (R/r)^{\ell+1} \Bigr]\,
Y^{\ell m}(\theta,\phi).
\label{U_resp}
\end{equation}
Equations (\ref{U_tidal}) and (\ref{U_resp}) are both solutions to Laplace's equation $\nabla^2 U = 0$.

The complete potential outside the shell includes tidal and response pieces from the perturbation, as well as an unperturbed, monopole piece. It is given by
\begin{equation}
U_{\rm out} = \frac{GM}{r} - \frac{1}{(\ell-1)\ell} \frac{GM}{R} \Bigl[ e^{\ell m}(t)\, (r/R)^\ell
+ 2 q^{\ell m}(t)\, (R/r)^{\ell+1} \Bigr]\, Y^{\ell m}(\theta,\phi).
\label{U_out}
\end{equation}
We write the potential inside the shell as
\begin{equation}
U_{\rm in} = \frac{GM}{R} \biggl[ 1 - \frac{1}{(\ell-1)\ell} u^{\ell m}(t)\, (r/R)^\ell\,
Y^{\ell m}(\theta,\phi) \biggr], 
\label{U_in}
\end{equation}
in terms of interior moments $u^{\ell m}(t)$. We omit a term proportional to $r^{-(\ell+1)}$ because it would diverge at $r=0$.

The potentials evaluated at the shell are given by
\begin{subequations}
\label{potential_S} 
\begin{align}
U_{\rm out}(\SSS) &= \frac{GM}{R} \biggl\{ 1 - \frac{1}{(\ell-1)\ell} \bigl[
e^{\ell m} + 2 q^{\ell m} + (\ell-1)\ell\, h^{\ell m} \bigl]\, Y^{\ell m}(\vartheta,\varphi) \biggr\},\\
U_{\rm in}(\SSS) &= \frac{GM}{R} \biggl\{ 1 - \frac{1}{(\ell-1)\ell}\, u^{\ell m}\,
Y^{\ell m}(\vartheta,\varphi) \biggr\},
\end{align}
\end{subequations}
where the spherical harmonics are now expressed in terms of the intrinsic angles $\vartheta^A$. To arrive at Eq.~(\ref{potential_S}) we made use of Eq.~(\ref{embedding_deformed}) and linearized all expressions with respect to the perturbation.

The normal component of the gravitational fields, also evaluated at the shell, are given by
\begin{subequations}
\label{field_S} 
\begin{align}
g_{\rm out}(\SSS) &= -\frac{GM}{R^2} \biggl\{ 1 + \frac{1}{(\ell-1)\ell} \Bigl[
\ell e^{\ell m} - 2 (\ell+1) q^{\ell m} - 2 (\ell-1)\ell\, h^{ \ell m} \Bigl]\,
Y^{\ell m}(\vartheta,\varphi) \biggr\},\\
g_{\rm in}(\SSS) &= -\frac{GM}{R^2} \biggl\{ -\frac{1}{(\ell-1)\ell} \Bigl[ \ell u^{\ell m} \Bigr]\,
Y^{\ell m}(\vartheta,\varphi) \biggr\},
\end{align}
\end{subequations}
with $g := n^a \partial_a U$.

\subsection{Fluid variables}
\label{subsec:variables} 

The mass density on the deformed shell is written as
\begin{equation}
\sigma = \frac{M}{4\pi R^2} \Bigl[ 1 + \sigma^{\ell m}(t)\, Y^{\ell m}(\vartheta,\varphi) \Bigr],
\label{sigma_pert}
\end{equation}
and the surface pressure is expressed as
\begin{equation}
p = \frac{GM^2}{16\pi R^3} \Bigl[ 1 + p^{\ell m}(t)\, Y^{\ell m}(\vartheta,\varphi) \Bigr]. 
\label{pressure_pert}
\end{equation}
The equation of state $p = p(\sigma,s)$ implies that the pressure and density moments are related by
\begin{equation}
p^{\ell m} = \Gamma\, \sigma^{\ell m},
\label{EoS}
\end{equation}
where $\Gamma$ is the adiabatic index of Eq.~(\ref{Gamma_N}). It is assumed that the tidal deformation is adiabatic, in the sense that it comes with no change in specific entropy.

The components of the fluid's velocity vector are obtained by differentiating the embedding relations of Eq.~(\ref{embedding_deformed}). We get
\begin{equation}
v_n = R\, \frac{d h^{\ell m}}{dt}\, Y^{\ell m}(\vartheta,\varphi), \qquad
v^A =\Omega^{AB} \biggl[\frac{d j^{\ell m}}{dt}\, Y_B^{\ell m}(\vartheta,\varphi)
+ \frac{d f^{\ell m}}{dt}\, X_B^{\ell m}(\vartheta,\varphi) \biggr].
\label{velocity}
\end{equation}
These expressions imply that
\begin{equation}
D_A v^A = -\ell(\ell+1) \frac{d j^{\ell m}}{dt}\, Y^{\ell m}(\vartheta,\varphi).
\label{div_v}
\end{equation}

\subsection{Dynamical equations}
\label{subsec:eqns}

With the results listed in the preceding sections, we find that the statement of mass conservation of Eq.~(\ref{mass_cons}) yields
\begin{equation}
\sigma^{\ell m} = -2 h^{\ell m} + \ell(\ell+1) j^{\ell m}, 
\end{equation}
and that the junction conditions of Eq.~(\ref{gravity_eqns}) produce
\begin{subequations}
\label{q_u} 
\begin{align}
q^{\ell m} &= -\frac{(\ell-1)\ell^2}{2(2\ell+1)} \Bigl[ h^{\ell m} + (\ell+1) j^{\ell m} \Bigr], \\
u^{\ell m} &= e^{\ell m} + \frac{(\ell-1)\ell(\ell+1)}{2\ell+1} \Bigl( h^{\ell m} - \ell j^{\ell m} \Bigr).
\end{align}
\end{subequations}
Making the substitutions in the fluid equations (\ref{fluid_eqns}), we arrive at differential equations for the remaining variables $h^{\ell m}$, $j^{\ell m}$, and $f^{\ell m}$, which appear in the embedding relations of Eq.~(\ref{embedding_deformed}). These are
\begin{subequations}
\label{dynamics1}
\begin{align}
0 &= \frac{R^3}{GM} \frac{d^2 h^{\ell m}}{dt^2}
+ \biggl[ \Gamma - \frac{2\ell^3-\ell^2+9\ell+6}{4(2\ell+1)} \biggr]\, h^{\ell m}
- \frac{1}{2} \ell(\ell+1) \biggl[ \Gamma - \frac{2(\ell+1)}{2\ell+1} \biggr]\, j^{\ell m} 
+ \frac{1}{\ell-1}\, e^{\ell m}, \\
0 &= \frac{R^3}{GM} \frac{d^2 j^{\ell m}}{dt^2}
- \frac{1}{2} \biggl[ \Gamma - \frac{2(\ell+1)}{2\ell+1} \biggr]\, h^{\ell m}
- \frac{1}{4} \ell(\ell+1) \biggl[ \Gamma - \frac{4}{2\ell+1} \biggr]\, j^{\ell m}
+ \frac{1}{(\ell-1)\ell}\, e^{\ell m}
\end{align}
\end{subequations}
for the even-parity sector, and
\begin{equation}
\frac{d^2 f^{\ell m}}{dt^2} = 0
\label{dynamics_odd} 
\end{equation}
for the odd-parity sector. As was anticipated previously, Eq.~(\ref{dynamics_odd}) leads to trivial consequences.  

In the following we shall focus on the even-parity sector of the perturbation equations. To simplify the equations we shall set $GM/R^3 \equiv 1$ by performing a rescaling of time. Henceforth, time will be measured in units of $(R^3/GM)^{1/2}$, and frequencies will be measured in units of $(GM/R^3)^{1/2}$.

It is helpful to package the dynamical variables within a two-dimensional vector
\begin{equation}
{\sf h} := \left(
\begin{array}{c}
  h^{\ell m} \\ j^{\ell m}
\end{array}
\right),
\end{equation}
and to group the various coefficients in front of $h^{\ell m}$ and $j^{\ell m}$ in Eq.~(\ref{dynamics1}) into a matrix ${\sf A}$. If we also write
\begin{equation}
{\sf S}_{\sf h} := \left(
\begin{array}{c}
  \frac{1}{\ell-1}\, e^{\ell m}
  \\
  \frac{1}{(\ell-1)\ell}\, e^{\ell m}
\end{array}
\right),
\end{equation}
then the dynamical system becomes
\begin{equation}
\ddot{{\sf h}} + {\sf A} {\sf h} + {\sf S}_{\sf h} = 0, 
\label{dynamics2}
\end{equation}
with an overdot indicating differentiation with respect to (dimensionless) time.

\subsection{Solution to the dynamical equations}
\label{subsec:diag}

Our task now is to integrate Eq.~(\ref{dynamics2}). We begin with a decoupling of the equations. Let $\lambda_\pm$ and ${\sf v}_\pm$ be the eigenvalues and eigenvectors of the matrix ${\sf A}$, respectively, so that
\begin{equation}
{\sf A} {\sf v}_\pm = \lambda_\pm {\sf v}_\pm.
\end{equation}
Let ${\sf U} := ({\sf v}_+, {\sf v}_-)$ be the matrix of eigenvectors, so that
\begin{equation}
{\sf U}^{-1}  {\sf A} {\sf U} = {\sf \Lambda} := \left(
\begin{array}{cc}
  \lambda_+ & 0 \\
  0 & \lambda_-
\end{array}
\right) 
\end{equation}
is the diagonal matrix of eigenvalues. Then the transformation ${\sf h} = {\sf U} {\sf x}$, where
\begin{equation}
{\sf x} := \left(
\begin{array}{c}
  x^{\ell m} \\ y^{\ell m}
\end{array}
\right)
\end{equation}
is a vector of new variables, brings the dynamical system to the decoupled form
\begin{equation}
\ddot{\sf x} + {\sf \Lambda} {\sf x} + {\sf S}_{\sf x} = 0,
\label{dynamics3}
\end{equation}
where ${\sf S}_{\sf x} := {\sf U}^{-1} {\sf S}_{\sf h}$ is a new source vector. Integration of Eq.~(\ref{dynamics3}) is then immediate. 

The eigenvalue problem for the matrix ${\sf A}$ produces the quadratic equation
\begin{equation}
\lambda^2 - \frac{1}{4} \bigl[ (\ell^2+\ell+4)\Gamma - (\ell^2+\ell+6) \bigr] \lambda
- \frac{(\ell-1)\ell(\ell+1)(\ell+2)}{16(2\ell+1)} \bigl[ (2\ell-3) \Gamma - 4 \bigr] = 0.
\end{equation}
This is the same equation as Eq.~(\ref{sigma0_eq}), in which we used the notation $\varsigma_0^2$ for $\lambda$. The solutions are $\lambda_\pm := \varsigma^2_{0\pm}$, as displayed in Eq.~(\ref{sigma0_slns}). Their most important property for our purposes is that
\begin{equation}
\lambda_+ > 0, \qquad \lambda_- < 0.  
\end{equation}
With an arbitrary choice of normalization we have that the eigenvectors are
\begin{equation}
{\sf v}_\pm := \left(
\begin{array}{c}
  v^1_\pm \\ v^2_\pm
\end{array}
\right)
\end{equation}
with
\begin{subequations}
\begin{align}
v_\pm^1 &= \frac{1}{(2\ell+1)(\lambda_+-\lambda_-)}, \\
v_\pm^2 &= \frac{1}{(2\ell+1)(\lambda_+-\lambda_-)}
\frac{4(2\ell+1)(\Gamma-\lambda_\pm) - (2\ell^3-\ell^2+9\ell+6)}{
  2\ell(\ell+1) [ (2\ell+1)\Gamma - 2(\ell+1)]}.
\end{align}
\end{subequations}
From these we can form the matrix ${\sf U}$, calculate its inverse ${\sf U}^{-1}$, and obtain the new source term ${\sf S}_{\sf x}$.

We introduce the notation
\begin{equation}
\omega_\ell := \sqrt{\lambda_+}, \qquad
\kappa_\ell := \sqrt{-\lambda_-}
\end{equation}
for the square root of the eigenvalues, and write the dynamical system of Eq.~(\ref{dynamics3}) in the explicit form
\begin{subequations}
\label{dynamics4}
\begin{align}
0 &= \ddot{x}^{\ell m} + \omega_\ell^2\, x^{\ell m}
- \frac{1}{4} (2\ell+1) \Bigl( 2\Gamma - \frac{4\kappa_\ell^2}{\ell-1}
+ \ell - 2 \Bigr) e^{\ell m}, \\
0 &= \ddot{y}^{\ell m} - \kappa_\ell^2\, y^{\ell m}
+ \frac{1}{4} (2\ell+1) \Bigl( 2\Gamma +\frac{4\omega_\ell^2}{\ell-1}
+ \ell - 2 \biggr) e^{\ell m}.
\end{align}
\end{subequations}
The original variables are then given by
\begin{subequations}
\label{h_vs_x}
\begin{align}
h^{\ell m} &= \frac{1}{(2\ell+1) (\omega_\ell^2 + \kappa_\ell^2)}\,
\Bigl\{ x^{\ell m} + y^{\ell m} \Bigr\}, \\
j^{\ell m} &= \frac{2}{\ell(\ell+1)(2\ell+1)[ (2\ell+1)\Gamma - 2(\ell+1)]
  (\omega_\ell^2 + \kappa_\ell^2)} \Bigl\{
\bigl[ (2\ell+1)(\Gamma - \omega_\ell^2) - \tfrac{1}{4} (2\ell^3-\ell^2+9\ell+6) \bigr] x^{\ell m}
\nonumber \\ & \quad \mbox{}
+ \bigl[ (2\ell+1)(\Gamma + \kappa_\ell^2) - \tfrac{1}{4} (2\ell^3-\ell^2+9\ell+6) \bigr] y^{\ell m}
\Bigr\}.
\end{align}
\end{subequations}

The general solution to Eq.~(\ref{dynamics4}) is
\begin{subequations}
\label{solutions_xy}
\begin{align}
x^{\ell m}(t) &= x^{\ell m}(0) \cos \omega_\ell t
+ \frac{\dot{x}^{\ell m}(0)}{\omega_\ell} \sin \omega_\ell t
\nonumber \\ & \quad \mbox{} 
+ \frac{1}{4} (2\ell+1) \Bigl( 2\Gamma - \frac{4\kappa_\ell^2}{\ell-1}
+ \ell - 2 \Bigr)
\frac{1}{\omega_\ell} \int_0^t e^{\ell m}(t')\, \sin \omega_\ell(t-t')\, dt', \\ 
y^{\ell m}(t) &= y^{\ell m}(0) \cosh \kappa_\ell t
+ \frac{\dot{y}^{\ell m}(0)}{\kappa_\ell} \sinh \kappa_\ell t
\nonumber \\ & \quad \mbox{} 
- \frac{1}{4} (2\ell+1) \Bigl( 2\Gamma + \frac{4\omega_\ell^2}{\ell-1}
+ \ell - 2 \Bigr)
\frac{1}{\kappa_\ell} \int_0^t e^{\ell m}(t')\, \sinh \kappa_\ell(t-t')\, dt'.
\end{align}
\end{subequations} 
Making the substitutions in Eq.~(\ref{h_vs_x}) we obtain the original variables $h^{\ell m}(t)$ and $j^{\ell m}(t)$. The shell's mass multipole moments are then given by Eq.~(\ref{q_u}), which becomes
\begin{align}
q^{\ell m} &= -\frac{(\ell-1)\ell}{ 4(2\ell+1)[(2\ell+1)\Gamma - 2(\ell+1)]
  (\omega_\ell^2 + \kappa_\ell^2) } \Bigl\{
\bigl[ 2(\ell+2) \Gamma - 4 \omega_\ell^2 - (\ell^2+\ell+6) \bigr] x^{\ell m}
\nonumber \\ & \quad \mbox{}
+ \bigl[ 2(\ell+2) \Gamma + 4 \kappa_\ell^2 - (\ell^2+\ell+6) \bigr] y^{\ell m} 
\Bigr\}
\label{q_vs_xy}
\end{align} 
in terms of the decoupled variables. 

The motion of the deformed shell is described by a linear superposition of the independent modes $x^{\ell m}(t)$ and $y^{\ell m}(t)$. While $x^{\ell m}$ is essentially a driven harmonic oscillator with a positive squared frequency $\omega_\ell^2 = \lambda_+ > 0$, $y^{\ell m}$ possesses a negative squared frequency $-\kappa_\ell^2 = \lambda_- < 0$. As a consequence, while the motion described by $x^{\ell m}(t)$ is bounded and oscillating in time, the one described by $y^{\ell m}(t)$ becomes unbounded as $t \to \infty$. {\it The tidal response of the thin shell necessarily features a dynamical instability.}

\subsection{Normal modes of vibration}
\label{subsec:normal}

The dynamical instability for any multipole order $\ell$ takes its origin in the existence of a normal mode of vibration with a negative eigenvalue. The mode equations correspond to Eq.~(\ref{dynamics4}) with the external source removed,
\begin{equation}
\ddot{x}_\ell + \omega_\ell^2\, x_\ell = 0, \qquad
\ddot{y}_\ell - \kappa_\ell^2\, y_\ell = 0. 
\label{mode_eqns}
\end{equation}
The general solutions are
\begin{equation}
x_\ell(t) = x_\ell(0) \cos \omega_\ell t
+ \frac{\dot{x}_\ell(0)}{\omega_\ell} \sin \omega_\ell t, \qquad 
y_\ell(t) = y_\ell(0) \cosh \kappa_\ell t
+ \frac{\dot{y}_\ell(0)}{\kappa_\ell} \sinh \kappa_\ell t,
\label{mode_slns}
\end{equation} 
and it is clear that while $x_\ell(t)$ is a stable mode, $y_\ell(t)$ is unstable. Any small perturbation will soon grow large, and its subsequent evolution will require a non-perturbative description.

Normal modes also exist for the special cases $\ell = 0$ and $\ell = 1$. The monopole case requires a separate treatment (not detailed here), and we find that there is only a single mode with an eigenvalue $\lambda = \Gamma - 3/2$. As we claimed previously, this mode is stable whenever $\Gamma \geq 3/2$. The dipole modes are recovered simply by setting $\ell = 1$ in the previous results. Here we get
\begin{equation}
\lambda_+ = \frac{1}{2} (3\Gamma - 4), \qquad \lambda_- = 0.
\end{equation}
We see that the first mode is stable when $\Gamma \geq 3/2$, and that the second mode is marginally stable, with a vanishing eigenvalue. It is easy to verify that the zero mode comes with $j_1^m = h_1^m$, and that $h_1^m(t)$ is linear in time. To understand the meaning of this, we select the axisymmetric mode with $m=0$, and write the embedding relations as
\begin{equation}
r = R(1 + h_1 \cos\vartheta), \qquad
\theta = \vartheta - h_1 \sin\vartheta, \qquad
\phi = \varphi,
\end{equation}
where $h_1 := h_1^0$. The Cartesian description of the deformed shell is then
\begin{equation}
x = R \sin\vartheta \cos\varphi + O(h_1^2), \qquad
x = R \sin\vartheta \sin\varphi + O(h_1^2), \qquad
z = R \cos\vartheta + R h_1 + O(h_1^2),
\end{equation}
and we see that the perturbation describes a translation by $R h_1(t)$ in the $z$-direction.

\subsection{Static tides}
\label{subsec:static}

When the tidal field is idealized as static, so that $e^{\ell m}$ does not depend on time, the solution to Eq.~(\ref{dynamics4}) is simply
\begin{subequations}
\begin{align} 
x^{\ell m} &= \frac{2\ell+1}{4\omega_\ell^2} \Bigl( 2\Gamma - \frac{4\kappa_\ell^2}{\ell-1}
+ \ell - 2 \Bigr) e^{\ell m}, \\
y^{\ell m} &= \frac{2\ell+1}{4\kappa_\ell^2} \Bigl( 2\Gamma + \frac{4\omega_\ell^2}{\ell-1}
+ \ell - 2 \Bigr) e^{\ell m}.
\end{align} 
\end{subequations}
Making the substitution in Eq.~(\ref{q_vs_xy}), we obtain
\begin{equation}
q^{\ell m} = k_\ell\, e^{\ell m}, \qquad
k_\ell = -\frac{\ell(\ell+1)(\ell+2)[ (\ell+2)\Gamma - (\ell+3)]}{8(2\ell+1) \omega_\ell^2 \kappa_\ell^2}
\label{q_static}
\end{equation}
for the dimensionless mass moments. We see that {\it the tidal constant $k_\ell$ is necessarily negative}; the factor $(\ell+2)\Gamma - (\ell+3)$ is larger than $\ell/2$ and therefore positive when $\Gamma \geq 3/2$.

Looking more deeply into Eq.~(\ref{q_static}), we observe that the sign of $k_\ell$ is directly tied to the sign of $\lambda_+ \lambda_- = -\omega_\ell^2 \kappa_\ell^2$. The fact that $\lambda_+ > 0$ while $\lambda_- < 0$ guarantees that $k_\ell < 0$. In other words, {\it the negative sign of the tidal constant is directly associated with the existence of an unstable normal mode.} This connection is interesting, and it echoes a similar relationship that applies to a three-dimensional fluid body, as was previously discussed in Sec.~\ref{subsec:tidal_const}. 

The tidal constants can be given a more explicit expression if we use the eigenvalue equation to write
\begin{equation}
\omega_\ell^2 \kappa_\ell^2 = \frac{(\ell-1)\ell(\ell+1) [ (\ell+2)(2\ell-3) \Gamma - 4(\ell-2)]}{16 (2\ell+1)}.
\end{equation}
Making the substitution in Eq.~(\ref{q_static}), we obtain
\begin{equation}
k_\ell = -\frac{2(\ell+2)}{\ell-1} \frac{ (\ell+2)\Gamma - (\ell+3) }{ (\ell+2)(2\ell-3) \Gamma - 4(\ell-2)}.
\label{love}
\end{equation}
The factor in the denominator, $(\ell+2)(2\ell-3) \Gamma - 4(\ell-2)$, is greater than or equal to $3\ell^2 - \frac{5}{2} \ell - 1$ when $\Gamma \geq 3/2$, and this is greater than or equal to 6 when $\ell \geq 2$; the factor is always positive. Equation (\ref{love}) agrees with Eq.~(\ref{kL_Newton}), which gives the Newtonian limit of the relativistic version of the even-parity tidal constant.

\begin{acknowledgments} 
This work was supported by the Natural Sciences and Engineering Research Council of Canada.  
\end{acknowledgments} 

\appendix
\section{List of coefficients}
\label{app:coefficients}

The coefficients that appear in Eq.~(\ref{Cn_structure}) are given by
\begin{subequations}
\begin{align}
c_1^0 &= \ell^2(\ell+1)^2 \Bigl\{ \bigl[ -144 \Gamma + 108 \bigr] {\cal C}^4
+ \bigl[ 24(\ell^2+\ell+4)\Gamma - 6(6\ell^2+6\ell+7) \bigr] {\cal C}^3
+ 2\bigl[ (\ell^2-2\ell-5)(\ell^2+4\ell-2) \Gamma
\nonumber \\ & \quad \mbox{}
+ (19\ell^2+19\ell-23) \bigr] {\cal C}^2
+ (\ell-1) (\ell+2) \bigl[ -3(\ell^2+\ell-4) \Gamma - 14 \bigr] {\cal C}
+ (\ell-1)(\ell+2) \bigl[ (\ell-1)(\ell+2) \Gamma + 2 \bigr] \Bigr\} \sqrt{1-2 {\cal C}}
\nonumber \\ & \quad \mbox{}
+ \ell^2(\ell+1)^2 \Bigl\{ 72 {\cal C}^4
+ \bigl[ -24(\ell-1)(\ell+2) \Gamma + 6(5\ell^2+5\ell-22) \bigr] {\cal C}^3
+ \bigl[ -4(\ell-1)(\ell+2)(\ell^2+\ell-8) \Gamma
\nonumber \\ & \quad \mbox{}
- 3(13\ell^2+13\ell-32) \bigr] {\cal C}^2
+ (\ell-1)(\ell+2) \bigl[ 2(2\ell^2+2\ell-7) \Gamma + 16 \bigr] {\cal C}
+ (\ell-1)(\ell+2) \bigl[ -(\ell-1)(\ell+2) \Gamma - 2 \bigr] \Bigr\}, \\
c_1^1 &= 2 \ell(\ell+1) \Bigl\{ \bigl[ 12(\ell-1)(\ell+2) \Gamma - 36 \bigr] {\cal C}^3
+ \bigl[ -2(\ell-1)(\ell+2)(2\ell^2+2\ell-7) \Gamma - 3(5\ell^2+5\ell-16) \bigr] {\cal C}^2
\nonumber \\ & \quad \mbox{}
+ 6(\ell-1)(\ell+2)\bigl[ (\ell^2+\ell-3) \Gamma + 2 \bigr] {\cal C}
- 2(\ell-1)(\ell+2) \bigl[ (\ell-1)(\ell+2) \Gamma + 1 \bigr] \Bigr\} \sqrt{1-2{\cal C}}
\nonumber \\ & \quad \mbox{}
+ 4 \ell(\ell+1) \Bigl\{ \bigl[-72\Gamma + 54 \bigr] {\cal C}^4
+ \bigl[ 12(\ell^2+\ell+4) \Gamma - 3(6\ell^2+6\ell+7) \bigr] {\cal C}^3
+ \bigl[ 2(2\ell^4+4\ell^3-12\ell^2-14\ell+11) \Gamma
\nonumber \\ & \quad \mbox{}
+ (19\ell^2+19\ell-23) \bigr] {\cal C}^2
+ (\ell-1)(\ell+2) \bigl[ -(4\ell^2+4\ell-11) \Gamma - 7 \big] {\cal C}
+ (\ell-1)(\ell+2) \bigl[ (\ell-1)(\ell+2) \Gamma + 1 \bigr] \Bigr\}, \\
c_1^2 &= \ell(\ell+1) \biggl\lgroup
\Bigl\{ \bigl[ 36\ell^2+36\ell-48 \bigr] {\cal C}^2
+ \bigl[ -4(\ell-1)^2(\ell+2)^2 \Gamma + (3\ell^4+6\ell^3-21\ell^2-24\ell+24) \big] {\cal C}
\nonumber \\ & \quad \mbox{}
+ (\ell-1)^2(\ell+2)^2 \bigl[ 2\Gamma - 1 \bigr] \Bigr\} \sqrt{1-2{\cal C}}
+ \Bigl\{ -144 {\cal C}^3
+ \bigl[ 24(\ell-1)(\ell+2) \Gamma - (6\ell^2+6\ell-156) \bigr] {\cal C}^2
\nonumber \\ & \quad \mbox{}
+ \bigl[ 4(\ell-1)(\ell+2)(\ell^2+\ell-5) \Gamma - (2\ell^4+4\ell^3-10\ell^2-12\ell+52) \bigr] {\cal C}
+ (\ell-1)^2(\ell+2)^2 \bigl[ -2\Gamma + 1 \bigr] \Bigr\} \biggr\rgroup, \\
c_1^3 &= \Bigl\{ \bigl[ 48(\ell-1)(\ell+2) \Gamma + 36\ell(\ell+1) \bigr] {\cal C}^2
+ \bigl[ -24(\ell-1)(\ell+2) \Gamma - (12\ell^4+24\ell^3+12\ell^2+24) \bigr] {\cal C}
\nonumber \\ & \quad \mbox{}
+ \bigl[ 4(\ell-1)\ell(\ell+1)(\ell+2) \bigr] \Bigr\} \sqrt{1-2{\cal C}}
+ \Bigl\{ \bigl[ -288\Gamma + 216 \bigr] {\cal C}^3
+ \bigl[ -(48\ell^2+48\ell-240) \Gamma + (48\ell^2+48\ell-168) \bigr] {\cal C}^2
\nonumber \\ & \quad \mbox{}
+ \bigl[ 24(\ell-1)(\ell+2) \Gamma + (8\ell^4+16\ell^3-32\ell^2-40\ell+24) \bigr] {\cal C}
- \bigl[ 4(\ell-1)\ell(\ell+1)(\ell+2) \bigr] \Bigr\}, \\
c_1^4 &= -2(\ell-1)\ell(\ell+1)(\ell+2) \bigl( 1 + 3\sqrt{1-2{\cal C}} \bigr), \\
c_1^5 &= 24 {\cal C} \bigl( 1 + 3\sqrt{1-2{\cal C}} \bigr), 
\end{align}
\end{subequations}
\begin{subequations}
\begin{align}
c_2^0 &= \ell(\ell+1)(1-2{\cal C}) \biggl\lgroup
\Bigl\{ \bigl[ 24(\ell-1)(\ell+2) \Gamma - 72\bigr] {\cal C}^3
+ \bigl[ -12(\ell-1)(\ell+2)(\ell^2+\ell-3)\Gamma - 18(\ell^2+\ell-4) \bigr] {\cal C}^2
\nonumber \\ & \quad \mbox{}
+ (\ell-1)(\ell+2) \bigl[ (\ell^4+2\ell^3+5\ell^2+4\ell-24)\Gamma + 6(\ell^2+\ell+2) \bigr] {\cal C}
\nonumber \\ & \quad \mbox{}
- (\ell-1)\ell(\ell+1) (\ell+2) \bigl[ (\ell-1)(\ell+2) \Gamma + 2 \bigr] \Bigr\} \sqrt{1-2 {\cal C}}
\nonumber \\ & \quad \mbox{}
+ \Bigl\{ \bigl[ -288\Gamma + 216 \bigr] {\cal C}^4
+ \bigl[ 96(\ell^2+\ell+1) \Gamma - (108\ell^2+108\ell+12) \bigr] {\cal C}^3
+ \bigl[ 12(\ell^4+2\ell^3-9\ell^2-10\ell+10)\Gamma
\nonumber \\ & \quad \mbox{}
+ 3(\ell^2+\ell+8)(3\ell^2+3\ell-4) \bigr] {\cal C}^2
- (\ell-1)(\ell+2) \bigl[ 2(\ell^2+\ell+4)(\ell^2+\ell-3) \Gamma + 4(2\ell^2+2\ell+3) \bigr] {\cal C}
\nonumber \\ & \quad \mbox{}
+ (\ell-1)\ell(\ell+1)(\ell+2) \bigl[ (\ell-1)(\ell+2) \Gamma + 2 \bigr] \Bigr\} \biggr\rgroup, \\
c_2^1 &= 2 \ell(\ell+1) \biggl\lgroup
\Bigl\{ \bigl[ 12(\ell-1)(\ell+2) \Gamma - 36 \bigr] {\cal C}^3
+ \bigl[ -2(\ell-1)(\ell+2)(2\ell^2+2\ell-7) \Gamma - (15\ell^2+15\ell-48) \bigr] {\cal C}^2
\nonumber \\ & \quad \mbox{}
+ 6(\ell-1)(\ell+2)\bigl[ (\ell^2+\ell-3) \Gamma + 2 \bigr] {\cal C}
- 2(\ell-1)(\ell+2) \bigl[ (\ell-1)(\ell+2) \Gamma + 1 \bigr] \Bigr\} \sqrt{1-2{\cal C}}
\nonumber \\ & \quad \mbox{}
+ \Bigl\{ \bigl[-144\Gamma + 108 \bigr] {\cal C}^4
+ \bigl[ 24(\ell^2+\ell+4) \Gamma - 6(6\ell^2+6\ell+7) \bigr] {\cal C}^3
+ \bigl[ 4(2\ell^4+4\ell^3-12\ell^2-14\ell+11) \Gamma
\nonumber \\ & \quad \mbox{}
+ 2(19\ell^2+19\ell-23) \bigr] {\cal C}^2
- 2(\ell-1)(\ell+2) \bigl[ (4\ell^2+4\ell-11) \Gamma + 7 \big] {\cal C}
+ 2(\ell-1)(\ell+2) \bigl[ (\ell-1)(\ell+2) \Gamma + 1 \bigr] \Bigr\} \biggr\rgroup, \\
c_2^2 &= \Bigl\{ -\bigl[ (96\ell^2+96\ell-480)\Gamma + (72\ell^2+72\ell+216) \bigr] {\cal C}^3 
+ \bigl[ (144\ell^2+144\ell-432)\Gamma
\nonumber \\ & \quad \mbox{}
+ (36\ell^4+72\ell^3+24\ell^2-12\ell+216) \bigr] {\cal C}^2
+ \bigl[ 4(\ell-1)(\ell+2)(\ell^4+2\ell^3-\ell^2-2\ell-12) \Gamma
\nonumber \\ & \quad \mbox{}
- (3\ell^6+9\ell^5+21\ell^4+27\ell^3-12\ell^2-24\ell+48) \big] {\cal C}
+ (\ell-1)^2 \ell(\ell+1)(\ell+2)^2 \bigl[ -2\Gamma + 1 \bigr] \Bigr\} \sqrt{1-2{\cal C}}
\nonumber \\ & \quad \mbox{}
+ \Bigl\{ \bigl[ 576\Gamma - 432 \bigr] {\cal C}^4 
+ \bigl[ (192\ell^2+192\ell-960) \Gamma - (168\ell^2+168\ell-552) \bigr] {\cal C}^3
\nonumber \\ & \quad \mbox{}
+ \bigl[ -(24\ell^4+48\ell^3+168\ell^2+144\ell-528) \Gamma
+  (30\ell^4+60\ell^3+126\ell^2+96\ell-264) \bigr] {\cal C}^2
\nonumber \\ & \quad \mbox{}
+ \bigl[ -4(\ell-1)(\ell+2)(\ell^4+2\ell^3-4\ell^2-5\ell-12) \Gamma
+ (2\ell^6+6\ell^5-18\ell^4-46\ell^3-20\ell^2+4\ell+48) \bigr] {\cal C}
\nonumber \\ & \quad \mbox{}
+ (\ell-1)^2 \ell (\ell+1) (\ell+2)^2 \bigl[ 2\Gamma - 1 \bigr] \Bigr\}, \\
c_2^3 &= 4\Bigl\{ \bigl[ 12(\ell-1)(\ell+2) \Gamma + 9\ell(\ell+1) \bigr] {\cal C}^2
+ \bigl[ -6(\ell-1)(\ell+2) \Gamma - (3\ell^4+6\ell^3+3\ell^2+6) \bigr] {\cal C}
\nonumber \\ & \quad \mbox{}
+ \bigl[ (\ell-1)\ell(\ell+1)(\ell+2) \bigr] \Bigr\} \sqrt{1-2{\cal C}}
+ 4\Bigl\{ \bigl[ -72\Gamma + 54 \bigr] {\cal C}^3
+ \bigl[ -(12\ell^2+12\ell-60) \Gamma + (12\ell^2+12\ell-42) \bigr] {\cal C}^2
\nonumber \\ & \quad \mbox{}
+ \bigl[ 6(\ell-1)(\ell+2) \Gamma + (2\ell^4+4\ell^3-8\ell^2-10\ell+6) \bigr] {\cal C}
- \bigl[ (\ell-1)\ell(\ell+1)(\ell+2) \bigr] \Bigr\}, \\
c_2^4 &= 2 \Bigl\{ -72 {\cal C}^2 + 24 {\cal C} + 3(\ell-1)\ell(\ell+1)(\ell+2) \Bigr\} \sqrt{1-2{\cal C}}
+ \Bigl\{ 96 {\cal C}^2 - 48 {\cal C} + 2(\ell-1)\ell(\ell+1)(\ell+2) \Bigr\}, \\ 
c_2^5 &= 24 {\cal C} \bigl( 1 + 3\sqrt{1-2{\cal C}} \bigr),
\end{align}
\end{subequations}
\begin{subequations}
\begin{align}
c_3^0 &= \ell(\ell+1) \bigl(\sqrt{1-2{\cal C}}-1\bigr)\sqrt{1-2{\cal C}} \biggl\lgroup
\Bigl\{ \bigl[ 144\Gamma - 108 \bigr] {\cal C}^3
+ \bigl[ -36(\ell^2+\ell+2)\Gamma + (36\ell^2+36\ell+78) \bigr] {\cal C}^2
\nonumber \\ & \quad \mbox{}
- \bigl[ 2(2\ell^2+2\ell-1)(\ell^2+\ell-8)\Gamma + (23\ell^2+23\ell+2) \bigr] {\cal C}
+ 2(\ell-1)(\ell+2) \bigl[ (\ell-1)(\ell+2) \Gamma + 1 \bigr] \Bigr\} \sqrt{1-2 {\cal C}}
\nonumber \\ & \quad \mbox{}
+ \Bigl\{ \bigl[ (24\ell^2+24\ell + 96)\Gamma - 180 \bigr] {\cal C}^3
- \bigl[ 8(\ell^4+2\ell^3-\ell+16)\Gamma - 6(\ell^2+\ell+29) \bigr] {\cal C}^2
\nonumber \\ & \quad \mbox{}
+ \bigl[ (8\ell^4+16\ell^3-18\ell^2-26\ell+56) \Gamma + (\ell^2+\ell-50) \bigr] {\cal C}
- 2(\ell-1)(\ell+2) \bigl[ (\ell-1)(\ell+2) \Gamma + 1 \bigr] \Bigr\} \biggr\rgroup, \\
c_3^2 &= 2 (\sqrt{1-2{\cal C}}-1)\sqrt{1-2{\cal C}} \biggl\lgroup
\Bigl\{ \bigl[ 72\Gamma - 54 \bigr] {\cal C}^2 
- \bigl[ 36\Gamma + 21(\ell-1)(\ell+2) \big] {\cal C}
+ \bigl[ \ell(\ell+1)(\ell^2 + \ell + 10) \big] \Bigr\} \sqrt{1-2{\cal C}}
\nonumber \\ & \quad \mbox{}
+ \bigl[ 24(\ell^2+\ell+1) \Gamma + 18(\ell^2+\ell -3) \bigr] {\cal C}^2
- \bigl[ 12(\ell^2+\ell+1) \Gamma
+ 3(2\ell^4+4\ell^3+9\ell^2+7\ell-10) \bigr] {\cal C}
\nonumber \\ & \quad \mbox{}
+ \bigl[ 3\ell(\ell+1)(\ell^2+\ell+2) \bigr] \Bigr\}, \\
c_3^4 &= 24\bigl(2-3{\cal C} + 2\sqrt{1-2{\cal C}} \bigr) \bigl(1-\sqrt{1-{\cal C}} \bigr) \sqrt{1-2{\cal C}}, 
\end{align}
\end{subequations}
\begin{subequations}
\begin{align}
c_4^0 &= \ell^2(\ell+1)^2(1-2{\cal C}) \biggl\lgroup
\Bigl\{ -\bigl[ 288\Gamma - 216 \bigr] {\cal C}^3
+ \bigl[ 96(\ell^2+\ell+1)\Gamma - (108\ell^2+108\ell+84) \bigr] {\cal C}^2
\nonumber \\ & \quad \mbox{}
- \bigl[ 24(2\ell^2+2\ell-1)\Gamma - (9\ell^4+18\ell^3+51\ell^2+42\ell-24) \bigr] {\cal C}
- \bigl[ 2(\ell-1)\ell(\ell+1)(\ell+2) \bigr] \Bigr\} (1-2 {\cal C}) 
\nonumber \\ & \quad \mbox{}
+ \Bigl\{ \bigl[ 144 \bigr] {\cal C}^3
- \bigl[ 48(\ell-1)(\ell+2))\Gamma - (36\ell^2+36\ell-144) \bigr] {\cal C}^2
\nonumber \\ & \quad \mbox{}
+ 12(\ell-1)(\ell+2) \bigl[ 2\Gamma - (\ell^2+\ell+2) \bigr] {\cal C}
+ \bigl[ 4(\ell-1)\ell(\ell+1)(\ell+2) \bigr] \Bigr\} \sqrt{1-2{\cal C}}
\nonumber \\ & \quad \mbox{}
+ \Bigl\{ \bigl[ 288\Gamma - 360 \bigr] {\cal C}^3
- \bigl[ 48(\ell^2+\ell+4)\Gamma - (72\ell^2+72\ell+228) \bigr] {\cal C}^2 
\nonumber \\ & \quad \mbox{}
+ \bigl[ 24(\ell^2+\ell+1)\Gamma + 3(\ell^4+2\ell^3-13\ell^2-14\ell-8) \bigr] {\cal C}
- \bigl[ 2(\ell-1)\ell(\ell+1)(\ell+2) \bigr] \Bigr\} \biggr\rgroup, \\
c_4^1 &= 4\ell(\ell+1) {\cal C} \biggl\lgroup
\Bigl\{ -\bigl[ 96(\ell^2+\ell+1) \Gamma - (54\ell^2+54\ell+180) \bigr] {\cal C}^2
+ \bigl[ 24(\ell^4+2\ell^3+3\ell^2+2\ell+4)\Gamma
\nonumber \\ & \quad \mbox{}
- (18\ell^4+36\ell^3+93\ell^2+75\ell+114) \bigr] {\cal C}
- \bigl[ 12(\ell^4+2\ell^3+\ell^2+2) \Gamma
- 2(\ell^2+\ell+1)(5\ell^2+5\ell+6) \bigr] \Bigr\} (1-2 {\cal C}) 
\nonumber \\ & \quad \mbox{}
+ \Bigl\{ \bigl[ 288\Gamma - 216 \bigr] {\cal C}^3
- \bigl[ 96(\ell^2+\ell+1)\Gamma - 12(12\ell^2+12\ell+1) \bigr] {\cal C}^2
- \bigl[ 24(\ell^2+\ell-5)(\ell^2+\ell-1)\Gamma
\nonumber \\ & \quad \mbox{}
- (6\ell^4+12\ell^3-102\ell^2-108\ell+96) \big] {\cal C} 
+ 4 (\ell-1)(\ell+2) \bigl[ 3(\ell-1)(\ell+2)\Gamma - (\ell^2+\ell-3) \bigr] \Bigr\} \sqrt{1-2{\cal C}} 
\nonumber \\ & \quad \mbox{}
+ \Bigl\{ -\bigl[ 288\Gamma - 216 \bigr] {\cal C}^3
+ \bigl[ 192(\ell^2+\ell+1)\Gamma - (198\ell^2+198\ell+192) \bigr] {\cal C}^2 
- \bigl[ (192\ell^2+192\ell-24)\Gamma
\nonumber \\ & \quad \mbox{}
+ (12\ell^4+24\ell^3+195\ell^2+183\ell+18) \bigr] {\cal C}
+ \bigl[ 24(2\ell^2+2\ell-1)\Gamma  - 6(\ell^4+2\ell^3+8\ell^2+7\ell-2) \bigr] \Bigr\} \biggr\rgroup, \\
c_4^2 &= 4\ell(\ell+1) \biggl\lgroup
\Bigl\{ \bigl[ 48(\ell^2+\ell+1) \Gamma - (72\ell^2+72\ell+132) \bigr] {\cal C}^2
- \bigl[ 4(\ell^4+2\ell^3+3\ell^2+2\ell+10)\Gamma
\nonumber \\ & \quad \mbox{}
- 3(\ell^2+\ell+2)(3\ell^2+3\ell+10) \bigr] {\cal C}
+ \bigl[ 2(\ell-1)^2(\ell+2)^2 \Gamma
- (5\ell^4+10\ell^3+5\ell^2+4) \bigr] \Bigr\} (1-2 {\cal C}) 
\nonumber \\ & \quad \mbox{}
+ \Bigl\{ -\bigl[ 288\Gamma - 360 \bigr] {\cal C}^3
- \bigl[ 24(\ell^2+\ell-8)\Gamma + (42\ell^2+42\ell+180) \bigr] {\cal C}^2
+ \bigl[ 4(\ell-1)(\ell+2)(2\ell^2+2\ell-1)\Gamma
\nonumber \\ & \quad \mbox{}
- (3\ell^4+6\ell^3-23\ell^2-26\ell+4) \big] {\cal C} 
- 2 (\ell-1)^2(\ell+2)^2 \bigl[ 2\Gamma - 1 \bigr] \Bigr\} \sqrt{1-2{\cal C}} 
\nonumber \\ & \quad \mbox{}
+ \Bigl\{ \bigl[ 288\Gamma - 360 \bigr] {\cal C}^3
- \bigl[ 24(\ell^2+\ell+10)\Gamma - (114\ell^2+114\ell+312) \bigr] {\cal C}^2 
- \bigl[ 4(\ell^2+\ell+1)(\ell^2+\ell-8)\Gamma
\nonumber \\ & \quad \mbox{}
+ (6\ell^4+12\ell^3+80\ell^2+74\ell+56) \bigr] {\cal C}
+ \bigl[ 2(\ell-1)^2(\ell+2)^2\Gamma
+ (3\ell^4+6\ell^3+11\ell^2+8\ell-4) \bigr] \Bigr\} \biggr\rgroup, \\
c_4^3 &= 16 \biggl\lgroup
\Bigl\{ -\bigl[ 12(\ell-1)(\ell+2) \Gamma + 9\ell(\ell+1) \bigr] {\cal C}^2
+ \bigl[ 6(\ell-1)(\ell+2) \Gamma + 3(3\ell^4+6\ell^3+7\ell^2+4\ell+2) \bigr] {\cal C}
\nonumber \\ & \quad \mbox{}
+ \bigl[ -6\ell^2(\ell+1)^2 \bigr] \Bigr\} (1-2 {\cal C}) 
+ \Bigl\{ \bigl[ 72\Gamma - 54 \bigr] {\cal C}^3
+ \bigl[ 12(2\ell^2+2\ell-7)\Gamma - (39\ell^2+39\ell-42) \bigr] {\cal C}^2
\nonumber \\ & \quad \mbox{}
- \bigl[ 12(\ell-1)(\ell+2)\Gamma
- (3\ell^4+6\ell^3+15\ell^2+12\ell-12) \big] {\cal C} \Bigr\} \sqrt{1-2{\cal C}} 
\nonumber \\ & \quad \mbox{}
+ \Bigl\{ -\bigl[ 72\Gamma - 54 \bigr] {\cal C}^3
- \bigl[ 12(\ell^2+\ell-5)\Gamma - (48\ell^2+48\ell-42) \bigr] {\cal C}^2 
+ \bigl[ 6(\ell-1)(\ell+2)\Gamma
\nonumber \\ & \quad \mbox{}
- 6(2\ell^4+4\ell^3+6\ell^2+4\ell-1) \bigr] {\cal C}
+ \bigl[ 6 \ell^2(\ell+1)^2 \bigr] \Bigr\} \biggr\rgroup, \\
c_4^4 &= 8\ell(\ell+1) \biggl\lgroup
\Bigl\{ 36{\cal C} - 3(\ell^2+\ell+6) \Bigr\} (1-2{\cal C})
+ \Bigl\{ 12{\cal C} + 2(\ell-1)(\ell+2) \Bigr\} \sqrt{1-2{\cal C}}
+ \Bigl\{ -48{\cal C} + (\ell^2+\ell+22) \Bigr\} \biggr\rgroup, \\
c_4^5 &= 96 {\cal C} \bigl( 1 - \sqrt{1-2{\cal C}} \bigr) \bigl(1 + 3\sqrt{1-2{\cal C}}\bigr), 
\end{align}
\end{subequations}
where ${\cal C} := M/R$ is the shell's compactness. 

\section{Asymptotics of hypergeometric functions}
\label{sec:asymptotics}

We derive the results listed in Eq.~(\ref{F1234_asymp}), involving the functions defined in Eq.~(\ref{F1234}). We rely on the wealth of identities satified by hypergeometric functions, as reviewed, for example, in Chapter 15 of the {\it NIST Handbook of Mathematical Functions} (NIST) \cite{NIST:10}, or Chapter 15 of Abramowitz and Stegun (AS) \cite{abramowitz-stegun:72}.   

We begin with 
\begin{equation}
\FF_2 := \mbox{}_2 F_1(-\ell-1,-\ell-1;-2\ell;1-F),
\end{equation}
which we wish to evaluate when $F \ll 1$. The hypergeometric series [NIST (15.2.1) or AS (15.1.1)] gives
\begin{equation}
\FF_2 = \sum_{n=0}^{\ell+1} (-1)^n \frac{(\ell+1)!^2}{(\ell-n+1)!^2}
\frac{(2\ell-n)!}{(2\ell)!\, n!}  (1-F)^n,
\end{equation}
and in this we insert the binomial expansion for $(1-F)^n$. This produces
\begin{equation} 
\FF_2 = \sum_{n=0}^{\ell+1} \sum_{p=0}^n (-1)^{n+p} 
\frac{(\ell+1)!^2}{(\ell-n+1)!^2} \frac{(2\ell-n)!}{(2\ell)!}
\frac{1}{(n-p)!\, p!}\, F^p, 
\end{equation} 
or 
\begin{equation}
\FF_2 = \frac{(\ell+1)!^2}{(2\ell)!} \sum_{p=0}^{\ell + 1} \frac{(-1)^p}{p!} S_p\, F^p
\end{equation}
after changing the order of summation, where
\begin{equation}
S_p := \sum_{n=p}^{\ell+1} (-1)^n \frac{(2\ell-n)!}{(\ell-n+1)!^2} \frac{1}{(n-p)!}.
\end{equation}
An explicit evaluation of the sum returns $S_0 = 0$, $S_1 = 0$, and $S_2 = 1$, and we arrive at Eq.~(\ref{F2_asymp}).

We obtain Eq.~(\ref{F1_asymp}) directly from Eq.~(\ref{F2_asymp}) by invoking the identity [NIST (15.5.1) or AS (15.2.1)]
\begin{equation}
\frac{d}{dz} \mbox{}_2 F_1(a, b; c; z) = \frac{ab}{c} \mbox{}_2 F_1(a+1, b+1; c+1; z),
\label{F_identity}
\end{equation}
in which we insert $a = b = -(\ell+1)$, $c = -2\ell$, and $z = 1-F$. 

Next we turn to 
\begin{equation}
\FF_4 :=\mbox{}_2 F_1(\ell, \ell; 2\ell+2;1-F). 
\end{equation}
To evaluate this for $F \ll 1$ we invoke [NIST (15.8.10) or AS (15.3.11)]
\begin{align}
\mbox{}_2 F_1(a, b; a+b+2; z) &= \frac{\Gamma(a+b+2)}{\Gamma(a+2) \Gamma(b+2)}
\sum_{n=0}^1 \frac{(a)_n (b)_n}{(-1)_n\, n!} (1-z)^n
\nonumber \\ & \quad \mbox{} 
- \frac{\Gamma(a+b+2)}{\Gamma(a) \Gamma(b)} (z-1)^2 \sum_{n=0}^\infty
\frac{(a+2)_n (b+2)_n}{n! (n+2)!} (1-z)^n
\nonumber \\ & \quad \mbox{} 
\times \bigl[ \ln(1-z) - \psi(n+1) - \psi(n+3) + \psi(a+n+2) + \psi(b+n+2) \bigr],
\end{align}
where $(a)_n := a(a+1) \cdots (a+n-1)$ is the Pochhammer symbol. We insert $a = b = \ell$, $z = 1-F$, keep only the term with $n=0$ in the infinite sum, and use the facts that $\psi(1) = -\gamma$ and $\psi(3) = -\gamma + \frac{3}{2}$. We arrive at Eq.~(\ref{F4_asymp}) after simplification.

We get Eq.~(\ref{F3_asymp}) directly from Eq.~(\ref{F4_asymp}) after making use of Eq.~(\ref{F_identity}), in which we now insert $a= b = \ell$ and $c = 2\ell+2$.

\bibliography{main}

\end{document}